\documentclass[12pt]{article}

\pdfoutput=1

\usepackage{jheppub}

\usepackage{amsfonts}
\usepackage{amsmath,amssymb,amsthm}
\usepackage{graphicx}
\usepackage{color}
\usepackage{setspace}
\usepackage{bbm} 

\def\no{\nonumber}

\newcommand{\p}{\partial}
\newcommand{\dd}{\delta}
\newcommand{\N}{\mathcal{N}}
\newcommand{\ti}{\widetilde}
\newcommand{\wat}{\widehat}

\newcommand{\tauc}{\chi}
\newcommand{\rhota}{\eta}
\newcommand{\Qb}{\mathfrak{s}}

\newcommand{\mm}{m}
\newcommand{\sss}{\mathfrak{s}}

\newcommand{\be}{\begin{equation}}
\newcommand{\ee}{\end{equation}}
\newcommand{\bea}{\begin{eqnarray}}
\newcommand{\eea}{\end{eqnarray}}
\newcommand{\ba}{\begin{aligned}}
\newcommand{\ea}{\end{aligned}}

\newcommand{\lp}{\left(}
\newcommand{\rp}{\right)}
\newcommand{\tr}{\textrm{Tr}}

\newcommand{\wt}{\widetilde}
\newcommand{\IZ}{\mathbb{Z}}

\newcommand{\IH}{\mathbb{H}}

\newcommand{\CH}{\mathcal{H}}

\newcommand{\CL}{\mathcal{L}}

\newcommand{\s}{\sigma}
\def\v{\varphi}

\def\t{\tau}

\def\l{\lambda}
\def\a{\alpha}
\def\zbar{\overline{z}}
\def\={\, = \,}
\def\+{\, + \,}

\newcommand{\cont}{\rho}

\newcommand{\rap}[2]
{\setbox1=\hbox{#1}%
\setbox2=\hbox to\wd1{\hss #2\hss}%
\mbox{\rlap{\box1}\box2}}

\newcommand{\sign}{\mathrm{sign}}
\newcommand{\ZZ}{\mathbb{Z}}

\newcommand{\bC}{\ensuremath{\mathbb{C}}}

\newcommand{\bH}{\ensuremath{\mathbb{H}}}

\newcommand{\bR}{\ensuremath{\mathbb{R}}}

\newcommand{\bZ}{\ensuremath{\mathbb{Z}}}

\newcommand{\scA}{\ensuremath{\mathcal{A}}}

\newcommand{\scC}{\ensuremath{\mathcal{C}}}
\newcommand{\scD}{\ensuremath{\mathcal{D}}}

\newcommand{\scF}{\ensuremath{\mathcal{F}}}

\newcommand{\scH}{\ensuremath{\mathcal{H}}}
\newcommand{\scI}{\ensuremath{\mathcal{I}}}

\newcommand{\scL}{\ensuremath{\mathcal{L}}}

\newcommand{\scP}{\ensuremath{\mathcal{P}}}

\newcommand{\scR}{\ensuremath{\mathcal{R}}}

\numberwithin{equation}{section}       

\title{Localization of supersymmetric field theories on non-compact hyperbolic three-manifolds}

\author{Benjamin Assel, Dario Martelli, Sameer Murthy, and Daisuke Yokoyama}
\affiliation{Department of Mathematics, King's College London \\
The Strand, London WC2R 2LS, U.K.}

\abstract{We study  supersymmetric gauge theories with an R-symmetry, defined on {\it non-compact}, hyperbolic, Riemannian three-manifolds, focusing on the 
case of a supersymmetry-preserving quotient of Euclidean AdS$_{3}$. 
We compute  the exact partition function in these theories, using the method of localization, thus reducing the problem to the computation
of one-loop determinants around a supersymmetric locus. We evaluate the one-loop determinants employing three different techniques: an 
index theorem, the method of pairing of eigenvalues, and the heat kernel method. Along the way, we discuss aspects of supersymmetry in 
manifolds with a conformal boundary, including supersymmetric actions and boundary conditions. }

\emailAdd{benjamin.assel@gmail.com, dario.martelli@kcl.ac.uk, \\
sameer.murthy@kcl.ac.uk, daisuke.yokoyama@kcl.ac.uk}

\begin{document}

\vspace*{-2cm} 
\begin{flushright}
{\tt  KCL-MTH-16-07 } 
\end{flushright}

\maketitle

\section{Introduction }
\label{sec:Introduction}

Supersymmetric field theories in flat space have been studied for decades, as a formidable arena  for advancing 
our theoretical understanding of quantum field theories. 
A systematic study of supersymmetric field  theories defined on curved manifolds was 
initiated in~\cite{Festuccia:2011ws}, where it was pointed out that a convenient  viewpoint on these theories 
is to construct them as a rigid limit of certain off-shell supergravities. The method of supersymmetric localization 
allows us to obtain exact results for supersymmetric observables on such curved 
manifolds~\cite{Nekrasov:2002qd,Pestun:2007rz}. 
So far attention has been devoted to compact curved Riemannian manifolds, where the compactness helps the convergence properties of the path integral and simplifies the analysis of the saddle point loci. Following \cite{Pestun:2007rz}, a plethora of localization computations on compact Riemannian manifolds have been performed, in dimensions ranging from one to seven. See  
\cite{Kapustin:2009kz,Benini:2012ui,Doroud:2012xw,Kallen:2012va,Hama:2012bg,Hori:2014tda,Minahan:2015jta} for a representative list of references.

In this paper we will turn attention to supersymmetric gauge theories defined on \emph{non-compact} curved 
Riemannian manifolds. 
Some aspects of such theories have been discussed in the literature before, in the seminal work on the Omega background for 4d N = 2 theories \cite{Nekrasov:2002qd, Nekrasov:2003rj}, and in the context of AdS2 geometries~\cite{Dabholkar:2010uh, Dabholkar:2011ec, Gupta:2012cy, Murthy:2013xpa, Dabholkar:2014wpa, Murthy:2015yfa, Gupta:2015gga, Murthy:2015zzy, Aharony:2015hix}. 
A motivation for considering these backgrounds is that they provide a natural framework for attempting holographic 
constructions~\cite{Aharony:2015zea}. The paper~\cite{Aharony:2015hix} discussed the supersymmetry algebras 
preserved in~AdS$_p\times S^q$ backgrounds  for various~$p,q$, and, in some (free) cases, the supersymmetric 
Lagrangians and boundary conditions on the fields. It raises the interesting question of whether it may be possible to 
obtain exact results for supersymmetric field theories in these backgrounds. In this paper we discuss the computation 
of the exact partition function for a  broad class of ${\cal N}=2$ three-dimensional supersymmetric gauge theories (with matter),
defined on  a quotient of Euclidean AdS$_3$ preserving supersymmetry.

As we shall see, the  challenges that arise  in carrying out this computation concern the presence
of a conformal boundary at infinity. In particular, this will lead to an interesting interplay between supersymmetry, boundary conditions, and boundary actions. We will embrace the point of view advocated in
\cite{DiPietro:2015zia},  namely we will add ``compensating'' boundary terms to the standard supersymmetric actions, such that their combined supersymmetry variations vanish independently of specific boundary conditions. See also the recent \cite{Aprile:2016gvn} for related discussions.

 Supersymmetry in compact curved manifolds with boundaries has been considered in previous works involving localization, see \emph{e.g.} \cite{Sugishita:2013jca,Honda:2013uca,Hori:2013ika,Yoshida:2014ssa}. However,  non-compactness of the space, equipped with a negatively curved metric, introduces a number of novelties. We will deal with the additional complications borrowing  ideas from holographic renormalization, although in this paper we will not discuss any concrete holographic interpretation of our results. 

After carrying out this preliminary analysis, we will see that the standard localization argument will go through, enabling us to reduce the computation of the exact partition function (and certain supersymmetric Wilson loops) of the theories of interest to the evaluation of one-loop super-determinants around a BPS locus. Of course, computing one-loop determinants on non-compact spaces is \emph{per se} a non-trivial problem. The main tool that has been used so far  to perform these computations is the method of the heat kernel; this was developed in the 90's in a series of papers by Camporesi and Higuchi 
\cite{Camporesi:1992tm,Camporesi:1994ga,Camporesi:1995fb} and 
extended to  (super)gravity in the background of AdS$_3$ in \cite{Giombi:2008vd,David:2009xg}. We should stress that the technique of the heat kernel, is \emph{not manifestly supersymmetric}, because it treats fermionic and bosonic fields independently. In addition to this intrinsic problem, it must also be emphasized that the background geometries considered in \cite{Giombi:2008vd,David:2009xg}, were \emph{thermal} quotients of AdS$_3$ and therefore manifestly  not supersymmetric\footnote{As we shall see in the next section, to preserve supersymmetry in a quotient of hyperbolic space, it is necessary to switch on a specific background R-symmetry gauge field.}, so that we could not compare directly our results with those presented in  \cite{Giombi:2008vd,David:2009xg}. In this paper we will propose some modifications of the heat kernel method, leading to a result for the one-loop super-determinant, that we will also derive employing two other methods.

The first method is formal, and consists in utilising  a  version of the fixed point 
theorem of Atiyah and Bott~\cite{Atiyah:1984px}.  This method has been applied 
in~\cite{Gomis:2011pf,Drukker:2012sr} to the calculation of one-loop determinant on spheres, and 
in~\cite{Murthy:2015yfa} in the context of AdS$_2$. 
As we shall see later in the paper, this gadget will output a result that receives the contribution from 
the ``center'' of Euclidean AdS$_3$, which is the fixed point of a certain symmetry acting on it. 
A proper treatment of this method would require a rigorous formulation of the index theorem 
in the non-compact spaces under consideration. 
Here we will  simply assume that the boundary conditions we will require on our fields ensure that the 
index theorem holds. It would be interesting to make this mathematically rigorous.

The second method is that of the (un)pairing of modes \cite{Hama:2011ea, Alday:2013lba}, that can be conveniently implemented through a set of twisted variables, analogous to those 
considered in \cite{Closset:2013sxa,Closset:2014uda}. This was previously used to compute one-loop determinant in compact spaces, but we will see that since this is based on a local analysis of the modes contributing to the determinants, it goes through for the case of interest, albeit with certain technical caveats that we will explain in 
Section~\ref{ssec:UnpairedEigenmodes}. We will show that there are large cancellations between bosonic and fermionic modes, and the remaining ``unpaired'' modes obey simple first order equations, that can be solved explicitly for their eigenvalues. These unpaired modes are not square integrable, but we need to assume that they contribute to the determinant in order for the result to be consistent with the other two methods. We show in Section 5.3 that this implies an asymmetric treatment of the fields $\phi$ and $\tilde \phi$, which are Hermitian conjugates in the Lorentzian theory. We believe that this could be justified by a first-principles treatment of the Euclidean supersymmetric theory. This is reminiscent of a similar phenomenon described in \cite{Keeler:2016wko}, wherein a non-standard analytic continuation from the Lorentzian theory is used to justify that non-square-integrable modes contribute to the AdS$_2$ functional determinant.

The rest of this paper is organised as follows. In Section~\ref{sec:Background} we will briefly describe the background geometry. In Section~\ref{sec:Supersymmetry} we write out the supersymmetry transformations and Lagrangians of the field theories of interest. We also introduce our twisted variables, that will be used extensively in the following sections. In Section~\ref{sec:Asymptotics} we discuss the asymptotic behaviour of the fields and actions and set up the localization computation of the partition function. Section~\ref{sec:OneLoopDet} contains the computations of the one-loop determinants around the localization locus, using three different methods. Our results are summarized in Section~\ref{sec:summary} and we conclude the paper with a discussion in 
Section~\ref{sec:Discussion}. Three appendices contain useful identities and some intermediate computations.

\section{Background geometry }
\label{sec:Background}

The main focus of this paper will be the study of certain supersymmetric gauge theories in a background geometry comprising a quotient of hyberbolic space, equipped 
with the standard negatively curved Einstein metric. 
We begin with the hyperbolic space~$\mathbb{H}^3$, with metric given by 
  \bea
ds^2 (\mathbb{H}^3)  & = & L^2 \left( \cosh^2 \rhota \, d\tauc^2 + d\rhota^2 + \sinh^2\rhota \, d\varphi^2\right) ~, 
\label{hypglobalmodes0}
\eea
where $\rhota \in [0, +\infty)$ and $\varphi \sim \varphi + 2\pi$. The coordinate  $\tauc \in \mathbb{R}$  and one can think of this as the analytic continuation of a time coordinate  
$\tauc_l  = i \tauc$ in AdS$_3$ space-time in Lorentzian signature. In our conventions this metric has constant negative curvature 
with Ricci scalar given by $R=-6/L^2$.

We regard the metric (\ref{hypglobalmodes0}) as a background 
solving the Killing spinor equations of three-dimensional Euclidean new minimal supergravity  \cite{Closset:2012ru}, namely 
\begin{align}
\nabla_{\mu} \zeta - i A_{\mu} \zeta &= - \frac{H}{2} \gamma_\mu \zeta - i V_\mu \zeta - \frac 12 \epsilon_{\mu\nu\rho} V^\nu \gamma^\rho \zeta ~,\\
\nabla_{\mu} \ti\zeta + i A_{\mu} \ti\zeta &= - \frac{H}{2} \gamma_\mu \ti\zeta + i V_\mu \ti\zeta + \frac 12 \epsilon_{\mu\nu\rho} V^\nu \gamma^\rho \ti\zeta \, , 
\end{align}
where $\zeta, \ti\zeta$  are complex two-component spinors and $A_\mu, V_\mu$ and $H$ are specific background fields. In particular, choosing the orthonormal frame
\begin{align}
e^1 &= L \, d\rhota \, , \quad e^2 = L\, \cosh\rhota \, d\tauc \, , \quad e^3 = L \sinh \rhota \, d\varphi \, ,
\label{mainframe}
\end{align}
and  $A_\mu=V_\mu=0$, and $H=\frac 1L$, we find the four Killing spinors 
\begin{align}
\zeta_+ &=  \ti\zeta_+  = \frac{1}{\sqrt 2}\,  e^{\frac{i \varphi}{2} + \frac{\tauc}{2}} \binom{e^{- \frac{\rhota}{2}}}{e^{\frac{\rhota}{2}} } \, , \quad
\zeta_- = \ti\zeta_- = \frac{1}{\sqrt 2}\,  e^{-\frac{i \varphi}{2}- \frac{\tauc}{2}} \binom{e^{- \frac{\rhota}{2}}  }{- e^{\frac{\rhota}{2}} } \, .
\label{globspinors}
\end{align}
The spinors~$\zeta_\pm$ have R-charge $+1$ and the spinors $\ti\zeta_\pm$ have  R-charge $-1$.

Next, we consider a quotient of this space, where we compactify the $\tauc$ direction,  and perform the quotient of $\bH^3$ by the identification
\begin{align}
(\tauc, \varphi) \sim   (\tauc + 2\pi\tau_2, \varphi + 2\pi \tau_1)  \, ,
\label{quotient}
\end{align}
with $\tau_1, \tau_2$ real and $\tau_2 > 0$. We define $\tau = \tau_1 + i \tau_2$ and denote the quotient space  as $\bH^3/\tau\bZ \equiv \bH^3_\tau$.
As can be seen from (\ref{globspinors}) the spinors are not well-defined in this case, and therefore supersymmetry is broken.
In order to preserve two supercharges parameterized by well-defined spinors $\zeta_+ \equiv \zeta$ and $\ti\zeta_- \equiv \ti\zeta$, we need to turn on the background gauge field
\begin{align}
A &= \lp \frac i2 - \frac{\tau_1}{2\tau_2} \rp  d\tauc \ = \ - \frac{\bar\tau}{2\tau_2} d\tauc  \, .
\label{BkgrdGaugeField}
\end{align}
The Killing spinors which are preserved by the quotient are then
\begin{align}
\zeta_+ &=  \frac{1}{\sqrt 2}\,  e^{\frac{i}{2} \lp \varphi - \frac{\tau_1}{\tau_2}\tauc \rp } \binom{e^{- \frac{\rhota}{2}}}{e^{\frac{\rhota}{2}} }  \ \equiv \ \zeta   \, , \no\\
 \ti\zeta_- &= \frac{1}{\sqrt 2}\,  e^{-\frac{i}{2} \lp \varphi - \frac{\tau_1}{\tau_2}\tauc \rp} \binom{e^{- \frac{\rhota}{2}}  }{- e^{\frac{\rhota}{2}} }  \ \equiv \ \ti\zeta   \, .
 \label{quotspinors}
\end{align}
Note that the spinors are anti-periodic around the $\varphi$-circle, which is the correct behavior for spinors around a contractible circle. They obey $\zeta\ti\zeta = -\ti\zeta\zeta=1$.
Throughout the paper we regard the Killing spinors $\zeta, \ti\zeta$ as commuting (Grassmann-even) spinors.

We now construct various bilinears with these spinors, which will be useful in the remainder of the paper. In particular, we have the  
three complex one-forms 
\begin{align}
K  &=  \zeta \gamma_a \ti\zeta \, e^a  ~,\qquad   P    \ = \ \zeta \gamma_a \zeta \, e^a  ~, \qquad 
\ti P \  = \  \ti\zeta \gamma_a \ti\zeta \, e^a ~,
\end{align}
that in the frame (\ref{mainframe}) read 
\begin{align}
K &=    \cosh\rhota \, e^2 - i \sinh\rhota e^3  \, , \\
P &= e^{i \lp \varphi - \frac{\tau_1}{\tau_2}\tauc \rp } \lp e^1 - \sinh\rhota \, e^2 + i \cosh\rhota \, e^3  \rp  \, , \\
\ti P &=    - \, e^{-i \lp \varphi - \frac{\tau_1}{\tau_2}\tauc \rp } \lp e^1 + \sinh\rhota \, e^2 - i \cosh\rhota \, e^3  \rp \, .
\end{align}
The one-forms $K, P$ and $\ti P$ carry R-charges $0, 2$ and $-2$ respectively. 
The dual complex vector fields read 
\begin{align}
K^\mu \p_\mu &= \frac 1L \lp \p_\tauc - i \p_\varphi \rp  \, , \\
P^\mu \p_\mu &=  \frac 1L  e^{i \lp \varphi - \frac{\tau_1}{\tau_2}\tauc \rp } \lp  \p_\rhota - \tanh\rhota \, \p_\tauc + i \coth\rhota \, \p_\varphi   \rp  \, , \\
\ti P^\mu \p_\mu &=   -\, \frac 1L  e^{-i \lp \varphi - \frac{\tau_1}{\tau_2}\tauc \rp } \lp \p_\rhota + \tanh\rhota \, \p_\tauc - i \coth\rhota \, \p_\varphi   \rp \, ,
\end{align}
and give rise to six independent real Killing vectors\footnote{A generic background of new minimal supergravity with two Killing spinors of opposite R-charge admits only the complex Killing vector $K$
\cite{Closset:2012ru}. However, in the special background studied in this paper, $P$ and $\ti P$  also give rise to Killing vectors.}. 
In particular, these  generate the $\mathfrak{sl}(2, \bC)$ algebra of isometries of $\bH^3$ given by
\begin{align}
\frac 12 \, [K, P] = \frac 1L P \, , \quad  \frac 12 \,  [K, \ti P] = -\frac 1L \ti P \, , \quad \frac 12 \,  [\ti P,  P] = \frac 2L K \, .
\end{align}
Moreover, denoting by $\scL_X = X^\mu D_\mu = X^\mu (\nabla_\mu - i q_R A_\mu)$  the R-symmetry covariant derivative\footnote{Later, when we introduce gauge symmetry, the derivative $D_\mu$ will be also covariant with respect to the gauge connection.}  along a  vector $X^\mu$, 
and $\scL_{[X,Y]} = \frac{1}{2} \lp  \scL_X \scL_Y - \scL_Y \scL_X \rp $, 
the following relations hold 
\begin{align}
\scL_{[K,P]} =  \frac 1L \scL_P \, , \quad  \scL_{[K, \ti P]} = -\frac 1L \scL_{\ti P}
\, , \quad  \scL_{[ \ti P ,  P]} = \frac 2L \scL_{K} \, .
\label{LieDerivativeRel}
\end{align}
Note that an arbitrary complex vector $X^\mu$ can be decomposed in the $(K^\mu, P^\mu, \ti P^\mu)$ basis as
 \begin{align}
 X^\mu &= (K^\nu X_\nu) K^\mu - \frac 12 (\ti P^\nu X_\nu) P^\mu - \frac 12 (P^\nu X_\nu) \ti P^\mu \,.
 \end{align}
Further useful relations among $(K^\mu, P^\mu, \ti P^\mu)$ are given in Appendix~\ref{conventions:appendix}.

We close this section by briefly discussing the almost contact structure and the associated transversely holomorphic foliation \cite{Closset:2012ru}.
In any generic background preserving at least one Killing spinor $\zeta$, 
\be
\cont_\mu = \frac{1}{\zeta^\dagger \zeta} \zeta^\dagger \gamma_\mu \zeta
\ee
defines an almost contact one-form\footnote{This is denoted $\eta_\mu$ in \cite{Closset:2012ru}.}, normalised so that $\cont^\mu\cont_\mu=1$. Together with the 
Hodge dual two-form $\Phi=*\cont$, these define an almost contact metric structure on the three-dimensional space. In the case of interest we 
may focus  on either of the two spinors preserved by our background. On picking, without loss of generality, $\zeta=\zeta_+$, we find 
\bea
\cont & = &  - L \left( \tanh \eta d\eta + d\tauc\right) ~,
\eea
which turns out to be closed, \emph{i.e.}~$d\cont=0$.  One can  define a local coordinate $\varsigma$ such that $\cont=d\varsigma$, 
given by $e^{-\varsigma/L}  =  e^\chi \cosh\eta$,
and a  complex coordinate $\mathfrak{z}$ on the transversally holomorphic foliation given by
$e^{\mathfrak{z}}  =  e^{i\varphi - \chi}\tanh \eta$.
In these coordinates the metric takes the  form 
\bea
ds^2 & = & (d\varsigma + h d\mathfrak{z} + \bar h \bar{\mathfrak{z}})^2 + c^2 d\mathfrak{z} d\bar{\mathfrak{z}}~,
\label{canonmetric}
\eea
where $h=\bar h = 0$ and $c^2=L^2\sinh^2\eta$. The transverse two-form is $\Phi=-\tfrac{i}{2}c^2d\mathfrak{z}\wedge d\bar{\mathfrak{z}}$. Notice however that the frame~(\ref{mainframe}) differs from the canonical frame associated 
to~(\ref{canonmetric}), and consequently our spinors (\ref{quotspinors}) do not take the form 
given in Equation~(4.21) of~\cite{Closset:2012ru}.


\section{Supersymmetry transformations and actions }
\label{sec:Supersymmetry}

 In this section we provide the supersymmetry transformations of the vector and chiral multiplets and the supersymmetric actions of $\mathcal{N}=2$ supersymmetric gauge theories on~$\bH^3_\tau$, extracted from \cite{Closset:2012ru}. 
Throughout this paper we work in Euclidean signature. 
 We show that the Yang-Mills vector multiplet Lagrangian and chiral multiplet Lagrangian are $Q$-exact for a certain supercharge $Q$, up to total derivatives that we will discuss carefully.

 To address the question of the supersymmetry of the action in the presence of a boundary, 
 we  introduce a radial cut-off at a finite distance from the center of $\bH^3_\tau$ and add boundary terms, which ensure that supersymmetry is preserved on the compact space, independently of a choice of boundary conditions for the fields. The supersymmetric action is then obtained by sending the radial cutoff to infinity. This analysis will be used in later sections when we evaluate the various actions, to implement a supersymmetric holographic renormalization method.

\medskip

We begin by providing the supersymmetry algebra, generated by the two complex supercharges $\delta_\zeta$ and $\delta_{\ti\zeta}$, parametrized by the Killing spinors $\zeta$ of R-charge $+1$ and $\ti\zeta$ of R-charge $-1$.  The super-algebra is
\begin{align}
&  \{ \delta_{\zeta} , \delta_{\zeta} \} = \{ \delta_{\ti\zeta} , \delta_{\ti\zeta} \}  = 0 \,, \no\\
& \{ \delta_{\zeta} , \delta_{\ti\zeta} \} = - 2i \, \mathsf{L}'_K  + 2 \delta_{\rm gauge}(\sigma + i K^{\mu}\scA_\mu ) + 2 i  \frac{q_R}{L} \,,
\label{susyalgebra}
\end{align}
where $\mathsf{L}'_K = \mathsf{L}_K -  i q_R K^{\mu} (A_\mu - \frac 12 V_\mu)$, with $\mathsf{L}_K$  the Lie derivative along the vector 
$K^\mu$, $q_R$ is the R-charge of the field, $\scA_\mu$ is the gauge field and $\sigma$ is the real scalar in the vector multiplet (see below). The variation~$\delta_{\rm gauge}(\Lambda)$ denotes
 the infinitesimal gauge transformation with gauge parameter $\Lambda$. For a $U(1)$ gauge group, this is simply $\delta_{\rm gauge}
 (\Lambda)  = i w \Lambda$ acting on a matter field of charge $w$, or $\delta_{\rm gauge}(\Lambda) = i [\Lambda, . \,]$ acting on an adjoint
  valued matter field. The gauge field also has, of course,  inhomogeneous terms in the gauge transformation. 

 When there is a flavor symmetry $U(1)_F$, the super-algebra is deformed by a central charge 
\be
 \{ \delta_{\zeta} , \delta_{\ti\zeta} \} = - 2i \, \mathsf{L}'_K  + 2 \delta_{\rm gauge}(\sigma + i K^{\mu}\scA_\mu ) + 2 i  \frac{q_R}{L} + 2 i q_F (m + iK^\mu v_\mu) \,,
\label{susyalgebrawithmass}
\ee
 where $v^\mu$ is a flavor background gauge field and $m$ is the real mass deformation, introduced by weakly gauging $U(1)_F$, and $q_F$ is the flavor charge.

We now provide the supersymmetry transformations and supersymmetric actions.

\subsection{Vector multiplet}
\label{ssec:SusyVec}

We consider a gauge group $G$ and the associated vector multiplet $(\scA_\mu, \sigma, \lambda, \ti\lambda, D)$ valued in the adjoint representation of the gauge algebra. Since we are in Euclidean signature the bosonic fields $\scA_\mu, \sigma, D$ are taken to be complex and the spinors $\lambda, \ti\lambda$ to be independent. When discussing partition functions we will have to choose reality conditions reducing the number of real independent fields to its canonical value, however when discussing the supersymmetries we do not impose such constraints.

The supersymmetry transformations parametrized by the spinors $\zeta$ and $\ti\zeta$ are given by \cite{Closset:2012ru} \footnote{The variation with respect to the supersymmetry $\dd_\zeta$ are obtained by setting $\ti\zeta =0$ and vice-versa. We thus have $\dd = \dd_\zeta + \dd_{\ti\zeta}$.}
\begin{align}
\dd \scA_{\mu}  &= \ - i \lp \zeta \gamma_{\mu} \ti\lambda + \ti\zeta \gamma_{\mu} \lambda \rp \,, \no\\
\dd \sigma &= \  - \zeta \ti\lambda + \ti\zeta \lambda  \,, \no\\
\dd\lambda  &= \  - \frac i2 \epsilon^{\mu\nu\rho} \gamma_\rho \zeta \scF_{\mu\nu} + i  \zeta (D + \sigma H) - \gamma^\mu \zeta (i D_\mu \sigma - V_\mu\sigma)  \,, \label{SusyTransfoVec} \\
\dd\ti\lambda  &= \  - \frac i2 \epsilon^{\mu\nu\rho} \gamma_\rho \ti\zeta \scF_{\mu\nu} - i  \ti\zeta (D + \sigma H) + \gamma^\mu \ti\zeta (i D_\mu \sigma + V_\mu\sigma) \,,  \no \\
\dd D &= \  D_\mu \lp \zeta \gamma^{\mu} \ti\lambda - \ti\zeta \gamma^{\mu} \lambda \rp  - i \, V_{\mu} \lp \zeta \gamma^{\mu} \ti\lambda + \ti\zeta \gamma^{\mu} \lambda \rp   - [\sigma, \zeta \ti\lambda] -   [\sigma, \ti\zeta\lambda] - H \lp \zeta \ti\lambda - \ti\zeta \lambda \rp \,, \no
\end{align} 
with $D_\mu = \nabla_\mu - i q_R \lp A_\mu - \frac 12 V_\mu \rp - i \scA_\mu$ and $\scF_{\mu\nu} = \p_\mu\scA_\nu - \p_\nu\scA_\mu - i [\scA_\mu, \scA_\nu]$. The R-charges $q_R$ of the fields $(\scA_\mu,\sigma, \lambda, \ti\lambda, D)$ are $(0,0,1,-1,0)$ respectively.

The supersymmetric Yang-Mills Lagrangian is 
\begin{align}
\scL_{\rm YM} = \frac{1}{2 e^2} \tr  \, \Big[ \,  & \frac 12 \scF^{\mu\nu} \scF_{\mu\nu} + D^\mu \sigma D_\mu \sigma + i \sigma \epsilon^{\mu\nu\rho} V_\mu \scF_{\nu\rho} - V^\mu V_\mu \sigma^2 - (D+ \sigma H)^2   \no\\
& - i \ti\lambda \gamma^\mu (D_\mu + \frac i2 V_\mu) \lambda  - i \lambda \gamma^\mu (D_\mu + \frac i2 V_\mu) \ti \lambda  - 2i \ti\lambda [\sigma,\lambda]  + i H \ti\lambda \lambda \,   \Big]  \, ,
\end{align}
where $\frac{1}{e^2}$ denotes the Yang-Mills coupling.  This Lagrangian is invariant under the above supersymmetry transformations up to boundary terms, which we will discuss in some detail in the following.

On the $\bH^3$ or $\bH^3_\tau$ backgrounds described in Section~\ref{sec:Background}, the Lagrangian reduces to
\begin{align}
\scL_{\rm YM} = \frac{1}{2 e^2} \tr  \, \Big[ \,  & \frac 12 \scF^{\mu\nu} \scF_{\mu\nu} + D^\mu \sigma D_\mu \sigma  - \lp D+ \frac{\sigma}{L} \rp^2   \no\\
& - i \ti\lambda \gamma^\mu D_\mu  \lambda  - i \lambda \gamma^\mu D_\mu \ti \lambda  - 2 i \ti\lambda [\sigma,\lambda]  + \frac iL \ti\lambda \lambda \,   \Big]  \, .
\label{LagrangianVec}
\end{align}
The reality conditions which make the bosonic action positive definite are $\scA$ and $\sigma$ hermitian and $D' \equiv D + \frac{\sigma}{L}$ anti-hermitian. 
 However, in our analysis of asymptotic boundary conditions in Section~\ref{sssec:BdryVec}, we will find natural to impose different reality conditions at infinity. For instance in a Chern-Simons theory we will be led to consider the gauge field component $\scA_\chi$ as purely imaginary asymptotically. 
 These reality conditions can be associated to the theory obtained by Wick rotation from Lorentzian signature.
Defining the combinations
\be
\scA_{z} = \frac 12 \lp \scA_{\varphi} - i \scA_{\chi} \rp \,, \quad   \scA_{\bar z} = \frac 12 \lp \scA_{\varphi} + i \scA_{\chi} \rp \,,
\label{AzAzbar}
\ee
we can then choose $\scA_z$ and $\scA_{\bar z}$ independent and hermitian. The reality conditions on the fields are then 
\be\ba
& \scA_{\mu}{}^{\dagger} = \scA_{\mu} \,, \quad   \mu = \eta,  z , \bar z \,,   \cr
& \sigma^{\dagger} = \sigma \,, \quad \lp D + \frac{\sigma}{L} \rp^\dagger = - \lp D + \frac{\sigma}{L} \rp \,.
\label{VecRealCond}
\ea\ee

In the pure Yang-Mills theory, our analysis in Section~\ref{sssec:BdryVec} will allow for different asymptotics, so we will restrain ourselves from giving an explicit reality condition for this case. In general one should consider a complex gauge field and path integrate over a middle-dimensional slice in this complexified space.

The reality conditions that we choose for the fermionic fields are more easily described in terms of the twisted fields that we introduce in the next section.

\subsubsection{Twisted fields}
\label{sssec:TVarVec}

It will be convenient to define the so-called {\it twisted variables} or {\it twisted fields}, which  re-express all the fields in the multiplet in terms of Grassmann-even and odd scalars. For bosons we define
\begin{align}
 X^+ &= - i P^\mu \scA_\mu \,, \quad X^- = - i \ti P^\mu \scA_\mu \,, \quad X^0 = i K^\mu \scA_\mu  - \sigma \,, \quad \Sigma =   i K^\mu \scA_\mu  + \sigma \,.
\end{align}
For fermions we define the Grassmann-odd scalar fields
\begin{align}
\Lambda^+ &= \zeta \lambda \,, \quad \Lambda^- = -\ti\zeta \ti\lambda \,, \quad \Lambda^0 = \zeta\ti\lambda - \ti\zeta\lambda \,, \quad \Theta = i\lp \zeta\ti\lambda + \ti\zeta\lambda \rp \,.
\end{align}
This map can be inverted as follows
\begin{align}
\scA_\mu &= - \frac i2 (X^0 + \Sigma) K^\mu - \frac i2 X^- P^\mu - \frac i2 X^+ \ti P^\mu \,, \quad \sigma = \frac 12 \lp\Sigma - X^0\rp \,, \no\\
\lambda &=  \frac 12 \lp  \Lambda^0 + i \Theta \rp \zeta + \Lambda^+ \ti\zeta \,, \quad  \ti\lambda = \frac 12 \lp  \Lambda^0 - i \Theta \rp \ti\zeta + \Lambda^- \zeta \,.
\end{align}
The supersymmetry transformations in terms of the twisted variables are given in 
Appendix~\ref{app:SusyTwistedFields}.
In later sections  we will use the supercharge $Q =\frac 12 \delta_{\zeta} + \frac 12 \delta_{\ti\zeta}$ to perform the localization computations. The reason for choosing this basis is that the fields are organized in pairs, each pair comprising one field and its $Q$-superpartner:
\begin{align}
Q X^+ &= \Lambda^+ \,, \quad Q X^- = \Lambda^- \,, \quad Q X^0 = \Lambda^0 \,, \quad 
Q \Sigma = 0 \,, \; \text{and} \quad Q \Theta = D^0 \,,
\end{align}
with $D^0 = D + (3\Sigma + X^0)\frac{1}{2L} - \frac 12 [X^+, X^-]- \frac 12 \wat\scL_P X^- + \frac 12 \wat\scL_{\ti P} X^+$. The hats on the derivatives, as in $\wat\scL_P = P^\mu \wat D_\mu$, denote the fact that the derivatives are not covariant with respect to the gauge field $\scA$, but only with respect to the R-symmetry connection~$A$.
The supersymmetry transformations of the fields $\Lambda^{0,\pm}$ and $D^0$ can be worked out from the super-algebra ($Q \Lambda^0 = Q^2 X^0$, etc).

We can express the reality conditions \eqref{VecRealCond} for bosons -- and define reality conditions on fermions -- in terms of the twisted fields, 
\be\ba
& (X^0)^\dagger = X^0 \,, \quad \Sigma^\dagger = \Sigma \,, \quad (X^\pm)^\dagger = X^\mp \,, \cr
& (\Lambda^0)^\dagger = \Lambda^0 \,, \quad \Theta^\dagger = \Theta \,, \quad (\Lambda^\pm)^\dagger = \Lambda^\mp \,.
\ea\ee
Note that~\eqref{VecRealCond} provides natural reality conditions on the twisted fields.

\subsubsection{Q-exact action}
\label{sssec:QactionVec}

The Lagrangian $\scL_{\rm YM}$ \eqref{LagrangianVec} can be written as a $\dd_{\ti\zeta}$-exact term or as a $\dd_{\zeta}$-exact term, up to total derivatives:
\begin{align}
& V^{(1)}_{\rm vec} = \dd_{\zeta} \, \tr \, \frac{1}{2 e^2} \lp  \ti\lambda \lambda  + 2i D \sigma \rp \,, \label{deltaV1vec} \\
& \dd_{\ti\zeta} V^{(1)}_{\rm vec} = \scL_{\rm vec} + \frac{1}{ e^2} \tr   \, D^{\mu} \Big( -\sigma D_{\mu} \sigma - i \sigma \scF_{\mu\nu} K^\nu + \sigma D K_\mu  - \frac i2 \ti\lambda \gamma_\mu \lambda -  i (\ti\lambda \gamma_\mu \zeta)(\ti\zeta \lambda) \Big) \,, \no \\
& V^{(2)}_{\rm vec} = - \dd_{\ti\zeta} \, \tr \, \frac{1}{2 e^2} \lp  \ti\lambda \lambda  + 2i D \sigma \rp \,,  \label{deltaV2vec} \\
& \dd_{\zeta} V^{(2)}_{\rm vec} = \scL_{\rm vec} + \frac{1}{ e^2} \tr   \, D^{\mu} \Big( -\sigma D_{\mu} \sigma - i \sigma \scF_{\mu\nu} K^\nu - \sigma D K_\mu + \frac i2 \ti\lambda \gamma_\mu \lambda + i (\lambda \gamma_\mu \ti\zeta)(\zeta \ti\lambda) \Big) \,. \no
\end{align}
A few intermediate computations leading to \eqref{deltaV1vec} are given in Appendix~\ref{app:susycomputations}.
The terms $V^{(1)}_{\rm vec}$ and $V^{(2)}_{\rm vec}$ obey the relation
\begin{align}
\dd_{\ti\zeta} V^{(1)}_{\rm vec} - \dd_{\zeta} V^{(2)}_{\rm vec} = -2i \scL_K \Big[ \tr \, \frac{1}{2 e^2} \lp  \ti\lambda \lambda  + 2i D \sigma \rp \Big] \,,
\end{align}
in agreement with the algebra relation \eqref{susyalgebra}.

In the localization computation we will use the supercharge $Q  \equiv \frac 12 \big( \delta_\zeta + \delta_{\ti\zeta} \big)$ and consider a modified Lagrangian $\wat\scL_{\rm vec}$ defined by
\begin{align}
\wat\scL_{\rm vec} &= Q V_{\rm vec}   \no\\
V_{\rm vec}  &=   V_{\rm vec}^{(1)} + V_{\rm vec}^{(2)} \, . 
\label{Vvec}
\end{align}
This new Lagrangian differs from the original one by a total derivative. We have\footnote{Note that $\delta_{\ti\zeta}^2 = 0$ implies $Q V_{\rm vec}^{(1)} = \frac 12 \delta_\zeta V_{\rm vec}^{(1)}$ and similarly $Q V_{\rm vec}^{(2)} = \frac 12 \delta_{\ti\zeta} V_{\rm vec}^{(2)}$. }
\begin{align}
\wat\scL_{\rm vec} &= Q \lp V_{\rm vec}^{(1)} + V_{\rm vec}^{(2)} \rp   \no\\
& = \scL_{\rm YM}  + \frac{1}{ e^2}  \nabla^\mu \, \tr   \,  \Big[ -\sigma D_{\mu} \sigma - i \sigma \scF_{\mu\nu} K^\nu + \frac{i}{2} (\lambda \gamma_\mu \ti\zeta)(\zeta \ti\lambda)  -  \frac{i}{2} (\ti\lambda \gamma_\mu \zeta)(\ti\zeta \lambda) \Big] \,. \label{Lvec}
\end{align}
Being $Q$-exact, this Lagrangian will be used in the localization procedure as our deformation term (see Section~\ref{ssec:Localization}).

\subsubsection{Radial cutoff and supersymmetry}
\label{sssec:BdryVec}

In order to regularize infrared divergences and to treat boundary conditions in a supersymmetric way, we will need to introduce a spatial cut-off or boundary at finite distance from the center of the space. 
We show here that the Lagrangian $\wat\scL_{\rm vec}$ \eqref{Lvec} preserves two supercharges in the presence of a boundary.

We introduce a cut-off at a finite radial distance $\eta=\eta_0>0$ from the center of $\bH^3_\tau$. The boundary of this  ``chopped" $\bH^3_\tau$ is a two-torus.
In this case the Killing vector $K^\mu$ is tangent to the boundary and total derivatives of the form $\scL_K ( \cdots)$ vanish.
From the algebra relations we have
\begin{align}
& Q \wat\scL_{\rm vec} = Q^2 V_{\rm vec} = -  \frac{i}{2} \scL_K  V_{\rm vec} \,,  \no\\
& Q \wat S_{\rm vec} = \int_{\rhota \le \rhota_0} d^3x \sqrt g \,  Q \wat\scL_{\rm vec} \ = \ - \frac{i}{2}\int_{\rhota \le \rhota_0} d^3x \sqrt g \,  \scL_K  V_{\rm vec} \ = \ 0 \,.
\end{align}
In the last equality we have used the fact that $K$ is a Killing vector tangent to the boundary.

In the above discussion we can consider a general supercharge $\delta_{u, \ti u} = u \, \delta_{\zeta} + \ti u \, \delta_{\ti\zeta}$, $u,\ti u \in \bC$. Using the facts that $V_{\rm vec}^{(1)}$ is $\delta_{\ti\zeta}$-exact and $V_{\rm vec}^{(2)}$ is $\delta_{\zeta}$-exact, we obtain
\be
\ba
\delta_{u, \ti u} \wat\scL_{\rm vec} &= \frac{u}{2} \delta_{\zeta} \delta_{\ti\zeta} V_{\rm vec} + \frac{\ti u}{2} \delta_{\ti\zeta} \delta_{\zeta} V_{\rm vec}  
 \, = \, \frac{u}{2} \{ \delta_{\zeta} , \delta_{\ti\zeta} \} V_{\rm vec}^{(2)} + \frac{\ti u}{2} \{ \delta_{\ti\zeta} ,\delta_{\zeta}\} V_{\rm vec}^{(1)} \cr
&  = -2i \scL_{K} \lp \frac{u}{2}  V_{\rm vec}^{(2)} + \frac{\ti u}{2}  V_{\rm vec}^{(1)} \rp \cr
\delta_{u, \ti u} S_{\rm vec} &= \int_{\rhota \le \rhota_0} d^3x \sqrt g \,  \delta_{u, \ti u} \wat\scL_{\rm vec} \cr
&  = \ - 2 i \int_{\rhota \le \rhota_0} d^3x \sqrt g \,  \scL_K   \lp \frac{u}{2}  V_{\rm vec}^{(2)} + \frac{\ti u}{2}  V_{\rm vec}^{(1)} \rp \, = \, 0 \,.
\ea
\ee
We conclude that the Lagrangian $\wat\scL_{\rm vec}$ is appropriate to preserve the two supercharges $\delta_{\zeta}, \delta_{\ti\zeta}$ in the presence of a $T^2$ boundary, independently of the boundary conditions on the fields.

We remark in passing that any choice of Lagrangian of the form $\scL_{\rm vec}^{(v, \ti v)} = v \delta_{\zeta} V_{\rm vec}^{(1)} +  \ti v \delta_{\ti\zeta} V_{\rm vec}^{(2)}$, with $v + \ti v =1$, would be equally good, being invariant under $\delta_{\zeta}$ and $\delta_{\ti\zeta}$. Any two Lagrangians in this family differ by a total derivative term. The choice of Lagrangian $\wat\scL_{\rm vec}$ corresponds to $v=\ti v = \frac 12$.


\subsection{Chiral multiplet}
\label{ssec:SusyChi}

The supersymmetry transformations of a chiral multiplet $(\phi, \psi, F)$  of R-charge $r$ coupled to the vector multiplet, in a representation $\scR$ of the gauge group, are given by
\begin{align}
& \delta\phi \ = \ \sqrt 2 \zeta \psi \no\\
& \delta \psi \ = \ \sqrt 2 \zeta  F + i \sqrt 2 (\sigma + r H)  \phi \ti\zeta - \sqrt 2 i \gamma^\mu \ti\zeta D_\mu \phi \\
& \delta F \ = \ - \sqrt 2 i \lp \sigma + (r-2) H \rp \ti\zeta \psi - \sqrt 2 i D_\mu (\ti\zeta \gamma^\mu \psi )  + 2 i \ti\zeta \ti\lambda \phi  \, , \no
\end{align}
with $D_\mu = \nabla_\mu - i q_R \lp A_\mu - \frac 12 V_\mu \rp - i \scA_\mu$. The R-charges $q_R$ of the fields $(\phi, \psi, F)$ are $(r, r-1, r-2)$. The vector multiplet fields $(\sigma, \scA_\mu, \ti\lambda)$ are given in the representation $\scR$ and the indices are contracted appropriately.

The supersymmetry transformations of an anti-chiral multiplet $(\ti\phi, \ti\psi, \ti F)$ of R-charge $-r$  in the hermitian conjugate representation $\bar\scR$ are given by
\begin{align}
& \delta\ti\phi \ = \ -\sqrt 2 \ti\zeta \ti\psi \no\\
& \delta \ti\psi \ = \ \sqrt 2 \ti\zeta  \ti F - i \sqrt 2 \ti\phi (\sigma + r H) \zeta + \sqrt 2 i \gamma^\mu \zeta D_\mu \ti\phi \\
& \delta  \ti F \ = \ - \sqrt 2 i \zeta \ti\psi \lp \sigma + (r-2) H \rp  - \sqrt 2 i D_\mu (\zeta \gamma^\mu \ti\psi ) - 2 i \ti\phi \zeta \lambda   \, . \no
\end{align}
Note that here $ D_\mu = \nabla_\mu - i q_R \lp A_\mu - \frac 12 V_\mu \rp + i \scA_\mu$.
The R-charges  of the fields $(\ti\phi, \ti\psi, \ti F)$ are $(-r, -r+1, -r+2)$.

The Lagrangian of the chiral multiplet  is given by \cite{Closset:2012ru}
\begin{align}
\scL_{\rm chi} &=  D^\mu \ti\phi D_\mu \phi +  \ti\phi  \lp D + \sigma H \rp \phi + 2(r-1)H \ti\phi \sigma \phi - \ti F F \no\\
& \quad + \ti\phi \lp \sigma^2 + \frac r4 R + \frac 12 \lp r - \frac 12 \rp V^\mu V_\mu + r \lp r- \frac 12 \rp H^2 \rp \phi \no\\
& \quad   - i \ti\psi \gamma^\mu D_\mu \psi - i \ti\psi \lp \sigma + \lp r - \frac 12 \rp H \rp \psi  + \sqrt 2 i \lp \ti\phi \lambda \psi - \ti\psi \ti\lambda \phi \rp  \, .
\end{align}
We consider the $\bH^3_\tau$ background described in Section~\ref{sec:Background}. The Ricci scalar is given by $R = - \frac{6}{L^2}$. Furthermore we consider, for simplicity, and because it will be sufficient for our analysis,  a chiral multiplet coupled to a gauge multiplet with only the gauge field turned on, 
\begin{align}
\scA  \,, \quad \sigma  =  0 \,, \quad D = 0 \,.
\end{align}
To allow for a real mass deformation, we also turn on a constant background flavor vector multiplet $(v_\mu ,\sigma_F, D_F)$ with
\begin{align}
v &= \beta_F d\chi \,, \quad \sigma_F =  - D_F L = \mm \,,
\end{align}
and we assume charge $q_F=1$ under the flavor symmetry.
In this set-up the Lagrangian for a chiral multiplet, in a representation $\scR$ of the gauge group and with R-charge $r$, is
\be\label{LagrangianChi}
\scL_{\rm chi} = D^\mu \ti\phi D_\mu \phi +  \lp \mm + \frac{r-1}{L}   \rp^2 \ti\phi \phi -\frac{1}{L^2} \ti\phi \phi  - \ti F F  - i \ti\psi \gamma^\mu D_\mu \psi - i \lp \mm + \frac{r - \frac 12}{L}\rp \ti\psi \psi\, ,
\ee
with the constant background R-symmetry, flavor symmetry vectors, $A_\mu$, $v_\mu$, and the gauge field in the representation $\scR$, $\scA^{\scR}_\mu$, appearing in covariant derivatives, for instance $D_{\mu}\phi = \big(\p_\mu - i r A_\mu - i v_\mu - i \scA^{\scR}_\mu \big) \phi$.

The partition function is defined as a path integral over fields configurations obeying the reality conditions 
\be
\ti\phi  = \phi^\dagger  \,, \quad \ti F = - F^\dagger \, .
\ee
We will give the reality conditions on the fermionic fields using the twisted variables.

\subsubsection{Twisted fields}
\label{sssec:TVarChi}

It will be convenient for the chiral multiplet as well to introduce a set of  twisted fields. In this case the bosons are already scalars and so we need only introduce twisted fields for the fermions, which we decompose as follows
\begin{align}
\psi &=  \zeta B +  \ti\zeta C \, , \quad 
B = -  \ti\zeta \psi \, , \quad  C =   \zeta \psi \,,  \no\\
\ti\psi &=  \ti\zeta \ti B +  \zeta \ti C \, , \quad 
\ti B =   \zeta \ti\psi \, , \quad  \ti C = -  \ti\zeta \ti\psi  \, ,
\end{align}
where we used $\zeta \ti\zeta = - \ti \zeta \zeta = 1$. The R-charges of $(B, C,\ti B, \ti C)$ are $(r-2 \, , \,  r \, , \,  -r + 2 \, , \, -r)$.
This change of variables has a Jacobian equal to $|\zeta \ti\zeta | =1$, so it does not change the measure of  path integrals.
The supersymmetry transformations for the twisted fields are given in Appendix~\ref{app:SusyTwistedFields}.

The supersymmetric Lagrangian takes the form
\be\ba
\scL_{\textrm {chi}} &= D^\mu \ti\phi D^\mu \phi + \lp \mm + \frac{r-1}{L}   \rp^2  \ti\phi \phi -\frac{1}{L^2} \ti\phi \phi  - \ti F F \cr
& \quad  - i  \lp  \ti B \scL_{K} B + \frac{3}{2L} \ti B B + \ti C \scL_{K} C - \frac{3}{2L} \ti C C + \ti B \scL_{\ti P} C + \ti C \scL_{P} B  \rp  \cr
& \quad  -  i \lp \mm + \frac{r - \frac 12}{L}\rp (-  \ti B B +  \ti C C ) \, .
\label{eq:DY1} 
\ea\ee
Using the Fierz identities \eqref{Fierz2}, one can prove the relation\footnote{This is valid for any pair $\zeta, \ti\zeta$  of Grassman-even spinors.}
\begin{align}
K^a K^b - \frac 12 \lp \ti P^a P^b + P^a \ti P^b \rp &=  (\zeta \ti\zeta)^2 \delta^{ab} \, , \quad a,b=1,2,3.
\end{align}
This allows us to write
\begin{align}
D_{\mu}\ti\phi D^\mu \phi =  (\zeta \ti\zeta)^2 \lp \scL_K \ti\phi \scL_K \phi - \frac 12 \scL_{\ti P} \ti\phi \scL_P \phi - \frac 12 \scL_P \ti\phi \scL_{\ti P} \phi \rp \, .
\label{casimir}
\end{align}
The chiral multiplet Lagrangian can then be expressed as
\begin{align}
\scL_{\textrm {chi}} &=  \scL_K \ti\phi \scL_K \phi - \frac 12 \scL_{\ti P} \ti\phi \scL_P \phi - \frac 12 \scL_P \ti\phi \scL_{\ti P} \phi +  \lp \mm + \frac{r-1}{L}   \rp^2 \ti\phi \phi -\frac{1}{L^2} \ti\phi \phi  - \ti F F   \no\\
& \quad - i  \lp  \ti B \scL_{K} B + \ti C \scL_{K} C + \ti B \scL_{\ti P} C + \ti C \scL_{P} B \rp \\
& \quad  + i  \lp \mm + \frac{r - 2}{L}\rp \ti B B  - i \lp \mm + \frac{r-2}{L}\rp\ti C C     \,, \no
\end{align}
where we used $\zeta \ti\zeta =1$.
The complete reality conditions on the chiral multiplet fields are taken to be
\be\ba
& \phi^\dagger = \ti\phi \,, \quad F^\dagger = - \ti F \,, \cr
& B^\dagger = - \ti B \,, \quad C^\dagger = \ti C  \,.
\label{RealityCondChi}
\ea\ee

\subsubsection{Q-exact action}
\label{sssec:QactionChi}

The chiral multiplet Lagrangian $ \scL_{\textrm {chi}} $ can be expressed as a $\delta_{\zeta}$-exact or as a $\delta_{\ti\zeta}$-exact term, up to total derivatives. We have the following identities:
\begin{align}
\delta_{\zeta} V_{\rm chi}^{(1)} &= \scL_{\textrm {chi}} + \frac 12 \scL_{P}( \scL_{\ti P} \ti\phi \, \phi ) - \frac 12 \scL_{\ti P}( \scL_{P} \ti\phi \, \phi )  +  \scL_{K} [ \lp \mm + \frac{r}{L} \rp \ti \phi \phi + i \ti B B ] + i \scL_{P}(\ti C B) \, ,  \no\\
V_{\rm chi}^{(1)} &=  \delta_{\ti\zeta} \lp -\frac 12 (\ti B B + \ti C C) + i \ti\phi \scL_{K}\phi  - \frac{i}{L} \ti\phi \phi \rp  \label{deltaV1} \, , \\
\delta_{\ti\zeta} V_{\rm chi}^{(2)} &= \scL_{\textrm {chi}} - \frac 12 \scL_{P}(  \ti\phi \scL_{\ti P}\phi ) + \frac 12 \scL_{\ti P}(  \ti\phi  \scL_{P}\phi )  + \scL_{K} [-\lp \mm + \frac{r}{L}\rp \ti \phi \phi + i \ti C C ] + i \scL_{P}(\ti C B)  \, , \no \\
V_{\rm chi}^{(2)} &= \delta_{\zeta} \lp \frac 12 (\ti B B + \ti C C) + i (\scL_{K}\ti\phi )\phi + \frac{i}{L} \ti\phi \phi  \rp \,. \label{deltaV2}
\end{align}
A few intermediate computations leading to \eqref{deltaV1} are given in Appendix~\ref{app:susycomputations}.
Moreover $V_{\rm chi}^{(1)}$ and $V_{\rm chi}^{(2)}$ are related by 
\begin{align}
 \delta_{\zeta} V_{\rm chi}^{(1)} -  \delta_{\ti\zeta} V_{\rm chi}^{(2)} &= \scL_{K} \Big[ 2\lp \mm + \frac{r-1}{L}\rp \ti \phi \phi + i (\ti B B - \ti C C) \Big] \,.
\end{align}
In the localization computation we will use the supercharge $Q \equiv \frac 12 \lp \delta_\zeta + \delta_{\ti\zeta} \rp$ and consider a modified Lagrangian $\wat\scL_{\rm chi}$ defined as
\begin{align}
\wat\scL_{\rm chi} &= Q V_{\rm chi} \, ,  \no\\
V_{\rm chi}  &=   V_{\rm chi}^{(1)} + V_{\rm chi}^{(2)} \, .
\label{Vchi}
\end{align}
This new Lagrangian differs from the original one by a total derivative. We have explicitly\footnote{Note that $\delta_{\ti\zeta}^2 = 0$ implies $Q V_{\rm chi}^{(1)} = \frac 12 \delta_\zeta V_{\rm chi}^{(1)}$ and similarly $Q V_{\rm chi}^{(2)} = \frac 12 \delta_{\ti\zeta} V_{\rm chi}^{(2)}$. }
\be\ba
\wat\scL_{\rm chi} &= Q \lp V_{\rm chi}^{(1)} + V_{\rm chi}^{(2)} \rp \\
& = \scL_{\textrm {chi}} + \frac 14 \lp \scL_{P}( \scL_{\ti P} \ti\phi \, \phi - \ti\phi \scL_{\ti P} \phi ) - \scL_{\ti P}( \scL_{P} \ti\phi \, \phi - \ti\phi \scL_{P} \phi) \rp  \\
& \quad +  \frac i2 \scL_{K} (  \ti B B + \ti C C ) + i \scL_{P}(\ti C B) \,.
\ea\ee
The terms $\scL_K(...), \scL_P(...)$ and $\scL_{\ti P}(...)$ are all total derivatives.
 Explicitly, $\wat\scL_{\rm chi}$ is given in terms of the twisted fields by
\begin{align}
\wat\scL_{\textrm {chi}} &=  \scL_K \ti\phi \scL_K \phi -  \scL_P \ti\phi \scL_{\ti P} \phi + \frac 1L \ti\phi \scL_K \phi - \frac 1L \scL_K \ti\phi \, \phi + \lp \Big( \mm + \frac{r-1}{L}   \Big)^2 - \frac{1}{L^2} \rp \ti\phi \phi - \ti F F   \no\\
& \quad + i  \left[ \ \frac 12 \scL_K\ti B \, B -\frac 12 \ti B \scL_{K} B  + \frac 12 \scL_K \ti C \, C - \frac 12 \ti C \scL_{K} C - \ti B \scL_{\ti P} C + \scL_P \ti C  B \right. \no\\
& \qquad\quad \left. + \lp \mm + \frac{r - 2}{L}\rp \ti B B  - \lp \mm + \frac{r-2}{L}\rp\ti C C  \ \right]  \, .
\label{LagrChiSusy}
\end{align}
This new Lagrangian is Q-exact and can be used as the deformation term in the localization procedure (see Section~\ref{ssec:Localization}).

\subsubsection{Radial cutoff and supersymmetry}
\label{sssec:BdryChi}

As we did for the vector multiplet, in order to deal with supersymmetry on a space with a boundary, we wish to consider the situation where  we 
introduce a radial cut-off at $\eta=\eta_0 >0$.  We show now that $\wat\scL_{\rm chi}$ is an appropriate choice of Lagrangian  on this ``chopped" $\bH^3_\tau$, in the sense that is preserve the supercharges in the presence of the torus boundary. The analysis is as in the vector multiplet case.

We consider the action of a supercharge $\delta_{u, \ti u} = u \, \delta_{\zeta} + \ti u \, \delta_{\ti\zeta}$, $u,\ti u \in \bC$. Using the facts that $V_{\rm chi}^{(1)}$ is $\delta_{\ti\zeta}$-exact and $V_{\rm chi}^{(2)}$ is $\delta_{\zeta}$-exact, we obtain
\begin{align}
\delta_{u, \ti u} \wat\scL_{\rm chi} &= \frac{u}{2} \delta_{\zeta} \delta_{\ti\zeta} V_{\rm chi} + \frac{\ti u}{2} \delta_{\ti\zeta} \delta_{\zeta} V_{\rm chi}  
 \, = \, \frac{u}{2} \{ \delta_{\zeta} , \delta_{\ti\zeta} \} V_{\rm chi}^{(2)} + \frac{\ti u}{2} \{ \delta_{\ti\zeta} ,\delta_{\zeta}\} V_{\rm chi}^{(1)} \no\\
&  = -2i \scL_{K} \lp \frac{u}{2}  V_{\rm chi}^{(2)} + \frac{\ti u}{2}  V_{\rm chi}^{(1)} \rp \\
\delta_{u, \ti u} S_{\rm chi} &= \int_{\rhota \le \rhota_0} d^3x \sqrt g \,  \delta_{u, \ti u} \wat\scL_{\rm chi} \, = \ - 2 i \int_{\rhota \le \rhota_0} d^3x \sqrt g \,  \scL_K   \lp \frac{u}{2}  V_{\rm chi}^{(2)} + \frac{\ti u}{2}  V_{\rm chi}^{(1)} \rp \, = \, 0 \,. \no
\end{align}
We conclude that the Lagrangian $\wat\scL_{\rm chi}$ is appropriate to preserve the two supercharges $\delta_{\zeta}, \delta_{\ti\zeta}$ in the presence of a $T^2$ boundary, independently of the boundary conditions on the fields.

Here as well we notice that any choice of Lagrangian for the chiral multiplet of the form $\scL_{\rm chi}^{(v, \ti v)} = v \delta_{\zeta} V_{\rm chi}^{(1)} +  \ti v \delta_{\ti\zeta} V_{\rm chi}^{(2)}$, with $v + \ti v =1$, would provide an equally good choice of Lagrangian, invariant under $\delta_{\zeta}$ and $\delta_{\ti\zeta}$.

\subsection{Other supersymmetric actions}
\label{ssec:OtherActions}

In this section we discuss other supersymmetric actions and the relevant boundary terms needed for supersymmetry in the presence of a boundary.

\subsubsection{Chern-Simons action}
\label{sssec:CSAction}

We can consider adding to the action a supersymmetric Chern-Simons term with level $k \in \bZ$:
\begin{align}
S_{\rm CS} = i \frac{k}{4\pi}\int d^3x \sqrt g  \, \tr \, \Big[ \epsilon^{\mu\nu\rho} \lp \scA_{\mu} \p_\nu\scA_\rho - \frac{2i}{3} \scA_\mu \scA_\nu \scA_\rho \rp   +2i D\sigma + 2 \ti\lambda \lambda \Big] \,.
\label{SCS}
\end{align}
In the presence of a boundary the CS action is not invariant under supersymmetry transformations, but rather it picks up a boundary term. We have
\begin{align}
\delta S_{\rm CS} = - \frac{k}{4\pi} \int d^3x \sqrt g \, \nabla_\mu \tr \, \Big[ \epsilon^{\mu\nu\rho}\scA_{\nu} (\zeta\gamma_\rho\ti\lambda + \ti\zeta\gamma_\rho \lambda) + 2 \sigma (\zeta\gamma^\mu \ti\lambda - \ti\zeta\gamma^\mu \lambda) \Big] \,.
\end{align}
This supersymmetry variation can be re-expressed in terms of the twisted fields for a generic supercharge $\delta_{u,\ti u} \equiv  u \delta_\zeta + \ti u\delta_{\ti\zeta}$, with $u,\ti u \in \bC$, as
\begin{align}
& \delta_{u,\ti u} S_{\rm CS} \no\\
&= \frac{k}{8\pi} \int d^3x \sqrt g \,   \nabla_\mu \tr \, \Big[  2 K^{\mu} \lp  (X^0 -\Sigma) \Big( \frac{u -\ti u}{2}\Lambda^0 -i \frac{u+\ti u }{2} \Theta \Big) + \ti u X^- \Lambda^+ - u X^+ \Lambda^-  \rp \no\\
& \phantom{= \frac{k}{8\pi} \int d^3x \sqrt g \,   \nabla_\mu \tr \, \Big[} -  P^{\mu} \lp u(3\Sigma - X^0) \Lambda^- - X^-\Big( \frac{u+\ti u}{2}\Lambda^0 + \frac{u-\ti u}{2i} \Theta \Big) \rp \no\\
& \phantom{= \frac{k}{8\pi} \int d^3x \sqrt g \,   \nabla_\mu \tr \, \Big[} + \ti P^{\mu} \lp \ti u(3\Sigma - X^0) \Lambda^+ - X^+\Big( \frac{u+\ti u}{2}\Lambda^0 + \frac{u-\ti u}{2i} \Theta \Big) \rp \Big]\,.
\end{align}
When evaluating this term on the $\bH^3_\tau$ space with a torus boundary at $\eta=\eta_0$ (radial cut-off) we obtain
\begin{align}
& \delta_{u,\ti u}S_{\rm CS} \no\\
& = \frac{k}{8\pi} \int_{T^2} d^2x \sqrt{g_2} \,   n^\mu \, \tr \, \Big[ -  P_{\mu} \lp u(3\Sigma - X^0) \Lambda^- - X^-\Big( \frac{u+\ti u}{2}\Lambda^0 + \frac{u-\ti u}{2i} \Theta \Big) \rp \no\\
& \phantom{= \frac{k}{8\pi} \int d^2x \sqrt{g_2} \,   n^\mu \tr \, \Big[} + \ti P_{\mu} \lp \ti u(3\Sigma - X^0) \Lambda^+ - X^+\Big( \frac{u+\ti u}{2}\Lambda^0 + \frac{u-\ti u}{2i} \Theta \Big) \rp \Big] \,,
\end{align}
where $n^\mu \p_\mu = \frac 1L \p_\eta$ is a unit vector normal to the boundary, $d^2x \sqrt{g_2} = d\chi d\varphi \cosh\eta_0 \sinh\eta_0$ is the determinant of induced metric on the boundary, and we have used $n_\mu K^\mu =0$.

Remarkably this supersymmetry variation can be canceled by adding the following boundary term to the CS action:
\begin{align}
& S_{\rm CS}^{\rm bdry} = \frac{k}{16\pi} \int_{T^2} d^2x \sqrt g_2 \,  \tr \, \Big[   (3\Sigma - X^0) \lp (n_\mu P^{\mu} X^-  - n_\mu \ti P^{\mu} X^+  \rp \Big]\,, \\
& \delta_{u,\ti u}(S_{\rm CS} + S_{\rm CS}^{\rm bdry}) = 0 \,.
\end{align}
The invariance under the supersymmetry transformation holds without imposing any boundary condition on the fields. Since $\delta_{u,\ti u}$ is a generic supercharge and the boundary term does not depend on $u,\ti u$, we end up with a total action which is invariant under the two supercharges $\delta_\zeta, \delta_{\ti\zeta}$.

In terms of the original variables and making the $\eta_0$ dependence explicit, the boundary term is given by
\begin{align}
S_{\rm CS}^{\rm bdry} &=  \frac{k}{2\pi } \int_{T^2} d\chi d\varphi \,  \tr \lp  \sigma L + \scA_{\bar z} \rp \lp \cosh(2\eta_0) \scA_{\bar z} + \scA_{z} \rp \,,
\label{SbdryCS}
\end{align}
with $\scA_z = \frac 12 \lp \scA_\varphi - i \scA_\chi \rp$, $\scA_{\bar z} = \frac 12 \lp \scA_\varphi + i \scA_\chi \rp$. 
In addition we are free to add an extra boundary term of the form
\begin{align}
S^{\rm bdry}_f &= \int_{T^2} d^2x \sqrt{g_2} \, \tr [f(\Sigma)] \,,
\label{SbdrySigma}
\end{align}
with $f$ an arbitrary  function. This term is invariant under the two supercharges, since $\delta_{u,\ti u}\Sigma =0$ (see Appendix~\ref{app:SusyTwistedFields}).
We will fix the choice of boundary term $S^{\rm bdry}_f$ by requiring finiteness of the action. Related to this issue we must address the questions of gauge invariance of the Chern-Simons action with boundary terms and boundary conditions on the fields. We will discuss these issues all together in Section~\ref{sssec:YMandCSBdyCond}.

\subsubsection{Fayet-Iliopoulos term}
\label{sssec:FITerm}

Another supersymmetric action in gauge theories is the Fayet-Iliopoulos term with parameter $\xi \in \bR$:
\begin{align}
S_{\rm FI} = \xi \int d^3x \sqrt g  \, \tr \, \Big[  D - \frac{\sigma}{L} \Big] \,.
\label{SFI}
\end{align}
On a space with a boundary $S_{\rm FI}$ is not supersymmetric, rather it picks a boundary term under a generic supersymmetry transformation $\delta_{u,\ti u}$
\begin{align}
\delta_{u,\ti u} S_{\rm FI} &= \xi \int_{\rm bdry} d^2x \sqrt{g_2} \, \tr \, \Big[ n^\mu \lp u \zeta\gamma_\mu\ti\lambda - \ti u \ti\zeta\gamma_\mu\lambda \rp  \Big] \,,
\end{align}
where $n^\mu$ is a unit vector normal to the boundary.
Picking the boundary to be the torus at $\eta=\eta_0$, this boundary term can be expressed in terms of the twisted variables
\begin{align}
\delta_{u,\ti u} S_{\rm FI} &= \xi \int_{T^2} d^2x \sqrt{g_2} \, \tr \, \Big[ u (n^\mu P_\mu) \Lambda^- - \ti u (n^\mu \ti P_\mu) \Lambda^+  \Big] \,,
\end{align}
where we have used $n^\mu K_\mu = 0$. 
Supersymmetry under the $\delta_{u,\ti u}$ transformation can be restored by adding the boundary term
\be\ba
& S_{\rm FI}^{\rm bdry} = - \frac{\xi}{2} \int_{T^2} d^2x \sqrt{g_2} \, \tr \, \Big[ (n^\mu P_\mu) X^- -  (n^\mu \ti P_\mu) X^+  \Big] \,, \cr
& \delta_{u,\ti u} (S_{\rm FI} + S_{\rm FI}^{\rm bdry} ) = 0 \,.
\label{SFI2}
\ea\ee
Note that by adding the boundary term $S_{\rm FI}^{\rm bdry}$, one is able to preserve both 
supersymmetries $\delta_\zeta, \delta_{\ti\zeta}$.
In terms of the original fields the boundary term is given by
\begin{align}
S_{\rm FI}^{\rm bdry} = -\xi L \int_{T^2} d\chi d\varphi \, \tr \, \Big[ \cosh(2\eta_0) \scA_{\bar z} + \scA_z  \Big] \,,
\end{align}
with $\scA_z, \scA_{\bar z}$ as in \eqref{SbdryCS}.

As a consequence of adding a boundary term, the FI term is not gauge invariant without specifying boundary conditions, nor is it finite without adding supersymmetric boundary terms of the form \eqref{SbdrySigma}. We address these questions in Section~\ref{sssec:YMandCSBdyCond}.

\subsubsection{Mixed gauge-R Chern-Simons term}
\label{sssec:MixedCS}

We can consider mixed Chern-Simons term, in particular a mixed gauge-R symmetry Chern-Simons term with parameter $k_{\rm gR} \in \bZ$, as discussed in \cite{Closset:2012vg,Closset:2012vp}:
\begin{align}
S_{\rm gR} = \frac{k_{\rm gR}}{2\pi L}\int d^3x \sqrt g  \, \tr \, \Big[  D - \frac{\sigma}{L} \Big] \,.
\end{align}
On the supersymmetric background that we consider, this turns out to be the same as an FI term with quantized parameter $\xi = \frac{k_{\rm gR}}{2\pi L}$ and the analysis of boundary terms is as above.

\subsubsection{Superpotential}
\label{sssec:SuperPot}

Finally we can add a superpotential term to the action. This is given in terms of a holomorphic function of the chiral multiplet scalar fields $W(\phi^j)$ and a holomorphic function of the anti-chiral multiplet scalars $\ti W(\ti\phi^j)$, of R-charge 2 and -2 respectively. The superpotential action is given by
\begin{align}
S_{W} &= \int d^3x \sqrt g  \, \lp  \sum_j F_j \p_j W - \frac 12 \sum_{i,j} \psi^i \psi^j \p_i\p_j W  + \sum_j \ti F_j \p_j \ti W - \frac 12 \sum_{i,j} \ti \psi^i \ti \psi^j \p_i\p_j \ti W \rp \,,
\label{spot}
\end{align}
where  $(\phi^j, \psi^j, F^j)$ and $(\ti\phi^j, \ti\psi^j, \ti F^j)$ are the usual components of the (anti-)chiral multiplets. 

On the chopped $\bH^3_\tau$, with torus boundary at $\eta=\eta_0$, the supersymmetry variation of $S_{W}$ under a generic supercharge $\delta_{u,\ti u}$ is given by the boundary term
\begin{align}
\delta_{u,\ti u} S_{W} &= \int d^2x \sqrt{g_2} \, (-i \sqrt 2) n^\mu \lp \ti u  \ti P_\mu \sum_j C^j \p_j W   + u P_\mu \sum_j \ti C^j \p_j \ti W \rp \,,
\label{deltaSspot}
\end{align}
where $C^j, \ti C^j$ refer to the twisted fields  in  the corresponding multiplets and we have used the fact that total derivatives of the form $\scL_K(...)$ vanish. In checking the supersymmetry of the action, in the presence of real mass deformations by a weakly gauged flavor symmetry $G_F$, one must make use of the following identities, for each $i$,
\be\ba\label{IdSpot}
 \sum_j r_j \phi^j\p_i \p_j W &= (2-r_i)\p_i W \,, \cr
 \sum_j w_{F,j} \phi^j\p_i \p_j W &= -w_{F,i} \p_i W \,, 
\ea\ee
where $r_j, w_{F,j}$ denote the R-charge and flavor charge of the scalar $\phi^j$. These identities follow from the small $\theta$ expansions of: 
\be
W(e^{i r_j\theta} \phi^j) = e^{2i\theta} W(\phi^j) \,, \quad  W(e^{i w_{F,j}\theta} \phi^j) = W(\phi^j) \,, \quad \theta \in \bR \,,
\ee
which are a consequence of the covariance, respectively invariance, of the chiral superpotential $W$ under R-symmetry, respectively flavor symmetry.
A similar discussion applies to the anti-chiral superpotential $\ti W$.
\medskip

As before we would like to add a boundary term to restore supersymmetry, however we cannot do it for a generic supercharge $\delta_{u,\ti u}$. The best we can do is to add a boundary term which preserves one supercharge of the form $\delta_\zeta + \alpha \delta_{\ti\zeta}$,
with~$\alpha \neq 0$. In particular, for~$\a=1$ we can preserve~$Q \sim \delta_\zeta +  \delta_{\ti\zeta}$, in which case the boundary term is 
\be\ba
& S_{\rm W}^{\rm bdry} = \int d^2x \sqrt{g_2} \, i \, n^\mu \lp \ti P_\mu  W +  P_\mu \ti W  \rp \,, \cr
& Q (S_{\rm W} + S_{\rm W}^{\rm bdry}) = 0 \,.
\ea\ee
More supersymmetry may be preserved by further imposing boundary conditions, however for the localization computation we will only require invariance under the supercharge $Q$ and the above analysis ensures that $Q$ is preserved.


\section{Asymptotics, observables, and localization}
\label{sec:Asymptotics}

 In this section we first discuss the asymptotic boundary conditions that we impose on the various fields, defining more precisely the partition function that we propose to compute. Then we analyze the localization locus, which are the saddle points of the path integral arising in the supersymmetric localization computation. We find that the exact partition function is expressed as a sum of contributions labeled by flat connections on $\bH^3_\tau$, which we subsequently analyze. We evaluate the classical supersymmetric actions entering in the final expressions. Finally we discuss the generalization to the exact computation of supersymmetric Wilson loops.

\subsection{Asymptotic boundary conditions}\label{ssec:BdryCond}

In order to define the theory on the hyperbolic space $\bH^3_\tau$ we need to specify the boundary conditions on the fields or, more precisely, their asymptotic behavior since $\bH^3_\tau$ is non-compact. 
To derive these asymptotics we chop the $\bH^3_\tau$ by introducing a radial cut-off at $\eta=\eta_0>0$ and consider the variational principle on the space with a boundary. We then consider asymptotic expansions of the fields and impose that the boundary contributions to the equations of motion vanish as $\eta_0$ is sent to infinity.
Moreover we require that the asymptotics preserve supersymmetry. 
The upshot of this analysis is that the asymptotics of all the fields are given by constant background values at infinity, some of them zero,  and that the subleading terms in the asymptotic expansion are square-normalizable fluctuating modes. 
One important consequence following from this discussion is that supersymmetric Yang-Mills and Chern-Simons gauge theories have qualitatively different asymptotics.

\medskip

Before applying this recipe, we  explain how we choose an asymptotic expansion and how to treat the non-normalizable modes which may appear, following the methods of holographic renormalization \cite{deHaro:2000vlm}.
We assume that the asymptotic expansion at large $\eta$ for a generic field $\Phi$ takes the form 
\be\ba
\Phi &= e^{-\Delta \eta} \big( \Phi^{(0)} + \Phi^{(2)} \, e^{-2\eta} + \Phi^{(4)} \, e^{-4\eta}  + \cdots \big) \,.
\label{FieldEtaExpansion}
\ea\ee
The leading exponent $\Delta$ is fixed by solving the equations of motion at leading order in the large $\eta$ expansion. Typically there are two solutions $\Delta_{\pm}$ and the expansion \eqref{FieldEtaExpansion} starts with the smaller of the two $\Delta \equiv \Delta_-$. The leading term $\Phi^{(0)}$ turns out to be a non-normalizable mode, in the sense that it makes the action diverge, and should be seen as a background field, or non-fluctuating field, so that it is not integrated over in a path integral formulation. 
In the expansion \eqref{FieldEtaExpansion}, some subleading terms $\Phi^{(n)}$ may also be non-normalizable. In that case they must be fixed in terms of the leading mode $\Phi^{(0)}$ by solving the equations of motion order by order at large $\eta$.
When $n \equiv \frac 12 (\Delta_+ - \Delta_- )$ is an  integer, the expansion \eqref{FieldEtaExpansion} picks an extra term linear in $\eta$,
\be
\Phi = e^{-\Delta \eta} \big( \Phi^{(0)} + \cdots + ( \eta \wat\Phi^{(2n)} + \Phi^{(2n)} ) \, e^{-2n\eta} + \cdots \big) \,.
\ee
This extra mode is normalizable, except in the special case when $n=0$, which concerns massless vector fields, as we shall see below.

The resulting asymptotic behavior is then given by the sum of a non-normalizable background part and a normalizable fluctuating part:
\be
\Phi = e^{-\Delta\eta} \big( \Phi^{(0)} + \cdots \big) + \text{normalizable} \,.
\ee
Since we want to preserve supersymmetry, we must require in addition that in the full theory the collection of background fields $(\Phi^{(0)} + \cdots)$ are invariant under the supercharges $\delta_\zeta$ and $\delta_{\ti\zeta}$. 
Then the final step will be to enforce the boundary variational principle asymptotically. For this we consider the variation of the action under an arbitrary fluctuation of all the fields and require that the boundary piece 
vanishes as $\eta_0$ is sent to infinity, thus further constraining the asymptotics of the fields.

The analysis for the fermionic fields is simpler, since the constraint of supersymmetry imposes that their backgrounds vanish. Their asymptotics are then simply given by normalizable fluctuactions and we only need to ensure that normalizability is enough to satisfy the boundary variational principle. We now derive the asymptotics of the bosonic fields.

\subsubsection{Yang-Mills and Chern-Simons theories}
\label{sssec:YMandCSBdyCond}

Let us now start the analysis with the vector multiplet fields in a Yang-Mills theory. 
Applying the above prescription we find that the bosonic fields $\sigma$ and $\scA_\mu$ have the expansions \footnote{These expansions follow from solving Maxwell's equations $d\star\scF = \star j$ at leading order in $e^{-2\eta}$, upon assuming that the current $j$ coming from couplings to matter fields is subdominant $j = o(e^{-2\eta})$. This is consistent with the analysis of the chiral multiplet asymptotics of the next section.}
\be\ba
\sigma  &= \sigma^{(0)} + O(e^{-2\eta}) \,, \cr
\scA_\mu &=  \eta \wat\scA^{(0)}_\mu + \scA_\mu^{(0)} + O(e^{-2\eta}) \,, \quad \mu = \chi, \varphi \,, \cr 
\scA_{\eta} & = \scA^{(0)}_\eta + O(e^{-2\eta})   \,.
\ea\ee
The fields $\sigma^{(0)}$, $\wat\scA^{(0)}_\mu$ and $\scA^{(0)}_\eta$ are non-normalizable (they make the Yang-Mills action diverge), so they must be fixed to chosen values at $\eta = \infty$, whereas the other gauge field components $\scA^{(0)}_{\chi}, \scA^{(0)}_{\varphi}$ are normalizable and therefore fluctuating degrees of freedom. The $O(e^{-2\eta})$ subleading terms are normalizable and will be unconstrained.  To solve the equations of motion at leading order in $e^{-\eta}$, we also need to impose that the backgrounds $\sigma^{(0)}$, $\scA^{(0)}_\eta$ are in the center of the gauge algebra $\sigma^{(0)}, \scA^{(0)}_\eta \in Z_G$, so that their commutators with $\scA^{(0)}_{\chi}, \scA^{(0)}_{\varphi}$ vanish.

We now consider an arbitrary variation of the action  defined by the Lagrangian 
$\wat\scL_{\rm vec}$ \eqref{Lvec} on the chopped $\bH^3_\tau$ with torus boundary at $\eta=\eta_0$. It is convenient to express the variation in terms of the untwisted variables $\sigma, \scA$ for the bosonic fields and the twisted variables $\Lambda^{0,\pm}, \Theta$ for the fermionic fields, 
\be\ba
\delta S_{\rm vec} &= - \frac{1}{e^2 L}\int_{T^2} d^2x \sqrt{g_2}  \, \tr \Big[ \sigma D_\eta  (\delta\sigma) - \scF_{\eta \nu} \delta \scA^\nu + i \scF_{\eta\nu} K^\nu \delta\sigma + i \sigma K^\nu \delta\scF_{\eta\nu}   \cr
& \phantom{= -\frac{1}{e^2 L}\int_{T^2} d^2x \sqrt{g_2}  \, \tr \Big[} + \frac{i}{2}  \ti P_\eta (\Lambda^0 - i \Theta)  \delta\Lambda^+  + \frac{i}{2} P_\eta  (\Lambda^0 - i \Theta) \delta\Lambda^- \Big]   \cr
& \qquad  \ + \int_{\eta < \eta_0} d^3x \,  (\, \cdots)   \,,
\label{VarPrinVec}
\ea\ee
where  the bulk term $(\, \cdots)$ vanishes upon imposing the equations of motion. 
Using the asymptotic expansions of the bosonic fields and requiring no constraint on the normalizable subleading modes of the fields, we find that $\delta S_{\rm vec}$ vanishes asymptotically if and only if the background asymptotics are taken to be
\be
\sigma^{(0)} = 0 \,, \qquad \p_\mu\scA^{(0)}_\eta = \wat\scA^{(0)}_\mu \,, \ \mu = \varphi, \chi \,. 
\ee
The fermionic piece in $\delta S_{\rm vec}$ vanishes upon imposing normalizability of the fermions $\Lambda^{0,\pm}, \Theta = o(e^{-\eta})$.
After setting $D^{(0)}=0$ for the auxiliary field, we obtain an asymptotic background invariant under the two supercharges $\delta_\zeta, \delta_{\ti\zeta}$. One should still solve for backgrounds $\scA^{(0)}_\eta, \wat\scA^{(0)}_\mu$ globally defined on $\bH^3_\tau$. However in the following we choose to restrict our analysis to setting  $\scA^{(0)}_\eta = 0$, leading to
\be
\sigma^{(0)} = 0 \,, \qquad  \scA^{(0)}_\eta =0  \,, \qquad  \wat\scA^{(0)}_\mu = 0  \,. 
\ee
With these asymptotics, the subleading components of the gauge field $\scA^{(0)}_\mu$ are unconstrained and therefore considered as fluctuating fields.

Furthermore we have the possibility to add supersymmetric boundary terms of the form \eqref{SbdrySigma} and solve the variational principle on the boundary with these extra terms. An interesting choice of possible boundary term is 
\be
S_{\rm bdry}^{\rm YM} =  \frac{1}{e^2} \int_{T^2} d^2x \sqrt{g_2} \tr \Big[ \Sigma \p_\eta\Sigma \Big] \,,
\ee
where we recall that $\Sigma = \sigma + 2\scA_{\bar z}$. Solving the variational principle on the boundary leads to the following constraints:
\be
\scA^{(0)}_{\bar z} = 0 \,, \quad  \p_{\bar z} \sigma^{(0)} =0 \,, \quad  \p_{\bar z} \scA^{(0)}_\eta =0  \,, \quad  \wat\scA^{(0)}_{\bar z} = 0  \,.
\ee
This implies that $\sigma^{(0)}$ and $\scA^{(0)}_{\eta}$ are constant along the torus. The simplest asymptotics which are supersymmetric, within this class of boundary conditions,  have $\scA^{(0)}_\eta =0 $, $\wat\scA^{(0)}_{\mu} =0$, leading to
\be
\scA^{(0)}_{\bar z} = 0 \,, \quad  \sigma^{(0)} = \sigma_0  \in Z_{\mathfrak{g}} \ \text{constant} \,, \quad  \scA^{(0)}_\eta =0  \,, \quad  \wat\scA^{(0)}_\mu = 0  \,. 
\ee
The boundary condition $\scA^{(0)}_{\bar z} = 0$ fixes half of the gauge field modes along the boundary, leaving $\scA^{(0)}_{z}$ fluctuating. This choice is possible with the reality conditions \eqref{VecRealCond}. We will find similar boundary condition for the Chern-Simons theories below. 

To summarize, for the bosonic fields in the vector multiplet in the supersymmetric Yang-Mills theory there are two possible interesting choices of asymptotics,
\be\ba
 & \lim_{\eta \to \infty} \sigma = \sigma_0  \,, \quad  \lim_{\eta \to \infty} \scA_{\eta} = 0 \,, \cr
& \lim_{\eta \to \infty} \scA_{\varphi} = \alpha  \,, \quad \lim_{\eta \to \infty} \scA_{\chi} = \beta  \,, \cr
 \text{with} \quad  & (1) \ : \ \left\lbrace
\begin{array}{c}
 \sigma_0 = 0 \,, \\
 \alpha, \beta \ \text{fluctuating} \,,
\end{array} \right. \cr
 \text{or} \quad & (2) \ : \ \left\lbrace
\begin{array}{cc}
 \sigma_0 \in Z_{\mathfrak{g}}\ \text{constant}   \,,  & \\
 \alpha + i \beta =0 \,, & \alpha - i \beta \ \text{fluctuating} \,.
\end{array} \right.
\label{AandsigmaExpansions}
\ea\ee

We now reconsider the asymptotics for a theory when we add a Chern-Simons term with level $k$. All the fields in the vector multiplet now become massive, with masses of order $k e^2$. 
The equations of motion of the massive fields yield different asymptotic expansions.
For the vector field the new equation of motion is $d\star \scF = - \frac{ik e^2}{\pi} \scF$ and is solved asymptotically, in terms of the $\scA_z, \scA_{\bar z}$ combinations, with
\be\ba
\scA_{z} &= e^{-\Delta_z \eta} (\scA^{(0)}_z + \cdots ) \,, \quad \Delta_z \in \{ 0 , \frac{ke^2}{\pi} \} \,, \cr
\scA_{\bar z} &= e^{-\Delta_{\bar z} \eta} (\scA^{(0)}_{\bar z} + \cdots ) \,, \quad \Delta_{\bar z} \in \{ 0 , \frac{-ke^2}{\pi} \} \,,
\ea\ee
where we choose to set to zero possible non-normalizable linear terms in $\eta$. For $k>0$, the $\scA_z$ expansion starts with a normalizable  $O(1)$ term , while the $\scA_{\bar z}$ expansion starts with an $O(e^{\frac{k e^2}{\pi}\eta})$ non-normalizable background. The situation is reversed for $k<0$. We will not consider such diverging backgrounds and simply consider expansions for both $\scA_z$ and $\scA_{\bar z}$ starting with normalizable order $1$ terms, 
\be\ba
\scA_{z} &= \scA^{(0)}_z + O(e^{-2\eta}) \,, \quad
\scA_{\bar z} =\scA^{(0)}_{\bar z} + O(e^{-2\eta})\,.
\ea\ee
The scalar field $\sigma$ acquires a mass term and its expansion begins with a diverging non-normalizable term.
 As we shall see when we analyze the localization locus equations, supersymmetry does not allow for such a background, therefore we can simply set it to zero. The asymptotics of $\sigma$ are then given by normalizable fluctuations only, \emph{i.e.}~$\sigma = o(e^{-\eta})$.

The variation of the supersymmetric Chern-Simons action \eqref{SCS}, with boundary term \eqref{SbdryCS}, under an arbitrary fluctuation of the fields, is
\begin{align}
\delta (S_{\rm CS} + S_{\rm CS}^{\rm bdry}) = \frac{k}{2\pi} \int d\chi d\varphi \, \tr \, \Big[ & \delta\sigma L (\scA_z + \cosh(2\eta_0) \scA_{\bar z} ) \no\\
&+ \delta\scA_{\bar z}  (\sigma L + 2\scA_{\bar z}) \cosh(2\eta_0)  + \delta\scA_{z}  (\sigma L + 2\scA_{\bar z}) \Big] \no\\
& + \int d^3x ( \, \cdots ) \,,
\end{align}
where $( \, \cdots )$ denotes bulk terms which vanish upon imposing the equations of motion. Using the expansions \eqref{AandsigmaExpansions} would constrain the remaining background asymptotics $\alpha, \beta$ to be set to zero. Instead we can relax this strong constraint by adding a supersymmetric boundary term of the form \eqref{SbdrySigma},
\begin{align}
S_{\rm CS}^{\rm bdry}{}' &=  -\frac{k}{4\pi} \int d^2x \sqrt{g_2} \, \tr \Big[ \Sigma^2 \Big] \ = \ -\frac{k}{8\pi}  \int d\chi d\varphi \,         \sinh(2\eta_0) \tr \Big[ (\sigma L + 2 \scA_{\bar z})^2  \Big] \,,
\end{align}
and define
\be
S^{\rm tot}_{\rm CS} = S_{\rm CS} + S_{\rm CS}^{\rm bdry} + S_{\rm CS}^{\rm bdry}{}' \,.
\ee
The variation of the total Chern-Simons action becomes
\begin{align}
\delta S^{\rm tot}_{\rm CS} = \frac{k}{2\pi} \int d\chi d\varphi \, \tr \, \Big[ & \delta\sigma L (- \frac 12 \sinh(2\eta_0)\sigma L + \scA_z + e^{-2\eta_0} \scA_{\bar z} ) \no\\
&+ \delta\scA_{\bar z}  (\sigma L + 2\scA_{\bar z}) e^{-2\eta_0}   + \delta\scA_{z}  (\sigma L + 2\scA_{\bar z}) \Big] + \int d^3x ( \, \cdots ) \,.
\end{align}
Using $\sigma= o(e^{-\eta})$ and the expansions described above, the leading term in the asymptotic expansion is
\be
\frac{k}{2\pi} \int d\chi d\varphi \, \tr \, \Big[ 2 \delta\scA^{(0)}_{z} \scA^{(0)}_{\bar z} \Big] \,.
\ee
We then find that the boundary variation vanishes as $\eta_0 \to \infty$ when 
\begin{align}
&  \scA^{(0)}_{\bar z} = 0 \,, \quad
\textrm{or}  \quad \scA^{(0)}_{z}  \ \textrm{fixed}  \,.
\end{align}
In addition we require $\scA^{(0)}$ to define a supersymmetric asymptotic background. 
For simplicity we will restrict the choice of boundary conditions for the second option to $\scA^{(0)}$ constant on the torus and in the center of the gauge algebra  $\scA^{(0)}_z \in Z_{\mathfrak{g}}$. We will therefore consider the two options
\begin{align}
& (1) \quad \scA^{(0)}_{\bar z} = \frac12  (\alpha + i \beta) =  0 \,, \label{VecBdyCond1} \\
\textrm{or} \quad  &  (2) \quad  \scA^{(0)}_{z} = \frac 12 (\alpha - i \beta) \in  Z_{\mathfrak{g}} \ \textrm{constant}  \label{VecBdyCond2} \,.
\end{align}
These boundary conditions are familiar in the Chern-Simons theory literature, for instance in the holographic duality context in \cite{Kraus:2006wn}, or in studies of supersymmetric Chern-Simons theories with a boundary \cite{Berman:2009kj}. Note that the modification of the reality condition, taking $\scA_\chi$ purely imaginary, and therefore $\scA_z$ and $\scA_{\bar z}$ real and independent, allows us to fix one component and let the other fluctuate.

To summarize, the boundary conditions in the theory with a Chern-Simons term are given by \eqref{VecBdyCond1} or \eqref{VecBdyCond2} for the vector field, together with  normalizable asymptotics for $\sigma$, \footnote{We remind the reader that $f(\eta) = o(e^{-\eta})$ is equivalent to $\lim_{\eta \to \infty} e^{\eta} f(\eta)=0$.}
\be
\sigma = o(e^{-\eta}) \,.
\ee
For the pure supersymmetric Chern-Simons theory, the analysis leads to the same asymptotics as for the Yang-Mills and Chern-Simons theory.

The same analysis, applied to the theory with an FI term, leads us to introduce an extra boundary term (in addition to \eqref{SFI2})
\be
S_{\rm FI}^{\rm bdry}{}' =  \xi \int d^2x \sqrt{g_2} \, \tr \big[ \Sigma \big] \ = \ \frac{\xi L}{2}  \int d\chi d\varphi \,     \sinh(2\eta_0) \tr \big[  \sigma L + 2\scA_{\bar z}  \big] \,.
\label{SFI3}
\ee
This boundary term is introduced to relax the constraint on $\tr \scA_{\bar z}^{(0)}$. In addition, one can show that it makes finite the total FI action $S_{\rm FI}^{\rm tot} = S_{\rm FI} + S_{\rm FI}^{\rm bdry} + S_{\rm FI}^{\rm bdry}{}'$, given by the sum of the three terms \eqref{SFI}, \eqref{SFI2}, \eqref{SFI3}.
The variation of the action then produces a boundary term, which vanishes when $\tr[\delta\sigma^{(0)}]=0$ and $\tr[\delta\scA_z^{(0)}]=0$. Therefore, in the presence of an FI term, we must impose, in addition to the previous constraints,
\be
\tr[\sigma^{(0)}] = \textrm{constant} \,, \quad \tr[\scA_z^{(0)}]  = \textrm{constant} \,.
\label{BdryCondFI}
\ee
We will see in later sections that, except for the pure Yang-Mills theory, the presence of an FI term does not affect the partition function.

\smallskip

Finally, there is a subtlety in 3d theories, that in the presence of  massive fermions, the bare Chern-Simons level $k_0$ gets 
shifted\footnote{\label{keff:foot} To be more precise the determinant of a spinor of mass $M$, coupled to a gauge field $A$, on a compact space, contains a factor $\exp[i\pi \sign(M)\eta[A] /2]$ \cite{AlvarezGaume:1984nf}, with $\eta[A]$ the APS eta invariant \cite{APS} whose variation with respect to the gauge field $A$ matches the variation of a properly quantized Chern-Simons term $(4\pi)^{-1} \int A dA$. The proper regularization of such a determinant involves adding a factor $\exp(\pm i\pi \eta[A] /2)$ as reviewed in \cite{Seiberg:2016rsg}. The extension of this discussion to non-compact spaces, such as the one we study, necessitates to include the effect of boundary conditions. It is not clear to us what the consequences can be regarding ``Chern-Simons level shifts". We thank Cyril Closset for discussions on this issue.}
in the one-loop effective action~\cite{Aharony:1997bx}. In the $\N=2$ supersymmetric theories, due to this shift, the effective Chern-Simons level is given by $k_{\rm eff}= k_0 + \frac{1}{2}\sum_I \text{sign}(m_I) q_I^2$, where~$I$ labels the chiral multiplets of (integer) gauge charge $q_I$  and real masses $m_I$. The Chern-Simons level relevant for the above analysis is $k=k_{\rm eff}$. To avoid subtleties related to the identification of $k_{\rm eff}$, for simplicity, 
we will refer to the ``pure Yang-Mills" theories as  theories with $k_{\rm eff}=k_0 = 0$. This is certainly true at least for ``non-chiral theories''.

\subsubsection{Chiral multiplet}
\label{sssec:ChiBdyCond}

We now consider the asymptotics of the chiral multiplet fields. 
Let us take a chiral multiplet of R-charge $r$ and gauge charge $w$ under an abelian gauge symmetry.
The asymptotic expansions of the complex scalar can be written 
\be
\phi = e^{-\Delta\eta} (\phi^{(0)} + e^{-2\eta} \phi^{(2)} + \cdots) \,,
\ee
with  $\Delta = 1 - \sqrt{1 + (\mm L + r)(\mm L + r-2)}$.
We use now a result that we will derive when analyzing the supersymmetric locus equations (see Section~\ref{sssec:ChiLocus}), which is that for $r > 0$, supersymmetric configurations are given by $\phi=0$. So, assuming a positive R-charge $r>0$, we conclude that the asymptotic background field $\phi^{(0)}$, as well as all the subleading non-normalizable modes in the $\phi$ asymptotic expansion, are vanishing.
We therefore impose normalizablity of the scalar field and their superpartners (the twisted fermions) as follows
\be
\phi = o(e^{-\eta}) \,, \quad C = o(e^{-\eta}) \,, \quad  B = o(e^{-\eta})  \,.
\label{NormalChi}
\ee
The variation of the action with respect to the fields, expressed in terms of the twisted variables, is given by
\begin{align}
\delta S_{\rm chi} &= \int_{\eta < \eta_0} d^3x  (\, \cdots)  -  \int_{T^2} d^2x \sqrt{g_2} \frac 1L \lp \ti P_{\eta}  (\scL_{ P} \ti\phi \, \delta \phi  + i \ti B \, \delta C)  + P_{\eta} ( \delta\ti\phi \, \scL_{\ti P} \phi - i \delta \ti C \, B )  \rp   \,,
\label{VarPrinChi}
\end{align}
where the bulk term $(\, \cdots)$ vanishes upon imposing the equations of motion. 
The normalizability conditions \eqref{NormalChi} automatically ensure that the boundary term \eqref{VarPrinChi} vanishes in the limit $\eta_0 \to \infty$.

\subsection{Supersymmetric localization}
 \label{ssec:Localization}

In this subsection we discuss the localization computations and localization locus, following the method developed in \cite{Pestun:2007rz}.

The supersymmetric localization technique proceeds by adding, to the action of the theory, a Q-exact deformation term  $t \int d^3x \sqrt g \, QV$, with $t\in \bR_{>0}$, for a certain supercharge $Q$ and fermionic term $V$. Being Q-exact, the deformation term does not modify the value of the partition function, which therefore can be evaluated 
in the limit $t\to\infty$. In this limit the partition function reduces to an integration over the saddle point configurations of $QV$, also called localization locus,
\be\ba
Z &= \int_{QV[\phi_0] =0} \scD\phi_0 \,  e^{-S_{\rm class}[\phi_0]} \, Z_{\rm one-loop}[\phi_0] \,, 
\label{Zformal}
\ea\ee
where $S_{\rm class}[\phi_0]$ is the evaluation of the classical action on the saddle point configuration $\phi_0$ and $Z_{\rm one-loop}[\phi_0]$ is the one-loop determinant of the deformation term $\int d^3x \sqrt g \, QV$ around $\phi_0$.

For our computation we choose the deformation term
\be\ba \label{QVdef}
 QV &= t QV_{\rm vec} + t' QV_{\rm chi} = t \wat\scL_{\rm vec} + t' \wat\scL_{\rm chi} \,, \quad t, t' \in \bR_{>0} \,,
\ea \ee
 which is a sum of two terms, both $Q$-exact with respect to the supercharge $Q = \frac 12 (\delta_\zeta + \delta_{\ti\zeta})$, so that the partition function is independent of both $t$ and $t'$.  The fermionic terms $V_{\rm vec}$ and $V_{\rm chi}$ have been defined in \eqref{Vvec} and \eqref{Vchi}. 
 We then send $t\to\infty$ and $t'\to \infty$ in turn, localizing the vector multiplet fields first and then the chiral multiplet fields.

We now determine the localization locus of the fields.

\subsubsection{Vector multiplet locus}
\label{sssec:VecLocus}

The bosonic part of the localizing term $QV_{\rm vec} = \wat\scL_{\rm vec}$, given in \eqref{Lvec}, is the sum of the bosonic part of the original Lagrangian \eqref{LagrangianVec} and a boundary term. With the asymptotics \eqref{AandsigmaExpansions}, the boundary term vanishes, so  it is enough to focus on the bosonic part of the original Lagrangian, which is 
\be
QV_{\rm bos} \,  \sim \,   \scF^{\mu\nu}\scF_{\mu\nu} + D^{\mu} \sigma D_{\mu} \sigma - \Big( D + \frac{\sigma}{L} \Big)^2 \,.
\ee
It trivially vanishes when
$\scF_{\mu\nu} =  D_{\mu} \sigma =  D + \frac{\sigma}{L} = 0 $.
Unfortunately the reality conditions favored by the boundary analysis \eqref{VecRealCond} are such that
that the bosonic action $\scL_{\rm bos}$ is complex and its real part is not positive definite \footnote{One may think of using an alternative deformation term $QV$ with $V =  (Q\lambda)^{\dagger} \lambda + (Q\ti\lambda)^{\dagger} \ti\lambda$, where $\dagger$ denotes hermitian conjugation, which is manifestly positive definite. The issue with such a term is that, with our choices of reality conditions, $V$ would be a function of the fields as well as their hermitian conjugates, on which there is no natural action of the supersymmetry. Based on this observation, we do not consider such a deformation term.}, so that we cannot \emph{a priori} rely on a minimization principle to find the saddle-point configurations.
Here we assume that the localization locus is given by field configurations invariant under the supercharge $Q$ used for the localization, and therefore solve the BPS equations.

The $Q$-supersymmetric configurations are the solutions of the following equations
\be\ba
0 &= \frac i2 \epsilon^{\mu\nu\rho} \scF_{\mu\nu} \gamma_\rho \zeta -  i (D + \frac{\sigma}{L}) \zeta + i D_\mu \sigma \gamma^\mu \zeta  \, , \cr
0 &= \frac i2 \epsilon^{\mu\nu\rho} \scF_{\mu\nu} \gamma_\rho \ti\zeta +  i (D + \frac{\sigma}{L}) \ti\zeta - i D_\mu \sigma \gamma^\mu \ti\zeta   \, .
\ea\ee
Contracting with $\zeta$ and $\ti\zeta $, these equations are equivalent to
\be\ba
K^\mu (\star \scF)_\mu &= - \big(D+ \frac{\sigma}{L}\big) \,, \quad  \scL_K \sigma = 0 \,, \cr
\scL_P \sigma &= - P^\mu (\star \scF)_\mu \,, \quad \scL_{\ti P} \sigma = \ti P^\mu (\star \scF)_\mu \,,
\label{BPSEqnsVec}
\ea\ee
with  $(\star\scF)^\mu = \frac 12 \epsilon^{\mu\nu\rho} \scF_{\nu\rho}$. 
Again we face a difficulty. For generic complex fields, these equations have solutions characterized by arbitrary functions and we do not know how to select the relevant solutions without specifying the reality conditions on the fields, and it is well-known that the saddle point configurations in Euclidean path integrals may lie outside of the initial contour of integration. To circumvent the difficulty we propose the following strategy. We can consider the Lorentzian theory, obtained by Wick rotation, for which the reality conditions are fixed by supersymmetry, and solve for the BPS locus. 
Then we can Wick-rotate back the solutions to Euclidean signature, to obtain the localization locus of our path integral. This is a priori different from working directly with Equations (4.32). It is not clear whether the two approaches lead to the same answer at the end of the day or not. We take encouragement from the fact that such a procedure was used in \cite{Murthy:2015yfa} to localize a functional integral on AdS$_2$ space and the result agreed in a non-trivial manner with considerations from microscopic string theory. These issues should be certainly cleared up using a first principles treatment of Euclidean supergravity\footnote{This problem is currently being addressed in \cite{deWitReys}}.

 The Lorentzian BPS equations are obtained by the Wick rotation, which acts as $\chi \to i \chi$, $\p_\chi \to -i \p_\chi$, $\epsilon_{\mu\nu\rho} \to -i \epsilon_{\mu\nu\rho}$ and $\scA_{\chi} \to -i \scA_{\chi}$. It also ensures $K^\ast = - K$ and $P^{\ast} = - \ti P$.
The equations read as in the Euclidean theory, but with the Wick rotated vectors $K, P, \ti P$. 
  In the Lorentzian theory the reality conditions are fixed and compatible with the supersymmetry transformation generated by Q. We have 
\be
\text{Lorentzian theory:} \qquad    \scA^{\dagger} = \scA \,, \quad \sigma^{\dagger} = \sigma  \,, \quad   \big(D+ \frac{\sigma}{L}\big)^{\dagger} = \big(D+ \frac{\sigma}{L}\big) \,,
\ee
so all the fields are hermitian. The first equation in \eqref{BPSEqnsVec} is
\be
K^\mu (\star \scF)_\mu = - \big(D+ \frac{\sigma}{L}\big) \,.
\ee
The left-hand side is now anti-hermitian, since $(K_\mu)^\ast = - K^\mu$, while the right-hand side is hermitian, so they must vanish separately,
\be
K^\mu (\star \scF)_\mu = 0 \,, \quad  D+ \frac{\sigma}{L} =0 \,.
\ee
Decomposing the one-form component $(\star \scF)_\mu$ along the vectors $K, P, \ti P$ and using the remaining BPS equations, we obtain
 \be
(\star \scF)_\mu = - \frac 12 \lp   \scL_{\ti P} \sigma P_\mu  - \scL_P \sigma  \ti P_\mu  \rp \,.
\ee
Again the left-hand side is hermitian, while the right-hand side is anti-hermitian, due to the relation $P_\mu^\ast = - \ti P_\mu$, so they must vanish separately. Combining all results, we obtain the BPS locus equations
\be\ba
\text{Lorentzian theory:} \qquad   \scF_{\mu\nu}  = 0 \,,  \quad  D_{\mu} \sigma =  0  \,, \quad   D + \frac{\sigma}{L}  =0 \,.
 \label{LocusVecLorentzian}
\ea\ee
After a Wick rotation, back to the Euclidean theory, these equations still describe BPS configurations and we assume that no other configurations will contribute to the localization locus. We therefore obtain,
\be\ba
\text{Localization locus:} \qquad \scF_{\mu\nu}  = 0 \,,  \quad  D_{\mu} \sigma =  0  \,, \quad   D + \frac{\sigma}{L}  =0  \,.
 \label{LocusVec}
\ea\ee

The locus configurations \eqref{LocusVec} are characterized by flat gauge connections, which must be considered up to gauge transformations and subject to the asymptotics described in Section~\ref{sssec:YMandCSBdyCond}. 
Flat connections on the solid torus $S^1 \times D$ have been studied in \cite{Kirk1990, Jeffrey1992}. 
This analysis depends only on the topology of the space and it can therefore  be applied to studying flat connections on $\bH^3_\tau$. The flat connections are characterized by the asymptotic value of the gauge field,
\be 
\scA^{(0)} = \alpha d\varphi + \beta d\chi \,.
\ee 
Using the flatness of $\scA^{(0)}$, we can choose a gauge where $\alpha$ and $\beta$ are constant and in the Cartan subalgebra $\mathfrak{t}$,
\be
\alpha = \sum_{i=1}^{r_G} a_i H_i \in \mathfrak{t} \,, \quad \beta = \sum_{i=1}^{r_G} b_i H_i \in \mathfrak{t} \,,
\ee
where $H_i$ are the generators of  $\mathfrak{t}$,  and $r_G$ is the  rank of the gauge group. 
Given these asymptotics the gauge field is fixed, up to gauge transformations leaving $\scA^{(0)}$ invariant, by solving the flatness condition in the bulk of $\bH^3_\tau$. However not all values of $\alpha, \beta$ lead to globally defined flat gauge fields.  

We now  flesh out this discussion by choosing specific gauge groups. We will analyze $U(N)$ and $SU(N)$ gauge theories. Generalizing to gauge groups which are products of $U(N_i)$ and $SU(N_i)$ factors is straightforward.

Let us start with the abelian theory. The flatness condition in the bulk implies $\alpha = \frac{1}{2\pi}\int_{\varphi} \scA^{(0)} = 0$. By a gauge transformation the flat connection can be set into the simple form
\be 
\underline{\text{$U(1)$ theory:}} \quad \scA = \beta d\chi \,,
\label{FlatConnectionU1}
\ee 
with $\beta$ constant. 

Let us now consider an $SU(2)$ gauge theory. The constant asymptotic gauge field is given by
\be
\alpha = \text{diag}(a,-a) = a \sigma_3 \,, \quad \beta = \text{diag}(b,-b) = b \sigma_3 \,.
\ee
We now use results presented in \cite{Kirk1990} \footnote{The results in \cite{Kirk1990} are given for a straight torus boundary, $\tau_1 =0$. The generalization to an arbitrary torus,  $\tau_1 \neq 0$, is achieved by replacing the angle $\varphi$ by $\ti\varphi = \varphi - \frac{\tau_1}{\tau_2}\chi$, so that $\ti\varphi, \chi$ parametrize a straight torus.}. The flatness condition requires a trivial holonomy around the contractible circle at infinity $e^{i\int \alpha d\varphi}=1$, leading to $a = n \in \bZ$.  For each pair $(n,b)$ one can construct a smooth flat connection on the whole $\bH^3_\tau$ in the form
\be\ba
& \scA = i U^{-1} dU \,, \cr
& U(\eta, \varphi, \chi) = f_n(\eta, \ti\varphi) e^{-i \ti b  \sigma_3 \chi} \,,
\ea\ee
where $\ti\varphi = \varphi - \frac{\tau_1}{\tau_2}\chi$, $\ti b = b +\frac{\tau_1}{\tau_2}n$ and $f_n: \bR_{\ge 0} \times [0,2\pi] \to SU(2)$ is a smooth function satisfying, for some $\eta_+ > \eta_- > 0$,
\be\ba
f_n(\eta, \ti\varphi) &= e^{-i n \sigma_3 \ti\varphi} \,, \quad \text{for} \quad  \eta > \eta_+ \,, \cr
f_n(\eta, \ti\varphi) &= \text{constant} \,, \quad \text{for} \quad  0 \le \eta <  \eta_- \,.
\ea\ee
This implies in particular 
\be
\underline{\text{$SU(2)$ theory:}} \quad 
\scA = \left\lbrace
\begin{array}{cc}
\ti b  \sigma_3 d\chi  \,,  &   \qquad \text{for} \quad  0 \le \eta <  \eta_- \,, \\
 n \sigma_3 d\varphi  + b \sigma_3 d\chi \,, &  \text{for} \quad  \eta > \eta_+ \,.
 \label{FlatConnectionSU2}
\end{array} 
\right.
\ee
The matrix-valued function $U(\eta, \varphi, \chi)$ is not globally well-defined for generic $\ti b$, but $\scA$ is globally defined.

The generalization to a $U(N)$ or $SU(N)$ theory is straightforward. Let us consider the $U(N)$ theory. The asymptotic values of the flat connection are given by the matrices
\be
\alpha = \text{diag}(a_1 ,a_2, \cdots, a_N) \,, \quad \beta = \text{diag}(b_1 ,b_2, \cdots, b_N) \,.
\ee
Flat connections have trivial holonomy around the $\varphi$-circle at infinity $e^{i\int \alpha d\varphi}=1$, leading to the quantization of the $a_i$. For the abelian part, we found before that the constraint is stronger, it imposes $\alpha_{\mathfrak{u}(1)}=0$. We obtain
 \be
\{ a_i\}_{1\le i \le N} \in \bZ^N \,, \quad \sum_{i=1}^N a_i = 0 \,. 
\label{aiConstraints}
\ee 
A flat connection with these asymptotics can be expressed in the form $\scA = i U^{-1} dU$ as in the $SU(2)$ case, with the behaviors in neighborhoods of the origin and infinity
\be
\underline{\text{$U(N)$ theory:}} \quad
\scA = \left\lbrace
\begin{array}{cc}
\lp \beta + \frac{\tau_1}{\tau_2} \alpha \rp d\chi  \,,  &   \qquad \text{for} \quad  0 \le \eta <  \eta_- \,, \\
 \alpha d\varphi  + \beta d\chi \,, &  \text{for} \quad  \eta > \eta_+ \,.
\end{array} 
\right.
\label{FlatConnUN}
\ee
For $SU(N)$ theories, we simply have the extra constraint $\sum_{i=1}^N b_i = 0$.
\medskip

Finally we need to quotient by the Weyl group, which acts as permutations of the $\{a_i\}$ and $\{b_i\}$ and brings a factor $\frac{1}{N!}$ in the partition function, and by  large gauge transformations, when these are preserved by the boundary conditions. 
Large gauge transformations for the $U(N)$ gauge group shift the parameters $a_i, b_i$, for each $i$, as 
\begin{align}
& (a_i, b_i) \sim \lp a_i + k_i, b_i - \frac{\tau_1 k_i}{\tau_2} \rp \,, \quad k_i \in \bZ \,, \label{largeGT1} \\
& (a_i, b_i) \sim \lp a_i, b_i + \frac{l_i}{\tau_2} \rp \,, \quad l_i \in \bZ \,. \label{largeGT2}
\end{align}
In the pure Yang-Mills theory, the boundary conditions preserve the large gauge transformations. Using \eqref{largeGT1}, we can go to a gauge with $a'_i=0$ and $b'_i =  b_i + a_i\t_1/\t_2$, then using \eqref{largeGT2} we can reduce to $b'_i \in [0,\frac{1}{\tau_2}]$. The resulting path integral is an integral over the complex parameters $x_i = e^{-2\pi i \tau_2 b'_i}$,
\be
\underline{\text{$U(N)$ Yang-Mills theory:}} \quad Z = \frac{1}{N!} \int_{\scC} \prod_{i=1}^N \frac{dx_i}{2\pi i x_i}  \, \scI(\{x_i\}) \,.
\ee
Here $\scI(\{x_i\}) = \Delta \, \scI'(\{b'_i\})$ is the product of the integrand $\scI'$ to be computed and the Vandermonde determinant $\Delta = \prod_{i<j}(b'_i-b'_j)^2$ coming from the diagonalization of the flat connection. We anticipate that $\scI$ will be a function of the complex parameters $x_i$.
 The integration contour $\scC = \prod \scC_i$ is naively composed of unit circles, however it may happen that the integration contour gets deformed  in the complex plane to take into account saddle points corresponding to complex flat connections. We do not explore this possibility here and refer to \cite{Beem:2012mb} for a more complete discussion on this issue.

In the theories with Chern-Simons terms, the boundary condition $\scA_{\bar z} = 0$ breaks large gauge transformations and sets instead $b_i = i a_i \in i\bZ$. 
This is compatible with the reality conditions \eqref{VecRealCond} at infinity. We obtain
\be
\underline{\text{$U(N)$ Chern-Simons theory:}} \quad Z = \frac{1}{N!} \sum_{\{n_i\} \in \bZ^N} \delta \big(\sum_i n_i \big) \,\scI(a_i = n_i; b_i = i n_i) \,,
\ee
where $\scI(a_i = n_i; b_i = i n_i) $ is the summand to be computed, multiplied by the Vandermonde determinant $\Delta = \prod_{i<j}[b'_i-b'_j ]^2 = \prod_{i<j}[(n_i-n_j)\t/\t_2 ]^2$. The constraint $\sum_{i=1}^N n_i = 0$ is implemented by the factor $\delta \big(\sum_i n_i \big) $.

The locus equations \eqref{LocusVec} for the scalar field $\sigma$, together with the vanishing asymptotics $\sigma^{(0)} =0$ imply 
\be
\sigma = -\frac{D}{L} = 0 \,,
\label{Locussigma}
\ee
so that the localization locus of the vector multiplet fields are only characterized by the flat connections described above.

\subsubsection{Chiral multiplet  locus}
\label{sssec:ChiLocus}

We now turn to the localization locus of the chiral multiplet. We must look at the saddle point configurations of the $Q$-exact deformation term $Q V_{\chi} = \wat\scL_{\rm chi}$.
The reality conditions are
\begin{align}
\ti\phi & = \phi^\dagger \, , \quad \ti F = - F^\dagger \,,
\label{RealCondChi}
\end{align}
and the asymptotic behavior of the fields are
\begin{align}
\lim_{\eta \to \infty} e^{\eta} \phi  = 0 \,, \quad  \lim_{\eta \to \infty} e^{\eta}  F = 0 \,.
\end{align}
The bosonic action in the localizing term $QV_{\rm chi} = \wat\scL_{\rm chi}$ is equal to the sum of the original (bosonic) Lagrangian \eqref{LagrangianChi} and a boundary term which vanishes with the chosen asymptotics. 
The real part of the bosonic action consists of a sum of squares as well as the mass term
\be
\frac{1}{L^2}\lp (\mm L + r-1 )^2  - \big(1 + \frac{r}{2} - i w\beta \big)^2 \rp \ti\phi \phi \,,
\ee
where $i\beta$ is real as per our choice of reality condition for the gauge field.
For sufficiently large $\mm$, the above mass term is positive and therefore the action is minimized by
\begin{align}
\phi = F = 0 \,.
\label{LocusChi}
\end{align}
For these values the whole bosonic action, including the imaginary part, vanishes.
However we wish to consider arbitrary $\mm$ and large $i\beta$, in which case the mass squared becomes negative and it is not easy to minimize the real part of the bosonic action. Moreover the action has an imaginary part and the meaning of minimizing the action is unclear.

We will therefore assume, as we did for the vector multiplet, that the saddle point configurations, or localization loci, are given by the field configurations invariant under the supercharge used for the localization, namely $Q$-supersymmetric configurations,
\be
QB = Q\ti B = QC = Q\ti C = 0 \,.
\label{QclosedConfig}
\ee
Let us then solve the supersymmetric equations \eqref{QclosedConfig}. We focus on the chiral multiplet with R-charge $r$ and charge $w$ under an abelian gauge symmetry, with the abelian vector multiplet fields frozen to a localization locus configuration $\scA =  \beta d\chi$, $\sigma = 0$, and flavor background $v = \beta_F d\chi$, $\sigma_F = \mm$. The BPS equations \eqref{QclosedConfig} are explicitly
\be\ba
F &= - i \scL_{\ti P} \phi  \,, \quad \ti F = - i \scL_P \ti\phi \,, \cr
\scL_K \phi &= (\mm + r) \phi \,, \quad \scL_K \ti\phi = -(\mm + r) \ti\phi \,, 
\ea\ee
where we have set $L=1$.
The equations on the first line solve for $F$ and $\ti F$ in terms of $\phi$ and $\ti\phi$.
The equation for $\phi$ on the second line can be re-expressed as
\begin{align}
\p_\chi \phi - i \p_\varphi \phi = m_0 \phi \,,
\end{align}
with
\begin{align}
m_0 &= 
 \mm + i \beta_F + i w \beta  - \frac{ir\tau}{2\tau_2}  \,.
\end{align}
This is solved by
\begin{align}
\phi &= f(\eta, z) e^{\frac{ i m_0}{2} \bar z} \ = \ g(\eta, z) e^{\frac{m_0}{2i} (z - \bar z )} \,,
\end{align}
where $z=\varphi + i\chi$, $f$ is a holomorphic function of~$z$, and the last equality defines $g$ in terms of $f$.
The functions~$\phi$ and~$g$ are periodic in~$\varphi$ and can therefore
be expanded in a   Fourier series. Taking into account the holomorphicity in $z$ we obtain: 
\begin{align}
g(\eta,z)= \sum_{k \in \bZ} a_k(\eta) e^{i k z} \,.
\end{align}
To be globally well-defined $\phi$ must satisfy a second periodicity condition $\phi(\eta, \chi + 2\pi\tau_2, \varphi + 2\pi\tau_1) = \phi(\eta, \chi, \varphi)$. This implies for $g$
\begin{align}
g(\eta, z + 2\pi \tau) =  e^{-2\pi m_0 \tau_2} \, g(\eta, z ) \,.
\end{align}
This in turn leads to $a_k(\eta)=0$ or $e^{2\pi i k \tau} = e^{-2\pi m_0 \tau_2}$ for each $k$. Then there is a non-zero solution for $g$ if and only if there exists $k \in \bZ$ such that
\begin{align}
e^{2\pi i k \tau} = e^{-2\pi m_0 \tau_2} \,.
\end{align}
This is rephrased as 
\begin{align}
m_0 \tau_2 = - i k \tau + i n \,,
\label{specials0}
\end{align}
for some $(k,n) \in \bZ^2$. For generic values of $m_0$ this equation does not have a solution and we 
thus conclude that the localization locus is simply
\begin{align}
\phi = 0 \,, \quad F=0 \,,
\label{LocusChi2}
\end{align}
as found with the naive minimization initially.
The locus equations for $\ti\phi$ are analogous and lead to $\ti\phi=0$ for  generic values of $m_0$. The analysis for a chiral multiplet coupled to a non-abelian gauge field goes along the same lines and leads to the same locus.

\medskip

At the special values \eqref{specials0} of $m_0$ the locus equations admit non-trivial (non-singular) solutions. For instance, this happens for an uncharged massless scalar with even integer R-charge, $w = 0$, $\mm + i \beta_F=0$, $ r \in 2\bZ_{\le 0}$. 
 In the following we assume that~$r > 0$ for all chiral multiplets, in which case the locus is simply given by \eqref{LocusChi2}.

\subsection{Classical contributions }
\label{ssec:Classical}

We discuss here the classical contribution $Z_{\rm class} \equiv e^{S_{\rm class}}$ to the localization formula \eqref{Zformal}, which comes from the evaluation of the various classical actions on the localization locus discussed in the previous section. 
The classical actions described in Section~\ref{ssec:OtherActions} can be evaluated on the chopped $\bH^3_\tau$ with a torus boundary at $\eta = \eta_0$, on the locus configurations. The final evaluation then requires taking $\eta_0 \to \infty$. 

We first provide the evaluations for an abelian gauge field. For simplicity we start with a flavor vector multiplet, a supersymmetric background is given by 
\be
v =  \beta_F d\chi \,, \quad  \sigma  = -\frac{D}{L} = \mm \,,
\ee
with $\beta_F$ and $m$ constant.
The various actions on the chopped $\bH^3_\tau$ evaluate to
\be\ba
& S_{\rm YM} = 0 \,, \quad   S_{\rm YM}^{\rm bdry} = 0  \,, \cr
& S_{\rm CS} = \pi k \tau_2  (\sinh\eta_0)^2 (\mm L)^2   \,, \cr
& S_{\rm CS}^{\rm tot} 
= - \frac{\pi}{2} k \tau_2  (\mm L  + i \beta_F )^2  + O(e^{-2\eta_0}) \,,\cr
& S_{\rm FI} = - 4\pi^2 \tau_2 \xi L (\sinh \eta_0)^2  \mm L  \,, \cr
& S_{\rm FI}^{\rm tot} = 2\pi^2 \tau_2 \xi L    (\mm L + i \beta_F )   + O(e^{-2\eta_0}) \,.
\ea\ee
Note that the combinations with the additional boundary terms conspire nicely to make the action finite in the $\eta_0 \to \infty$ limit, regularizing the infinite volume of the space,
\be
Z_{\rm YM, flavor} = 1 \,, \quad  Z_{\rm CS, flavor} = e^{- \frac{\pi}{2} k \tau_2  (\mm L  + i \beta_F )^2} \,, \quad Z_{\rm FI, flavor} = e^{2\pi^2 \tau_2 \xi L    (\mm L + i \beta _F) } \,.
\label{ZclassFlavor}
\ee
 For a dynamical vector multiplet, the localization locus is given by the configurations
\be
A =  \beta d\chi \,, \quad  \sigma  = -\frac{D}{L} = 0 \,,
\ee
with $\beta$ constant (equal to zero in the Chern-Simons theory). The results are the same as for the flavor background, with $\beta_F$ replaced by $\beta$ and $m$ by zero,
\be
\underline{\rm Abelian \ theory:} \quad Z_{\rm YM, locus} = 1 \,, \quad  Z_{\rm CS, locus} = e^{\frac{\pi}{2} k \tau_2  \beta^2} =1 \,, \quad Z_{\rm FI, locus} = e^{2\pi^2 i \tau_2 \xi L   \beta  } =1 \,.
\label{ZclassAb}
\ee
In the Chern-Simons theory, or more generally with the asymptotic boundary conditions~\eqref{VecBdyCond1}, all these terms evaluate to one, since $\alpha=0$ in the abelian theory and $\beta = i \alpha = 0$. Similarly in the pure Yang-Mills theory with an FI term, the asymptotic constraints~\eqref{BdryCondFI} require $\beta= - i \alpha = 0$.

We now turn to the non-abelian gauge theory and again we focus on the $U(N)$ or $SU(N)$ vector multiplet. The localization locus \eqref{FlatConnUN} is 
\be
\scA = \left\lbrace
\begin{array}{cc}
\lp \beta + \frac{\tau_1}{\tau_2} \alpha \rp d\chi  \,,  &   \qquad \text{for} \quad  0 \le \eta <  \eta_- \,, \\
 \alpha d\varphi  + \beta d\chi \,, &  \text{for} \quad  \eta > \eta_+ \,.
\end{array} 
\right.
\ee
and $ \sigma = -\frac{D}{L} = 0$. The evaluation of the non-abelian Chern-Simons term on the flat connection is given in \cite{Kirk1990} when $\tau_1=0$. Reproducing their computation we find 
\be
\frac{i k}{4\pi} \int \tr \big[ \scA \wedge d\scA - \frac{2i}{3} \scA \wedge \scA \wedge \scA \big] \Big|_{\tau_1 =0} 
=  \pi  i k \tau_2 \tr [\alpha \beta] \,.
\ee
To include the $\tau_1$ dependence, we can go to the coordinates $(\chi, \ti\varphi) \equiv (\chi, \varphi -\frac{\tau_1}{\tau_2} \chi )$ which obey the periodicities of the straight torus, $(\chi, \ti\varphi) \sim (\chi, \ti\varphi + 2\pi) \sim (\chi + 2\pi\tau_2, \ti\varphi)$, and use the computation at $\tau_1=0$,
\be
\frac{i}{4\pi} \int \tr \big[ \scA \wedge d\scA - \frac{2i}{3} \scA \wedge \scA \wedge \scA \big] 
=  \pi  i \tau_2 \tr [\ti\alpha \ti\beta]  = \pi  i \tau_2 \tr \big[\alpha \big( \beta + \frac{\tau_1}{\tau_2}\alpha \big)\big]  \,,
\ee
where we inserted the flat connection parameters $\ti\alpha, \ti\beta$ in the $(\chi, \ti\varphi) $ coordinates.

The classical actions and their associated boundary terms, are easily evaluated:
\begin{align}
& S_{\rm YM} = 0 \,, \quad   S_{\rm YM}^{\rm bdry} = 0  \,, \\
& S_{\rm CS} = \pi  i k \tau_2 \tr \big[\alpha \big( \beta + \frac{\tau_1}{\tau_2}\alpha \big)\big]  \,, \\
& S_{\rm CS}^{\rm tot} 
= \pi k \tau_2 \tr \Big[ -\frac 12 ( \alpha + i \beta )^2 +  2\alpha ( \alpha + i \beta ) + \frac{i\tau}{\tau_2} \alpha^2 \Big] + O(e^{-2\eta_0}) \,,\\
& S_{\rm FI} = 0 \,, \\
& S_{\rm FI}^{\rm tot} = 4\pi^2 \tau_2 \xi L \, \tr \Big[ \frac 12  (\alpha + i \beta ) - \alpha \Big] + O(e^{-2\eta_0}) \,.
\end{align}
The classical contributions with the constraint on asymptotics \eqref{VecBdyCond1}, $2\scA_{\bar z}^{(0)} = \alpha + i\beta =0$, become
\be\ba
& \underline{\text{Non-abelian theory, } \alpha + i\beta =0  :} \cr
& \qquad  Z_{\rm YM, locus} = 1 \,, \quad  Z_{\rm CS, locus} = e^{i \pi k \tau  \tr [\alpha^2]} \,, \quad Z_{\rm FI, locus} = e^{-4\pi^2 \tau_2 \xi L  \tr [\alpha]  } = 1 \,,
\label{ZclassNonAb}
\ea\ee
where in the last equality we have used that $\alpha$ is traceless, which is part of the locus conditions.

We can also consider the second choice of asymptotics \eqref{VecBdyCond2}, that we encountered in our analysis, namely $2\scA_{z}^{(0)} = \alpha - i\beta = C$,  with $C$ a constant Cartan-valued algebra element. For simplicity we consider $C=0$. The classical contribution simplifies in this case as well:
\be\ba
& \underline{\text{Non-abelian theory, } \alpha - i\beta =0  :} \cr
& \hspace{2cm}  Z_{\rm YM, locus} = 1 \,, \quad  Z_{\rm CS, locus} = e^{i \pi k \bar\tau  \tr [\alpha^2]} \,, \quad Z_{\rm FI, locus} = 1 \,. \qquad
\label{ZclassNonAb2}
\ea\ee
For the chiral multiplet, the localization locus is
\be
\phi = F = 0 \,, 
\ee
and the classical action \eqref{LagrChiSusy} and the superpotential terms \eqref{spot} are vanishing
\be
S_{\rm chi} = 0 \,, \quad Z_{\rm chi, locus} =1 \,.
\ee

\subsection{BPS Wilson loops}
\label{ssec:WilsonLoops}

Our set-up easily generalizes to the computations of (the vev of) supersymmetric Wilson loop operators, which are defined in terms of a representation $\scR$ of the gauge group as,
\be
W_{\scR} = \tr_\scR \, P \exp{\oint_{\scC} dt \big( i \scA_\mu \dot x^\mu + \sigma L |\dot x| \big) } \,,
\ee
where $\tr_\scR$ the trace in the representation $\scR$, and $\scC$ is a closed integration cycle parametrized by $t \in [0, 2\pi]$.
The Wilson loop can preserve some supercharges when the integration cycle $\scC$ is embedded appropriately in the bulk geometry.
The supersymmetry invariance under the localization supercharge $Q$ leads to the constraints
\be\ba
0 &= Q \big( i \scA_\mu \dot x^\mu + \sigma L |\dot x| \big) = \frac 12 \lp \Lambda^0 ( \dot K + L |\dot x| ) + \Lambda ^- \dot P + \Lambda^- \dot{\ti P}  \rp \,, \cr
& \Rightarrow  \quad \dot K - L |\dot x| = \dot{P} = \dot{\ti P} = 0 \,, 
\ea\ee
where $\dot X \equiv \dot x^\mu X_\mu$ for $X= K, P, \ti P$. These constraints are solved if and only if the loop is placed at the origin of the $\bH^3_\tau$ space,
\be
\scC = \{ (\eta =0, \chi = \tau_2 t) ; \, t \in [0, 2\pi]  \} \,.
\ee
The vev of the BPS Wilson loop can be computed using the supersymmetric localization technique, in the same way as the partition function, with the extra insertion in the integrand of the evaluation of the loop on the localization locus,\footnote{We choose not to normalize by the partition function.}
\be
\langle W_{\scR} \rangle = \int_{QV[\phi_0] =0} \scD\phi_0 \,  W_{\scR}[\phi_0] \,  e^{-S_{\rm class}[\phi_0]} \, Z_{\rm one-loop}[\phi_0] \,.
\label{ZformalWilson}
\ee
On the vector multiplet locus \eqref{LocusVec}, \eqref{FlatConnUN},  the Wilson loop evaluates to
\be\ba
W_{\scR}[\alpha, \beta] \, = \, \tr_{\scR} \, e^{ 2\pi i  \big( \t_1 \alpha + \t_2\beta \big)  } \, = \, \sum_{w \in \scR} e^{2\pi i \big( \tau_1 w.a + \t_2 w.b \big)} \,,
\label{WilsonLoopFactor}
\ea\ee
where the sum is over the weights $w$ of the $\scR$, which is a representation of $U(N)$ here, and $w.a \equiv \sum_{i=1}^N w_i a_i$ and similarly for $w.b$.


\section{One-loop determinants }
\label{sec:OneLoopDet}

The remaining piece to compute in our formula \eqref{Zformal},\eqref{ZformalWilson} for the exact functional integral
is the one-loop determinant
of the deformation operator~$\int d^3x \sqrt g \, QV$ \eqref{QVdef}   
around the localization locus. 
We will focus on the chiral multiplet one-loop determinant and present three methods to compute it. We will consider the vector multiplet more briefly at the end of the section.
These methods have been used in computations of one-loop determinants on compact space, but their application to the non-compact space $\bH^3_\tau$ encounters some obstacles, that will require extra assumptions or prescriptions for each of them.

The first method relies on an index theorem. The second method is based on boson-fermion mode cancellations, 
counting the contribution of unpaired modes. 
The complications in both these approaches come from the non-compactness of the space. The third method relies 
on heat kernel determinant computations which is well-suited to the Euclidean~AdS$_{3}$ background. 
In this case the difficulty arises from the fact that the method does not preserve supersymmetry manifestly.
 We will provide a  regularization prescription, which we argue is compatible with supersymmetry. In the end we obtain the same final answer for the one-loop determinant from all three methods.

The one-loop determinant is computed for the $U(N)$ gauge theory, around the localization locus \eqref{LocusChi}, \eqref{LocusVec}, \eqref{Locussigma}, 
\be
\phi = F = 0 \,, \quad \sigma = D = 0 \,, \quad \scF_{\mu\nu} =0 \,,
\ee
and vanishing fermions. The flat connections on $\bH^3_\tau$ are given by \eqref{FlatConnUN} and depend on the asymptotic data $\alpha, \beta$.
In order to simplify the computation, we choose in this section a specific gauge for the background gauge field (or localization locus gauge field configuration), namely
\be
\scA = \big( \beta + \frac{\tau_1}{\tau_2}\alpha \big) d\chi \,.
\label{HKgauge}
\ee
To reach this gauge from the flat connection \eqref{FlatConnUN}, one has to perform a gauge transformation whose effect is to send $\eta_-$ to infinity. This is not compatible with the asymptotics of the gauge field that we discussed in Section~\ref{sec:Asymptotics} (in the Chern-Simons theory) and such a gauge transformation is not allowed in the full theory, however for the purpose of computing the one-loop determinant of the chiral multiplet, we can safely ignore the asymptotics chosen for the vector multiplet fields and simply regard $\scA$ as a background. The chiral multiplet Lagrangian $QV_{\rm chi}$ is invariant under gauge transformations, even those affecting the asymptotics, so we can go to the gauge where the flat connection is given in \eqref{HKgauge}.

We denote the quadratic fluctuations of the fields around their locus by the same name ($\phi, B, C, \cdots$).
The fluctuations of a chiral multiplet transforming in the representation $\scR$ of the gauge group can be decomposed into a sum of components labeled by the weights of $\scR$, and
the one-loop determinant  factorizes into a product over the weights $w$ of $\scR$,
\be
Z^{\rm one-loop}_{\rm chi}  = \prod_{w\in \scR} Z^{\rm one-loop}_{\rm abelian}( w.\scA) \,,
\ee
where $Z^{\rm one-loop}_{\rm abelian}( w.\scA)$ is the determinant of a chiral multiplet charged under the abelian gauge field $w.\scA \equiv w\alpha \,  d\varphi  + w\beta \, d\chi \equiv (\sum_{i=1}^N  w_i a_i ) d\varphi  + (\sum_{i=1}^N  w_i b_i ) d\chi $.   
We will therefore concentrate on the computation of the contribution of a single weight $w$, that we denote more simply $Z^{\rm one-loop}_{\text{chi}, \, w}$. As before we will assume R-charge $r$ and real mass $m$.

\subsection{Gauge-fixing}

The index theorem that we wish to use to compute the one-loop determinant relies on the supersymmetry algebra. So far we have only given the algebra \eqref{susyalgebra}, in which the vector multiplet fields appear on the right-hand side. The proper super-algebra arises only after introducing the ghosts for the gauge fixing and combining the supersymmetry transformations with BRST transformations.

This gauge fixing procedure is performed on fluctuations of the fields around a given localization locus configuration $\scA = \scA^{\rm loc}$ \eqref{HKgauge} and $\sigma =0$.
Following standard recipes
(see \emph{e.g.} \cite{Pestun:2007rz, Hama:2012bg, Drukker:2012sr}) we introduce the Grassmann-odd scalar fields $c,\ti c$ 
and the Grassmann-even scalar $b$, valued in the gauge algebra $\mathfrak{g}$. They are all assigned vanishing R-charge.
We then define the BRST transformation $Q_B$ by
\begin{align}
Q_B c &= i c^2 = \frac{i}{2} [c, c]  \,, \quad Q_B \ti c = b \,, \quad Q_B b = 0 \,, \no\\
Q_B{\bf X} &=  \delta_{\rm gauge}(c) \, {\bf X} \,,
\end{align}
where ${\bf X}$ denote a generic field of the vector or chiral multiplet and $\delta_{\rm gauge}(c)$ is the infinitesimal gauge transformation parametrized by $c$, for instance $\delta_{\rm gauge}(c) \sigma = i [c, \sigma]$,  $\delta_{\rm gauge}(c) \scA_\mu = D_\mu c$ \footnote{On a fermionic adjoint valued field we have $\delta_{\rm gauge}(c) \lambda = i [c, \lambda] \equiv  i c_\alpha \lambda_\beta [t^\alpha, t^\beta]$, with $t^{\alpha}$ the $\mathfrak{g}$ generators.}.
The transformations of the ghosts fields under the supercharge $Q$ are chosen to be
\be\ba
Qc &= \frac 12 \big[- \sigma + i K^\mu (\scA^{\rm loc}_\mu- \scA_\mu )\big]  = \frac 12 (\Sigma^{\rm loc} - \Sigma) \,, \cr
Q\ti c &= 0 \,, \quad Qb =  -\frac i2 \scL_K^{\rm loc} \ti c   \,,
\ea\ee
where  $\scL_K^{\rm loc} = K^\mu D^{\rm loc}_\mu$ is covariantized with respect to  $\scA^{\rm loc}$ instead of $\scA$ (and covariant with respect to the R-symmetry connection as before).
These choices ensure that the new supercharge $\wat Q \equiv Q + Q_B$ used for localization obey the modified algebra relation
\begin{align}
\wat Q^2 &= - \frac i2 \, \scL_K^{\rm loc}  + \frac i2  \frac{q_R}{L} \,,
\end{align}
on all fields ${\bf X}' \equiv {\bf X} - {\bf X}^{\rm loc}$, fluctuating around the localization locus, ghosts included.
The $\wat Q$ transformation of the ghosts fields $c, \ti c$ are  
\begin{align*}
\wat Q c &= -\frac 12(\Sigma - \Sigma^{\rm loc}) + i c^2 \,, \quad \wat Q \ti c = b \,.
\end{align*}

The localization computation is accordingly modified by replacing the  deformation term $QV$ by the $\wat Q$-exact term
\be\ba
& \wat Q (V + V_{\rm ghosts} ) = Q V + \wat Q V_{\rm ghosts} \,, \cr
& V_{\rm ghosts} = \int d^3x \sqrt g \, \tr \Big[ \ti c \lp G(\scA- \scA^{\rm loc}) + \frac{\kappa}{2} b  \rp \Big] \,, \cr
& \wat Q V_{\rm ghosts} = \int d^3x \sqrt g \, \tr \Big[ b \lp G(\scA- \scA^{\rm loc}) + \frac{\kappa}{2} b \rp + {\rm ferm.} \Big] \,,
\ea\ee
where $G(\scA- \scA_0)$ is the gauge fixing functional. A standard choice is  $G(\scA - \scA^{\rm loc}) = D^{\text{loc}\,  \mu} (\scA_\mu- \scA^{\rm loc}_{\mu})$, and $\kappa$ is an arbitrary positive number. The result of the path integral does not depend on the choice of $\kappa$ or $G$ \cite{Pestun:2007rz}.

It will be useful to provide the final algebra, including the central deformation by a flavor background,  for a scalar field of gauge charge~$w$, R charge~$q_R$ and flavor charge $q_F$, fluctuating  around the localization locus \eqref{HKgauge} given by $\scA^{\rm loc} = \big( \beta + \frac{\t_1}{\t_2}\alpha \big) d\chi$:
\be
\wat Q^{2} \=  \frac{1}{L} \big[- 2 \partial_{\zbar} 
 + i 2 q_G.\scA^{\rm loc}_{\zbar} +  q_R \frac{\t}{2\t_{2}} +  i q_{F}(mL  + 2 v_{\zbar}) \big] \, \equiv \, \CH \, , 
\label{susyalgebraInd}
\ee
where $q_G, q_R$ and $q_F$ are the gauge, R-symmetry and flavor charges respectively, and~$u_{\zbar}:=\frac12(u_{\v}+iu_{\chi})$ for any vector~$u_{\mu}$. 
In the gauge \eqref{HKgauge} and with $2v_{\zbar}=i\beta_F$, we have
\be
\CH = \frac{1}{L} \big[- 2 \partial_{\zbar}  + i q_G.\big(\beta + \frac{\t_1}{\t_2}\alpha \big) +  q_R \frac{\t}{2\t_{2}} + i q_{F}(mL  + i\beta_F)  \big] \,.
\label{H}
\ee
In the following we analyze the one-loop determinant of the chiral multiplet and we will simply use the notation $\scA$ for $\scA^{\rm loc}$, as in \eqref{HKgauge}, and $Q$ for $\wat Q$.

\subsection{Index theorem}
\label{ssec:IndexTheorem}

In this subsection we evaluate the super-determinant of the operator~$\int d^3x \sqrt g \, QV$ defined in 
Equation~\eqref{QVdef} using the Atiyah-Bott fixed point theorem. 
The content of this theorem~\cite{Atiyah:1974, Berline:1982, Atiyah:1984px},~\cite{Pestun:2007rz} is that, using the supersymmetry 
algebra~\eqref{susyalgebraInd}, one can reduce the one-loop calculation to the  
spacetime fixed points of the operator~$\CH$. 
Writing the quadratic part of the operator~$Q V$ as a sum of bosonic and fermionic terms with quadratic 
operators~$K_{b}$ and~$K_{f}$ respectively, we want to compute the quantity
\be \label{Z1loopFormal}
Z_\text{1-loop} \= \left(\frac{\det K_{f}}{\det K_{b}} \right)^{\frac12} \,. 
\ee
The square root appears because we take the determinant over real degrees of freedom. This will be important, since in the following,  we  will regard the fields $\phi$, $\ti\phi$ (and $B, \ti B$, \dots) as independent. 
We focus on the chiral multiplet determinant in this subsection and we deal with the vector multiplet 
determinant in a later subsection.

The first step 
is to organise the fields of the chiral multiplet into two sets: the elementary 
fields~$({\bf X}_{\rm bos}; {\bf X}_{\rm ferm}) = (\phi, \ti\phi; B, \ti B)$, 
and their Q-superpartners~$(Q{\bf X}_{\rm bos}; Q{\bf X}_{\rm ferm})$. 
The deformation term $V_{\rm chi}$ can be written as follows:
\begin{align}
V_{\rm chi} &= - \lp Q\ti B - \sqrt{2} i \scL_{P} \ti\phi \rp B - \ti B \lp QB - \sqrt{2} i \scL_{\ti P} \phi  \rp  \no\\
& \quad + \lp Q\ti C + \sqrt{2} i \scL_{K} \ti\phi - \sqrt{2}\frac{i}{L} \ti\phi \rp C + \ti C \lp QC + \sqrt{2} i \scL_{K} \phi  + \sqrt{2}\frac{i}{L} \phi \rp \, ,
\end{align}
where we have used the twisted fermionic variables~$(B, \ti B; C, \ti C)$ of the chiral multiplet 
as discussed in Section~\ref{sssec:TVarChi}. 
In terms of the elementary fields and their~$Q$-superpartners, we have:
\be
V_{\rm chi} = (Q{\bf X}_{\rm bos}, {\bf X}_{\rm ferm}) 
\lp
\begin{array}{cc}
D_{00} & D_{01} \\
D_{10} & D_{11} \\
\end{array}
\rp
\binom{{\bf X}_{\rm bos}}{Q{\bf X}_{\rm ferm}}  \,, \label{Vcoh}
\ee
with
\begin{align}
D_{00} = \lp
\begin{array}{cc}
0 & i\scL_K - i \big(\mm + \frac{r-1}{L}\big) \\
 i\scL_K + i \big(\mm + \frac{r-1}{L}\big) & 0 \\
\end{array}
\rp \,, \quad &
D_{01} = \lp
\begin{array}{cc}
0 & 0 \\
0 & 0 \\
\end{array}
\rp \,, \no\\
D_{10} = \lp
\begin{array}{cc}
0 & i \sqrt 2 \scL_{P} \\
 i \sqrt 2 \scL_{\ti P} & 0 \\
\end{array}
\rp \,, \quad &
D_{11} = \lp
\begin{array}{cc}
0 & -1 \\
-1 & 0 \\
\end{array}
\rp \,. \label{D10def}
\end{align}
The above expression of $V_{\rm chi}$ leads directly to
\be\ba
QV_{\rm chi} &= ({\bf X}_{\rm bos}, Q{\bf X}_{\rm ferm}) 
\lp
\begin{array}{cc}
\CH & 0 \\
0 & 1 \\
\end{array}
\rp
\lp
\begin{array}{cc}
D_{00} & D_{01} \\
D_{10} & D_{11} \\
\end{array}
\rp
\binom{{\bf X}_{\rm bos}}{Q{\bf X}_{\rm ferm}}  \cr
& - (Q{\bf X}_{\rm bos}, {\bf X}_{\rm ferm}) 
\lp
\begin{array}{cc}
D_{00} & D_{01} \\
D_{10} & D_{11} \\
\end{array}
\rp
\lp
\begin{array}{cc}
1 & 0 \\
0 & \CH \\
\end{array}
\rp
\binom{Q{\bf X}_{\rm bos}}{{\bf X}_{\rm ferm}}  \,.
\label{QVcoh}
\ea\ee
The ratio of the determinants of the kinetic operators of the fermions and bosons
can be expressed in terms of a similar ratio of the operator~$\CH$: 
\be \label{detratio}
\frac{\det K_{f}}{\det K_{b}} \= \frac{\det_{{\bf X}_{\rm fer}} \CH}{\det_{{\bf X}_{\rm bos}} \CH} 
\= \frac{\det_{{\rm Coker}_{D_{10}}} \CH}{\det_{_{{\rm Ker}_{D_{10}}}} \CH} \, .
\ee
The first equality  follows from \eqref{QVcoh}.
The second equality 
is a consequence of the fact that the operator~$D_{10}$ pairs all modes of the elementary 
fields~$({\bf X}_{\rm bos}; {\bf X}_{\rm ferm})$ with non-zero eigenvalues. 
The right-hand side of~\eqref{detratio} can be computed easily from the $\CH$-equivariant 
index (the variable~$t$ below is purely auxiliary and the result does not depend on it):  
\be \label{indD10} 
\text{ind} (D_{10})(t) := \text{Tr}_{{\rm Ker}_{D_{10}}}  \, e^{-i\CH t} - \text{Tr}_{{\rm Coker}_{D_{10}}}  \, e^{-i\CH t} \, . 
\ee
Indeed, the expansion of the index 
\be
\text{ind} (D_{10})(t) = \sum_{n} a(n) \, e^{-i \lambda_{n} t} 
\ee 
contains the eigenvalues~$\l_{n}$ of~$\CH$ and their indexed degeneracies~$a(n)$, and 
we can read off the ratio of determinants in \eqref{detratio} as:
\be \label{detratio1}
\frac{\det_{{\bf X}_{\rm fer}} \CH}{\det_{{\bf X}_{\rm bos}} \CH} = \prod_{n} \, \l_{n}^{-a(n)} \, . 
\ee
This infinite product, of course, is understood to be regulated, as we discuss below.

So far the discussion was general. Now we specify to the case of interest in this paper which is a three-dimensional 
background with a~$U(1) \times U(1)$ action. 
We use the Atiyah-Bott theorem for the case that there is a $G=H\times K-$action on the space with~$H$ acting freely.
(\cite{Atiyah:1974}, Section~3). This has been discussed recently in~\cite{Drukker:2012sr} for the case~$S^{2} \times S^{1}$. 
Here our space of interest is~$X:=\IH_{3}/\IZ$ with the metric~\eqref{hypglobalmodes0} and the identifications~\eqref{quotient}.
In terms of the coordinates ($\chi'=\chi$, $\v' = \v- \frac{\t_{1}}{\t_{2}} \chi$), 
the periodicity conditions are:
\be
(\chi',\v') \sim (\chi',\v'+2 \pi n), \qquad (\chi',\v') \sim (\chi'+2 \pi n \t_{2}, \v'), \qquad n \in \IZ \, .
\ee
The action of~$H$ and~$K$ on the coordinates are generated by:
\be
H: \; \p_{\chi'} = \p_{\chi} + \frac{\t_{1}}{\t_{2}} \p_{\v} \, , \qquad K: \; \p_{\v'} = \p_{\v} \, . 
\ee
The group~$H$ acts freely on~$X$. 
The right-hand side of the supersymmetry algebra~\eqref{susyalgebraInd} is a combination of generators 
of~$H$ and $K$, and gauge, R-symmetry, and flavor symmetry transformations.

There are two technical points that are important in this discussion of the index theorem. 
Firstly, our space is non-compact with the fields reaching all
the way to the conformal boundary at infinity as discussed in Section~\ref{ssec:BdryCond}. It is not clear that the index theorem as stated 
in~\cite{Atiyah:1974} applies as such to our case\footnote{There is a small discussion of non-compact spaces in~\cite{Atiyah:1974}, 
Chapter 3, but this involves a situation where one can excise a region of the space so as to be left with an effectively compact space. 
This is not, \emph{a priori}, our situation, and we clearly need to deal with the boundary conditions on the various fields carefully. 
Here we use boundary conditions consistent with supersymmetry, as discussed at length in Section~\ref{ssec:BdryCond}. 
We then apply the compact version of the index theorem as such, \emph{assuming} that there are no contributions associated to the boundary. This issue clearly 
needs a more rigorous mathematical treatment.}. The second point is that the operator~$D_{10}$, whose index we compute,
should be transversally elliptic on~$X$ with respect to the~$H$-action.
This means that the determinant of the symbol~$\s(D_{10})$, obtained by replacing the partial derivatives~$\p_{\mu} \to i p_{\mu}$, 
should not vanish for non-zero momenta transverse to the vector field generated by~$H$. 
Such an operator should reduce to an elliptic operator on the quotient space~$X/H$.
That this is true can be verified from the expression~\eqref{D10def} for~$D_{10}$. Upon replacing the partial 
derivatives~$\p_{\mu} \to i p_{\mu}$, we find that~$- \det(\s(D_{10})) = p_{\eta}^{2} + (\coth{\eta} \, p_{\v}+ i \tanh{\eta} \, p_{\chi})^{2}$. 
The determinant vanishes when~$p_{\eta}=0$ and~$\coth{\eta} \, p_{\v}+ i \tanh{\eta} \, p_{\chi} =0$. This shows that 
it is not elliptic, because at $\eta=0$ the equation is satisfied for arbitrary $p_\chi$. When the momentum parallel to the~$H$-action vanishes, \emph{i.e.}~$p_{\chi'}=p_{\chi}+\frac{\t_{1}}{\t_{2}}p_{\v}=0$, 
we have that~$- \det(\s(D_{10})) = p_{\eta}^{2} + (\coth{\eta} - i\frac{\t_{1}}{\t_{2}} \tanh{\eta})^{2} \, p_{\v}^{2}$.
This determinant vanishes only when~$p_{\eta}=p_{\v}=0$. In other words, the operator indeed reduces to an elliptic operator 
on the quotient space~$X/H$. 

In such cases the index is equal to a sum over representations of~$H$:
\be
\text{ind}(D_{10}) = \sum_{\text{R = Rep}(H)} \, \text{ind}(D^{\text{R}}) \,  \chi_{\text{R}}(h) \, , 
\ee
where~$D^{\text{R}}$ is the operator 
that is induced by~$D_{10}$ on~$(X \times \text{R})/H$, and~$\chi_{\text{R}}$ is the character 
of the representation. 
Noting that the radius of the non-contractible circle is~$R_{\chi'}=\t_{2} L$, 
we can identify the group element~$h=\exp(-it/\tau_{2}L)$.

Now,  the operator~$D^{\text{R}}$ is independent of the representation~$\text{R}$, since the group~$H$ is abelian,
and so we denote it by~$D^{\text{R}}=D'$. 
The representations of~$H$ are labelled by~$n_{2} \in \IZ$. Thus we have:
\be \label{DD'rel}
\text{ind}(D_{10}) = \sum_{n_{2} \in \IZ} \, \text{ind}(D') \,  h^{n_{2}} \, . 
\ee
Having thus factored out the~$H$-dependence, the problem reduces to computing the equivariant 
index of the operator~$D'$ on the space~$X/H$, with respect to the following combined action of~$K$ 
and the internal symmetries:
\be
\CH' \, \equiv \, \CH -  \frac{(-i)}{L} \p_{\chi'} \, = \, \frac{1}{L} \bigl(i\frac{\t}{\t_{2}}\p_{\v} 
+ i 2 q_G.\scA_{\zbar} +  \frac{q_{R}\t}{2\t_{2}} +  i q_{F}(mL  + i\beta_F) \bigr) \, .
 \ee

The operator~$\CH'$ acts on the quotient space~$X/H$ as a translation of~$\v$. 
This action, that we denote by~$x \mapsto \wt x = e^{-i\CH' t} x$, has a fixed point
at the center~$\eta=0$. The index of the operator~$D'$ reduces to the fixed points of the manifold~$X/H$ 
under the action of~$\CH'$:
\be \label{ASindthm}
\text{ind} (D')(t) = \sum_{\{x \mid \wt x = x\}} \frac{\bigl( \text{Tr}_{{\bf X}_{\rm bos}} \, - \, \text{Tr}_{{\bf X}_{\rm fer}} \bigr)  \, 
e^{-i\CH' t }}{\det (1- \p \wt x/\p x)} \, .
\ee

The calculation is simplified by going to complex coordinates in which the metric on the space~$X/H$ is
\begin{equation}\label{metriccomplex}
    ds^2 = L^{2} (d \eta^{2} + \sinh^{2} \eta \, d\v^{2})  = L^{2} \frac{4 dw d\bar w}{(1- w \bar w)^2}  \, .
\end{equation}
At the fixed point~$w=0$, the action of the operator~$e^{-i\CH' t}$ on the spacetime coordinates 
is~$w \mapsto \exp(\frac{t\tau}{L\t_2}) w$.
Therefore, the determinant factor in the denominator of~\eqref{ASindthm} is, with~$p = e^{- i t/L}$:
\be
\det (1- \p \wt x/\p x) = (1-p^{-i\t/\t_{2}}) \, (1-p^{i\t/\t_{2}}) \, . 
\ee

We now need the charges of the elementary fields $({\bf X}_{\rm bos}; {\bf X}_{\rm ferm}) = (\phi, \ti\phi; B, \ti B)$
at the fixed point under the operator~$\CH'$, which reduces to 
\be
\CH'\mid_{\eta=0}  \=  \frac{1}{L} \Bigl(i\frac{\t}{\t_{2}}\p_{\v} 
 - q_G.\bigl(\beta + \frac{\tau_1}{\tau_2} \alpha \bigr)  +  \frac{q_{R}\t}{2\t_{2}} +  i q_{F}(mL  + i\beta_F)   \Bigr) \, .
 \ee
All the fields in the twisted variables are scalars so they are neutral under the first term~$\p_{\v}$. 
The charges of~$(\phi, \wt \phi)$  are~$\mp (i\mu + \frac{r\t}{2\t_{2}})/L $, 
and those of~$(B, \wt B)$ are~$\mp (i\mu + \frac{(r-2)\t}{2\t_{2}})/L$, respectively,
where\footnote{We remind the reader that we are focusing here on the contribution to the one-loop determinant of the fields with gauge charge/weight $w$.}
\be
\mu= \mm L + i\beta_F + i w.\bigl(\beta + \frac{\tau_1}{\tau_2} \alpha \bigr) \,.
\label{mu}
\ee 
The index of the operator~$D'$ is thus:
\bea \label{indD'ans}
\text{ind}(D')(t) & = & 
\frac{p^{\mu - \frac{i r \tau}{2\tau_2}} +p^{-\mu + \frac{i r \tau}{2\tau_2}} -p^{\mu - \frac{i (r-2) \tau}{2\tau_2}} - p^{-\mu + \frac{i (r-2) \tau}{2\tau_2}}}{(1-p^{-i\t/\t_{2}})(1-p^{i\t/\t_{2}})} \cr
& = & \frac{p^{\mu - \frac{i r \tau}{2\tau_2}} }{1-p^{-i\t/\t_{2}}}  - \frac{p^{-\mu + \frac{i (r-2) \tau}{2\tau_2}} }{1-p^{-i\t/\t_{2}}} \cr
& = & \sum_{n_{1} =0}^{\infty} \, p^{\mu - i (r/2 + n_1)\t/\t_2 }  \, -\, p^{-\mu  - i(n_{1}+1-r/2)\t/\t_{2} }  \, ,
\eea
where we expanded in powers of $p^{-i\t/\t_{2}} = e^{ -\frac{t \t}{L \t_2 }}$.
Putting together Equations~\eqref{DD'rel} and \eqref{indD'ans} we obtain the result:
\be
\text{ind}(D_{10}) (t) = \sum_{n_{2} \in \IZ \atop n_{1} \ge 0} \, 
e^{(-\frac{it}{L})(\mu -i \frac{(n_{1}+r/2)\t}{\t_{2}}+ i \frac{n_{2}}{\t_{2}})}  
\, -\, e^{(-\frac{it}{L})(-\mu  -i \frac{(n_{1}+1- r/2)\t}{\t_{2}} + i \frac{n_{2}}{\t_{2}})} \, . 
\ee
From this expression, we read off the one-loop determinant:
\bea \label{ZIndexchiNA}
[1] \qquad Z_{\text{chi},\, w}^\text{1-loop} & = & \Big( \prod_{n_{2}\in \IZ \atop n_{1} \ge 0} 
\frac{-(n_{1}+1- \frac r2)\frac{i\t}{\t_{2}} + i \frac{n_{2}}{\t_{2}} - \mu }{- (n_{1} + \frac r2 ) \frac{i\t}{\t_{2}}
+ i \frac{n_{2}}{\t_{2}} + \mu } \, \Big)^{1/2} \,.
\eea 

Finally we make some comments about the expansion of expression~\eqref{indD'ans} for the index. In the above  
treatment we expanded it in powers of~$p^{-i\t/\t_{2}} = e^{-\frac{t \t}{L \t_2 }}$. Alternatively, if we  expand in powers of $p^{i\t/\t_{2}} = e^{ \frac{t \t}{L \t_2 }}$, we would obtain:
\be \label{indD'ans2}
\text{ind}(D')(t) 
 =  - \frac{p^{\mu - \frac{i (r-2) \tau}{2\tau_2}} }{1-p^{i\t/\t_{2}}}  + \frac{p^{-\mu + \frac{i r \tau}{2\tau_2}} }{1-p^{i\t/\t_{2}}} 
 =  \sum_{n_{1} =0}^{\infty} \,  - p^{\mu + i (n_1 + 1 - r/2)\t/\t_2 }  \, + \, p^{-\mu  + i(n_1 + r/2 )\t/\t_{2} }  \, ,
\ee
leading to
\bea \label{ZIndexchiNA2}
[2] \qquad  Z_{\text{chi},\, w}^\text{1-loop} & = & \Big( \prod_{n_{2}\in \IZ \atop n_{1} \ge 0} 
\frac{(n_{1}+1- \frac r2)\frac{i\t}{\t_{2}} - i \frac{n_{2}}{\t_{2}} + \mu }{(n_{1} + \frac r2 ) \frac{i\t}{\t_{2}}
- i \frac{n_{2}}{\t_{2}} - \mu } \, \Big)^{1/2} \, ,
\eea
which is actually the same as \eqref{ZIndexchiNA}.
If we try instead a mixed (and \emph{a priori} unnatural) expansion in $p^{i\t/\t_{2}}$ and $p^{-i\t/\t_{2}}$ for the two 
terms in~\eqref{indD'ans}, for instance
 \be \label{indD'ans3}
\text{ind}(D')(t) 
 =  \frac{p^{\mu - \frac{i r \tau}{2\tau_2}} }{1-p^{-i\t/\t_{2}}}  + \frac{p^{-\mu + \frac{i r \tau}{2\tau_2}} }{1-p^{i\t/\t_{2}}} 
 =  \sum_{n_{1} =0}^{\infty} \,   p^{\mu - i (n_1 + r/2)\t/\t_2 }  \, + \, p^{-\mu  + i(n_1 + r/2 )\t/\t_{2} }  \, ,
\ee
we obtain
\bea \label{ZIndexchiNA3}
[3] \qquad  Z_{\text{chi},\, w}^\text{1-loop}& = &  \prod_{n_{2}\in \IZ \atop n_{1} \ge 0} 
\frac{1}{(n_{1} + \frac r2 ) \frac{i\t}{\t_{2}}
+ i \frac{n_{2}}{\t_{2}} - \mu }  \, ,
\eea
up to an irrelevant overall phase. The remaining possible option is 
 \be \label{indD'ans4}
\text{ind}(D')(t) 
 =  -\frac{p^{\mu - \frac{i (r-2) \tau}{2\tau_2}} }{1-p^{i\t/\t_{2}}}  - \,  \frac{p^{-\mu + \frac{i (r-2) \tau}{2\tau_2}} }{1-p^{-i\t/\t_{2}}}  =   \sum_{n_{1} =0}^{\infty} \, -  p^{\mu + i (n_1 +1 - r/2)\t/\t_2 }  \, - \, p^{-\mu  - i(n_1 +1 - r/2 )\t/\t_{2} }  \,,
\ee
which gives:
\bea \label{ZIndexchiNA4}
[4] \qquad  Z_{\text{chi},\, w}^\text{1-loop} & = &  \prod_{n_{2}\in \IZ \atop n_{1} \ge 0} 
\big( (n_{1}+1- \frac r2)\frac{i\t}{\t_{2}} + i \frac{n_{2}}{\t_{2}} + \mu \big) \, .
\eea
These last two choices give answers different from~\eqref{ZIndexchiNA}, with only bosonic or only 
fermionic net contributions to the one-loop determinant. As it is natural to expand a given meromorphic function 
at one point only in one expansion parameter, we arrive at the conclusion that 
only $p^{i\t/\t_{2}}$ or only $p^{-i\t/\t_{2}}$ are correct, leading both to~\eqref{ZIndexchiNA}. 
We will see that this ambiguity of the four different expansions exists also in the method that we will discuss in the next section,
and there is a different principle which singles out  the same result~\eqref{ZIndexchiNA}.

\subsection{Unpaired eigenmodes}
\label{ssec:UnpairedEigenmodes}

The one-loop determinant can be alternatively computed via the unpaired eigenmodes method, which can be seen as a complementary point of view on the result of the index computation. The idea is 
to exploit the large cancellation between fermion and boson eigenvalues, leaving only the contributions of unpaired eigenvalues.  It has been used for instance in \cite{Hama:2011ea, Alday:2013lba, Closset:2013sxa}. The advantage of this method is that it gives us the knowledge of the actual modes which contribute to the final result, and therefore gives us more insights into the physics behind the computation.

The first part of the calculation is identical to that of the previous computation, and therefore we will take as our starting point the formula \eqref{detratio}
\be 
Z_{\text{chi},\, w}^\text{1-loop}  =  \Big( \, \frac{\det_{{\bf X}_{\rm fer}} \CH}{\det_{{\bf X}_{\rm bos}} \CH} \, \Big)^{1/2} = \Big( \, \frac{\det_{B, \ti B} \CH}{\det_{\phi, \ti\phi} \CH} \, \Big)^{1/2} \,,
\label{Z1loopUnpaired}
\ee
where the determinants are over all bosonic modes $\phi, \ti\phi$ and fermionic modes  $B, \ti B$ respectively. We wish now to work out the pairing between bosonic and fermionic modes, which results in eigenvalues cancellations between the numerator and the denominator. Importantly, we will consider the fields $\phi, B$ and $\ti \phi, \ti B$ as independent and treat them separately.

First we observe that the operators $\scL_{P}$ and $\scL_{\ti P}$ commute with $\CH$:
\be
[\CH, \scL_{P}] = 0 \,, \quad [\CH, \scL_{\ti P}] = 0 \,.
\label{CommutationRelH}
\ee
This follows from writing $\CH$ as $\CH = -i ( \scL_K -  Q_R \frac{1}{L} - Q_{F} m  )$, with $Q_F$ the flavor 
charge operator, the commutation relations~\eqref{LieDerivativeRel}, and the commutation relations 
with the R-charge operator: $[Q_R,  \scL_{P}] = 2  \scL_{P}$, $[Q_R,  \scL_{\ti P}] = -2  \scL_{\ti P}$. 
The commutation relations \eqref{CommutationRelH} imply that $\scL_{P}$ and $\scL_{\ti P}$ can be used as operators pairing bosonic and fermionic modes of appropriate R-charges with the same $\CH$ eigenvalues.\footnote{We remind the reader that the fields $(\phi, \ti\phi, B, \ti B)$ have R-charge $(r, -r, r-2, 2-r)$ respectively, and that the $\scL_P$ and $\scL_{\ti P}$ operators raise, and respectively lower, the R-charge by 2.} For instance, for a mode $\phi$ with eigenvalue $\lambda$, $\CH \phi = \lambda\phi$, we have a corresponding mode
\be\ba
B = \scL_{\ti P} \phi , \quad \CH B = \lambda B \,.
\ea\ee
The contribution of this pair $(\phi, B)$ to the one-loop determinant \eqref{Z1loopUnpaired} is trivial (equal to one), since the fermionic and bosonic eigenvalues cancel each other. The net contribution to \eqref{Z1loopUnpaired} is reduced to the $\phi$ modes which have no fermionic partner, namely the modes obeying $\scL_{\ti P} \phi =0$, and the $B$ modes which have no bosonic partner, namely the modes $B \neq \scL_{\ti P} \phi$ for any $\phi$. These are the definitions of the kernel and cokernel of $\scL_{\ti P}$ respectively. The net contribution to the one-loop determinant of the $\phi$ and $B$ modes is then 
\be
[a] \qquad  (\phi, B) \quad \rightarrow \quad  \lp \frac{\prod_{B \in {\rm Coker} \scL_{\ti P}} \lambda_B }{\prod_{\phi \in {\rm Ker} \scL_{\ti P}} \lambda_\phi} \rp^{1/2} \,.
\ee
Alternatively we can think of pairing the $\phi$ and $B$ fields using the $\scL_P$ operator, and associate to a mode $B$ with eigenvalue $\lambda$, the mode
\be\ba
\phi = \scL_{ P} B , \quad \CH \phi = \lambda \phi \,.
\ea\ee
This leads to a net contribution to the one-loop determinant of $\phi$ and $B$ of the form
\be
[b] \qquad  (\phi, B) \quad \rightarrow  \quad \lp \frac{\prod_{B \in {\rm Ker} \scL_{P}} \lambda_B }{\prod_{\phi \in {\rm Coker} \scL_{P}} \lambda_\phi} \rp^{1/2} \,.
\ee
Similarly we can pair the fields $\ti \phi$ and $\ti B$ using either $\scL_{\ti P}$ or $\scL_{P}$, namely $\ti\phi = \scL_{\ti P} \ti B$ or $\ti B = \scL_{ P} \ti \phi$, leading to a net contribution to the one-loop determinant of $\ti\phi$ and $\ti B$ of the form
\be
[a'] \qquad  (\ti\phi, \ti B) \quad \rightarrow  \quad \lp \frac{\prod_{\ti B \in {\rm Ker} \scL_{\ti P}} \lambda_{\ti B} }{\prod_{\ti\phi \in {\rm Coker} \scL_{\ti P}} \lambda_{\ti\phi}} \rp^{1/2} \,,
\ee
or 
\be
[b'] \qquad  (\ti\phi, \ti B) \quad \rightarrow  \quad \lp \frac{\prod_{\ti B \in {\rm Coker} \scL_{P}} \lambda_{\ti B} }{\prod_{\ti\phi \in {\rm Ker} \scL_{P}} \lambda_{\ti\phi}} \rp^{1/2} \,.
\ee
In previous studies of one-loop determinants from the unpaired eigenmodes method, for instance for the one-loop determinant on the three-sphere \cite{Hama:2011ea, Alday:2013lba}, the pairing operators $\scL_P$ and $\scL_{\ti P}$ are the adjoint of each other, implying
the identities $\text{Coker} \scL_P = \text{Ker} \scL_{\ti P}$ and $\text{Coker} \scL_{\ti P} = \text{Ker} \scL_{P}$. The two choices of pairing $[a]$ and $[b]$ described above are then equivalent, and similarly for the two choices $[a']$ and $[b']$. In addition, the eigenvalues $\lambda_\phi$ of the $\phi$ modes in $\text{Ker} \scL_{\ti P}$ are paired with the opposite egenvalues $\lambda_{\ti\phi} = -\lambda_\phi$ of the complex conjugate $\ti\phi$ modes in $\text{Ker} \scL_{P}$, so that, up to a sign, we have $\prod_{\phi \in {\rm Ker} \scL_{\ti P}} \lambda_{\phi} = \prod_{\ti\phi \in {\rm Ker} \scL_{P}} \lambda_{\ti\phi}$.  The final one-loop determinant in these cases reduces to $Z = (\det_{\text{Ker} \scL_{\ti P}} \CH / \det_{\text{Ker} \scL_{P}} \CH )$.

To get an idea of which modes contribute to the one-loop determinant in our case, we study now the kernels and cokernels of $\scL_P$ and $\scL_{\ti P}$ explicitly.

A straightforward computation shows that the modes $X$ satisfying $\scL_{\ti P} X =0$ and the modes $Y$ satisfying $\scL_{P} Y = 0$, of R-charge $q_R$, flavor charge $q_F$ and gauge charge $w$, and their $\CH$ eigenvalues,
are locally given by\footnote{The $X$ modes in Ker $\scL_{\ti P}$ can be modes of the fields $\phi$ or $\ti B$. The $Y$ modes in Ker $\scL_{P}$ can be modes of the fields $\ti\phi$ or $B$.  }
\be\ba\label{UnpairedModes}
\underline{\scL_{\ti P}X = 0:} \quad  & X(\rhota, \tauc, \varphi) =  \, e^{-i n_1 \varphi} \, e^{ i p_X  \tauc} \, (\sinh\rhota)^{n_1} \, (\cosh\rhota)^{- i (p_X - \Qb )} \, ,   \cr
&\textrm{with} \quad   \lambda_X  \ = \   p_X - \Qb + i (  q_F\mm + q_R + n_1 )  \, ,  \\[3mm]
\underline{\scL_{P}Y=0:} \quad  &  Y(\rhota, \tauc, \varphi) = e^{i n_1 \varphi} \,e^{ i  p_Y  \tauc} \, (\sinh\rhota)^{n_1} \, (\cosh\rhota)^{i (p_Y - \Qb)} \, , \cr
&\textrm{with} \quad   \lambda_Y  \ = \   p_Y - \Qb + i  (  q_F\mm + q_R - n_1 )  \, , 
\ea\ee
where we have set $L=1$, and with $\Qb \equiv q_F\beta_F +w.\big( \beta + \alpha \frac{\t_1}{\t_2} \big) - q_R \frac{\bar\tau}{2\tau_2}$.
Periodicity in the $\varphi$ direction then requires $n_1 \in \bZ$.
Periodicity under the quotient \eqref{quotient} imposes, for a given $n_1 \in \bZ$, the quantization of $p_X$ and $p_Y$\footnote{Here the integers $n_1,n_2$ characterizing the mode $X$ are unrelated to the integers $n_1,n_2$ characterizing the mode $Y$. }
\begin{align}
p_X &= \frac{n_2 +  n_1 \tau_1}{\tau_2}  \, , \quad n_2 \in \bZ \, ,  \quad p_Y =  -\frac{n_2 + n_1 \tau_1}{\tau_2}  \, , \quad n_2 \in \bZ \,.
\end{align}
We therefore obtain modes labeled by two integers $(n_1,n_2) \in \bZ^2$. Regularity (or normalizability) of the modes at the origin of the space $\eta=0$ leads us to exclude the modes with $n_1 <0$. If we wish to exclude also the modes which are not square normalizable, $\int |X|^2 < \infty$, $\int |Y|^2 < \infty$, then, for the R-charge $r$ lying in a canonical range $0<r<2$, we would exclude all the modes with $n_1 \geq 0$, leaving no modes at all in the kernels.
On the other hand if we allow for the modes diverging at infinity, then the kernels of $\scL_{\ti P}$ and $\scL_P$ are spanned by the $X_{n_1,n_2}$ modes and $Y_{n_1,n_2}$ modes respectively, with $(n_1,n_2) \in \bZ_{\geq 0}\times \bZ$. This leads to
\be\ba
& \prod_{\phi \in {\rm Ker}' \scL_{\ti P}} \lambda_\phi  = \prod_{n_{2}\in \IZ \atop n_{1} \ge 0} 
\big( (n_{1}+\frac r2)\frac{\t}{\t_{2}} + \frac{n_{2}}{\t_{2}} + i \mu \big)  = \prod_{\ti\phi \in {\rm Ker}' \scL_{P}} (-\lambda_{\ti\phi})  \,, \cr
& \prod_{B \in {\rm Ker}' \scL_{P}} (- \lambda_{B})  = \prod_{n_{2}\in \IZ \atop n_{1} \ge 0} 
\big( (n_{1}+ 1 -\frac r2)\frac{\t}{\t_{2}} + \frac{n_{2}}{\t_{2}} - i \mu \big)  = \prod_{\ti B \in {\rm Ker}' \scL_{\ti P}}  \lambda_{\ti B}  \,,
\label{Kernels}
\ea\ee
where ${\rm Ker}'$ indicates that we count unpaired modes which diverge at infinity.

We now study the cokernels and ask first the question whether there are modes which cannot be written in the form $\scL_{\ti P} X$. It is enough to focus on a basis of fields with a given momentum $(n_1,n_2) \in \bZ^2$, $-i \p_\varphi X_{n_1,n_2} = n_1 X_{n_1,n_2}$, $-i \p_\chi X_{n_1,n_2} = p_X X_{n_1,n_2}$, with $p_X = -\frac{n_2 +  n_1 \tau_1}{\tau_2}$.   
 We find that for any field $\check X_{n_1,n_2}$, there is a corresponding local mode $X_{n_1+1,n_2}$ such that $\check X_{n_1,n_2} = \scL_{\ti P} X_{n_1+1,n_2}$, given by
 \be\ba
 X_{n_1+1,n_2}(\eta) &= -   \int_{\eta_0}^\eta d\eta' \bigg(\frac{\sinh\eta'}{\sinh\eta}\bigg)^{n_1} \bigg(\frac{\cosh\eta'}{\cosh\eta}\bigg)^{i (p_X - \Qb)} e^{i \lp \varphi - \frac{\t_1}{\t_2}\chi \rp} \check X_{n_1,n_2}(\eta') \,.
 \ea\ee
 The above mode may not be well-defined, or normalizable, at $\eta=0$ for any choice of $\eta_0$. Since $\check X_{n_1,n_2}$ is a normalizable mode at $\eta=0$, we have $\check X_{n_1,n_2}(\eta) \sim \eta^{x}$, as $\eta \to 0$, with $x >-1$.
An analysis of the behavior $X_{n_1,n_2}$ near $\eta=0$ reveals that, for $n_1 + x \geq -1$, the choice $\eta_0=0$ leads to a well-defined/normalizable mode at the origin, while for $n_1 + x < -1$, any constant $\eta_0 >0$ is enough to enforce normalizability at the origin.

If we choose to include non-normalizable modes of $X$ in the allowed set of modes, then we conclude that there is always a choice of $\eta_0$ such that $ X_{n_1+1,n_2}$ is well-defined and therefore the cokernel of $\scL_{\ti P}$ is empty.  On the other hand if we require $X_{n_1+1,n_2}$ to be also normalizable at infinity $\eta \to \infty$, then we would find modes $\check X_{n_1,n_2}$ for which $X_{n_1+1,n_2}$ is not well-defined, and the cokernel of $\scL_{\ti P}$ would not be empty. Exhibiting a basis of modes of the cokernel is this case requires more work.

Similarly we find that if we accept non-normalizable modes (at infinity) in the spectrum, then the cokernel of $\scL_{P}$ is empty, and if we do not accept these modes, then it is not empty. 

Gathering all the results, we find that depending on which normalizablity condition (at infinity) that we impose on the various fields, we may obtain different answers. To select these normalizability conditions, we assume that the following identities hold:
\be\ba
& \text{Coker}_{X} \scL_P = \text{Ker}_{X} \scL_{\ti P} \,, \quad  \text{Coker}_{X} \scL_{\ti P} = \text{Ker}_{X} \scL_{P} \,.
\label{KerCokerRel}
\ea\ee
 Here (Co)Ker$_X$ denotes the (co)kernel over the field $X$. These identities are true in the standard situation when $\scL_P$ and $\scL_{\ti P}$ are adjoint operators\footnote{In our situation the operators $\scL_P$ and $\scL_{\ti P}$ are not the adjoint of each other, at least under the naive hermitian conjugation. One may define a $\ddagger$ operation relating $\scL_P$ and $\scL_{\ti P}$, inherited from the Wick rotation of the Lorentzian theory. Such a complex conjugation would act on the coordinate $\chi$, as well as on the parameter $\tau_1$, as if they were purely imaginary. It may be possible to use these observations to justify \eqref{KerCokerRel} more rigorously.}. More importantly, they ensure that the contribution from $\phi, B$ computed in the methods $[a]$ and $[b]$ give the same answer, and similarly for the contribution of $\ti\phi, \ti B$ computed with the methods $[a']$ and $[b']$.

This leaves us with four possible choices of normalizability:
\begin{itemize}
\item \underline{$\phi, \ti B$ non-normalizable, $\ti\phi, B$ normalizable:} \\
We have $\text{Ker}_\phi \scL_{\ti P} \neq \emptyset$, since $\phi$ can be non-normalizable, $\text{Ker}_{\ti\phi} \scL_{P} = \emptyset$, since $\ti\phi$ is normalizable, $\text{Coker}_{B} \scL_{\ti P} = \emptyset$, since we allow for pairing $B$ with non-normalizable $\phi$, and $\text{Coker}_{\ti B} \scL_P \neq \emptyset$, since we only allow for pairing $\ti B$ with normalizable $\ti\phi$, leaving some modes unpaired in the cokernel. Similarly we also have $\text{Ker}_B \scL_{P} = \emptyset$, $\text{Ker}_{\ti B} \scL_{\ti P} \neq \emptyset$, $\text{Coker}_{\phi} \scL_{P} \neq \emptyset$, $\text{Coker}_{\ti\phi} \scL_{\ti P} = \emptyset$. This is compatible with \eqref{KerCokerRel} for each field. Assuming that \eqref{KerCokerRel} holds, we obtain, from any possible choice of pairings, using \eqref{Kernels} for non-empty kernels, 
\be
Z_{\text{chi},\, w}^\text{1-loop} = \lp  \frac{\prod_{\ti B \in {\rm Ker'} \scL_{\ti P}} \lambda_{\ti B} }{\prod_{\phi \in {\rm Ker'} \scL_{\ti P}} \lambda_\phi} \rp^{1/2} = \ \Big(  \prod_{n_{2}\in \IZ \atop n_{1} \ge 0} 
\frac{(n_{1}+ 1 -\frac r2)\frac{\t}{\t_{2}} + \frac{n_{2}}{\t_{2}} - i \mu }{ (n_{1}+\frac r2)\frac{\t}{\t_{2}} + \frac{n_{2}}{\t_{2}} + i \mu }   \Big)^{1/2} \,,
\label{ZIndexchiUnpaired}
\ee
up to an overall phase $\prod (\pm i)$ which we drop. 
\item \underline{$\ti\phi, B$ non-normalizable, $\phi, \ti B$ normalizable:} \\
This is the inverse case. The non-empty spaces are $\text{Ker}_{\ti\phi} \scL_{P}$, $\text{Coker}_{B} \scL_{\ti P}$, $\text{Ker}_B \scL_{P}$ and$\text{Coker}_{\ti\phi} \scL_{\ti P}$, which is compatible with \eqref{KerCokerRel}. Assuming that \eqref{KerCokerRel} holds, and using  \eqref{Kernels}, the partition function evaluates to 
\be
Z_{\text{chi},\, w}^\text{1-loop} = \lp  \frac{\prod_{B \in {\rm Ker'} \scL_{P}} \lambda_{B} }{\prod_{\ti \phi \in {\rm Ker'} \scL_{P}} \lambda_{\ti \phi}} \rp^{1/2} = \ \Big(  \prod_{n_{2}\in \IZ \atop n_{1} \ge 0} 
\frac{(n_{1}+ 1 -\frac r2)\frac{\t}{\t_{2}} + \frac{n_{2}}{\t_{2}} - i \mu }{ (n_{1}+\frac r2)\frac{\t}{\t_{2}} + \frac{n_{2}}{\t_{2}} + i \mu }   \Big)^{1/2} \,,
\ee
up to an overall phase. This is the same answer as with the first choice of normalizability.
\item \underline{$\phi, \ti\phi$ non-normalizable, $B, \ti B$ normalizable:} \\
The non-empty spaces are $\text{Ker}_{\ti\phi} \scL_{P}$, $\text{Ker}_{\phi} \scL_{\ti P}$, $\text{Coker}_{\phi} \scL_{P}$ and $\text{Coker}_{\ti\phi} \scL_{\ti P}$. In this case the one-loop determinant receives contributions only from the bosonic modes,
\be
Z_{\text{chi},\, w}^\text{1-loop} = \frac{1}{\big( \prod_{\phi \in {\rm Ker'} \scL_{\ti P}} \lambda_{\phi} \prod_{\ti \phi \in {\rm Ker'} \scL_{P}} \lambda_{\ti \phi} \big)^{1/2}}  = \  \prod_{n_{2}\in \IZ \atop n_{1} \ge 0} 
\frac{1 }{ (n_{1}+\frac r2)\frac{\t}{\t_{2}} + \frac{n_{2}}{\t_{2}} + i \mu }  \,.
\ee
\item \underline{$B, \ti B$ non-normalizable, $\phi, \ti\phi$ normalizable:} \\
The non-empty spaces are $\text{Ker}_{B} \scL_{P}$, $\text{Ker}_{\ti B} \scL_{\ti P}$, $\text{Coker}_{\ti B} \scL_{P}$ and $\text{Coker}_{B} \scL_{\ti P}$. In this case the one-loop determinant receives contributions only from the fermionic modes,
\be
Z_{\text{chi},\, w}^\text{1-loop} =  \big( \prod_{\ti B \in {\rm Ker'} \scL_{\ti P}} \lambda_{\ti B} \prod_{B \in {\rm Ker'} \scL_{P}} \lambda_{B} \big)^{1/2}  = \  \prod_{n_{2}\in \IZ \atop n_{1} \ge 0} 
\big( (n_{1}+1 -\frac r2)\frac{\t}{\t_{2}} + \frac{n_{2}}{\t_{2}} - i \mu \big)  \,.
\ee
\end{itemize}
Notice that the four results given above are precisely matching the four results \eqref{ZIndexchiNA}, \eqref{ZIndexchiNA2}, \eqref{ZIndexchiNA3}, \eqref{ZIndexchiNA4}, obtained from the four possible expansions of the index. We can therefore associate to each index result a certain choice of normalizability for the fields $\phi, \ti\phi, B, \ti B$, and provide the unpaired modes contributing to the one-loop determinant in each case. 

In the index computation, we discarded the last two results above, selecting purely bosonic or purely fermionic unpaired modes, on the basis that they required unnatural expansions of the index. Here we may discard these choices of normalization on the physical ground that allowing both $\phi$ and $\ti\phi$ non-normalizable modes (divergent at infinity), or $B$ and $\ti B$ non-normalizable modes, would make the action of the theory diverge. Instead, allowing diverging modes for $\phi$, but not $\ti\phi$, or diverging modes of $B$, but not $\ti B$, does not lead to an obvious contradiction, since the diverging modes do not have conjugate modes and therefore do not seem to appear in the action. 

We conclude that only the two first choices above are physical. They  both lead to the same result \eqref{ZIndexchiUnpaired}, in agreement with the index computation \eqref{ZIndexchiNA}. The main difference with previous unpaired eigenmodes computations in the literature is the asymmetric treatment of the fields $\phi$ and $\ti\phi$, or $B$ and $\ti B$. The final result  \eqref{ZIndexchiNA} contains ill-defined infinite products, that we need to regularize.

\subsection{Regularization}

To regularize the infinite product~\eqref{ZIndexchiNA} we notice that it has zeros at $\mu = \frac{i n_2}{\tau_2} - \frac{i \tau}{\tau_2} (n_1+1 - \frac r2)$, with $n_2\in \bZ$ and $n_1 \in \bZ_{\ge 0}$, and it has poles at $\mu= -\frac{i n_2}{\tau_2} + \frac{i \tau}{\tau_2} (n_1 + \frac r2)$, with $n_2\in \bZ$ and $n_1 \in \bZ_{\ge 0}$. A regularized expression is then given by
\be
\ba
Z^{\rm one-loop}_{\text{chi}, \, w} &=  e^{\scF} \Big( \prod_{n \ge 0} \frac{1 - e^{2\pi (\tau_2 \mu + i (n+1-\frac r2)\tau)}  }{1 - e^{-2\pi (\tau_2 \mu - i (n + \frac r2)\tau)} } \, \Big)^{1/2} \ = \  e^{\scF} \Big( \prod_{n \ge 0} \frac{1 - q^{ n+1- \frac{r}{2}} y^{-w} y_F^{-1} }{1 - q^{ n + \frac{r}{2}} y^w y_F}  \, \Big)^{1/2}   \,, \cr
&=  e^{\scF} \frac{(q^{1 - \frac{r}{2}} y^{-w} y_F^{-1}; q )^{1/2}_{\infty}}{(q^{\frac{r}{2}} y^{w} y_F; q )^{1/2}_{\infty}} \,,
\ea\ee
with $q= e^{2\pi i \tau}$, $y = e^{-2\pi i (\tau_1\alpha + \tau_2 \beta)}$, $y_F= e^{-2\pi \tau_2(\mm + i \beta_F)}$ and $e^\scF$ a function of $\mu$ without poles nor zeroes. The $q$-Pochhammer symbol appearing in the above 
formula is defined by 
\be
(x,q)_\infty \equiv \prod_{n\geq 0} (1-x q^n)~.
\ee

Let us see in more detail how to regularize the infinite product and then determine the factor $\scF$.
We regularize the product over $n \in \bZ$ with the following formula\footnote{This formula involves using $\prod_{n=0}^\infty n^2 = e^{-2 \zeta'(0)} =  2\pi$.}
\begin{align}
 &\prod_{n \in \ZZ} \left(i n +  x \right) = 2\sinh (\pi x)
 =  e^{\pi x} \left(1-e^{-2\pi x} \right)  \,  .
\end{align}
We obtain
\be\ba
Z^{\rm one-loop}_{\text{chi}, \, w} &= \Big( \prod_{n_1 \ge 0} \prod_{n_2 \in \bZ}  \frac{  i n_2-   i \tau \lp n_1 +1 - \frac r2  \rp - \tau_2 \mu }{  i n_2  - i \tau  (n_1 + \frac r2) +  \tau_2 \mu } \, \Big)^{1/2} \cr
& = \Big( \prod_{n_1 \ge 0}   \frac{ e^{-\pi (i(n_1+1- \frac r2)\tau + \tau_2 \mu )} \left(1-e^{2\pi (i (n_1+1 - \frac r2)\tau + \tau_2 \mu) } \right)  }{ e^{\pi (i (n_1+ \frac r2) \tau - \tau_2 \mu) } \left(1- e^{2\pi (i (n_1 + \frac r2) \tau - \tau_2 \mu)} \right)  } \, \Big)^{1/2}  \cr
&= e^{\scF} \Big( \prod_{n \ge 0}   \frac{ 1-e^{2\pi (i (n+1- \frac r2)\tau + \tau_2 \mu)}  }{  1- e^{2\pi (i (n+\frac r2) \tau - \tau_2 \mu)}   }  \, \Big)^{1/2}\,,
\ea\ee
with
\begin{align}
\scF &= - \frac{\pi i \tau}{2}  \lp \sum_{n \ge 0} \lp n + 1- \frac r2 - \frac{i \tau_2 \mu}{\tau}\rp -   \sum_{n \ge 0} \lp n  + \frac r2 + \frac{i \tau_2 \mu}{\tau}\rp\rp \,.
\label{Fthermalsums}
\end{align}
Following \cite{Assel:2015nca}, we regularize the two infinite sums separately, using the Hurwitz zeta function $\sum_{n \ge 0} \lp n + x \rp = \zeta_H(-1,x) = -\frac 12 \lp x^2 - x + \frac 16 \rp$. This leads to
\begin{align}
  \scF = 0 \,.
\end{align}
The Hurwitz zeta function regularization of infinite sums has been used also in \cite{Yoshida:2014ssa}. We believe that this regularization is compatible with supersymmetry as in \cite{Assel:2015nca}, but we do not have a first principle derivation.\footnote{The alternative regularization using the Riemann zeta function leads to $\scF = \frac{\pi i \tau}{4} (1 - r) +  \frac{\pi}{2} \t_2 \mu$. } 
The final regularized result, for the chiral multiplet of R-charge $r$ and abelian flavor charge $q_F$, in the representation $\scR$ of the gauge group, is 
\be
Z^{\rm one-loop}_{\text{chi}} 
=  \prod_{w \in \scR} \frac{(q^{1 - \frac{r}{2}} y^{-w} y_F^{-q_F}; q )^{1/2}_{\infty}}{(q^{\frac{r}{2}} y^{w} y_F^{q_F}; q )^{1/2}_{\infty}} \,,
\label{ZchiFinal}
\ee
with $q= e^{2\pi i \tau}$, $y = e^{-2\pi i (\tau_1\alpha + \tau_2 \beta)}$, $y_F= e^{-2\pi \tau_2(\mm + i \beta_F)}$.
The result is invariant under the shifts \eqref{largeGT1} and \eqref{largeGT2} due to large gauge transformations. 

Finally we can make some comments about the ``unphysical" results \eqref{ZIndexchiNA3} and \eqref{ZIndexchiNA4}. After regularization, they lead to one-loop determinants $Z \sim (q^{\frac{r}{2}} y^{w} y_F; q )^{-1}_{\infty}$ and $Z \sim (q^{1 - \frac{r}{2}} y^{-w} y_F^{-1}; q )_{\infty}$, respectively. We observe that, after setting $\tau_1=0$, these results match the one-loop determinants of a chiral multiplet on the compact space $S^1\times D^2$ computed by localization in \cite{Yoshida:2014ssa}, with Neumann and Dirichlet boundary condition, respectively. Our interpretation of this result is that \eqref{ZIndexchiNA3} and \eqref{ZIndexchiNA4} correspond to the one-loop determinant on the chopped $\bH^3_\tau$ space, namely the space truncated at a given radial distance $\eta$, with Dirichlet or Neumann boundary conditions. This space is compact and topologically equivalent to $S^1\times D^2$, it is therefore plausible that the one-loop determinants on the two spaces are identical. To confirm this picture more rigorously, we would need to revisit the boundary conditions and one-loop determinant computations for the chopped $\bH^3_\tau$. The one-loop determinant on the non-compact $\bH^3_\tau$ is however different from those.

\subsection{Heat kernel}
\label{ssecHeatKernel}

In order to confirm the result obtained from the index theorem and unpaired eigenmodes methods, which required some extra assumptions, we provide in this section an alternative derivation, using heat kernels.

The eigenvalues and determinant of the Laplacian can be computed using the heat kernel on the space of interest. In \cite{Camporesi:1992tm,Camporesi:1994ga} the heat kernels and the associated zeta functions  of spin $s$ fields on hyperbolic spaces were computed.  The heat kernel on thermal AdS was found in \cite{Giombi:2008vd, David:2009xg}, relying on group-theoretic techniques and using the method of images. Thermal AdS$_3$ is the same geometry as $\bH^3_\tau$, however it does not include the  R-symmetry background gauge field necessary to preserve supersymmetry, nor the background flat connection corresponding to the locus configuration.
In this section we use the results of \cite{David:2009xg} and modify them to include the effect of the background R-symmetry connection and flat gauge connection, to compute the one-loop determinant of the chiral multiplet in a third way.
We will find that the result \eqref{ZchiFinal} is recovered when the heat kernel method is used with a specific regularization scheme, which we interpret as a scheme preserving supersymmetry.

To compute the determinant and eigenvalues spectrum of the spin $s$ Laplacian operator on $\bH^3$, which we denote $\Delta_{(s)}$, one can consider the heat kernel $K^{(s)}(t, x^\mu, y^\mu)$, which is defined by the equation
\begin{align}
(\p_t - \Delta_{(s)}) K^{(s)}(t,x^\mu,y^\mu) = 0 \,,
\label{HKdefeqn}
\end{align}
where the Laplacian acts on the $x^\mu$ variables, and with the boundary condition at $t=0$,
\be
K^{(s)}(0,x^\mu,y^\mu) = \delta^{3}(x^\mu,y^\mu) \,.
\ee
The eigenvalues $\lambda_n$ of the Laplacian deformed by a mass term, $-\Delta_{(s)} + M^2$, are encoded in the zeta function
\begin{align}
\zeta_s(M,z) &= \sum_n \frac{d_n}{(\lambda_n)^z} \,,
\end{align}
where $n$ labels the eigenvalues and $d_n$ is the degeneracy of $\lambda_n$. The zeta function is defined for $z$ such that the series converges and analytically continued over the complex plane.  On $\bH^3$ the eigenvalues are labeled by a continuous parameter $\lambda$, as well as discrete parameters, and the sum over $n$ is replaced by an integration over $\lambda$ with an appropriate Plancherel measure \cite{Camporesi:1992tm,Camporesi:1994ga}. The zeta function can be computed in terms of the heat kernel evaluated at coincident points by the formula
\begin{align}
\zeta_s(M,z) &= \frac{1}{\Gamma(z)} \int_0^\infty dt \int_{\bH_3} d^3x \sqrt{g} \,t^{z-1} \,  K^{(s)}(t,x,x) \, e^{- M^2 t}\,,
\label{zetaHKformal}
\end{align}
and is related to the determinant by
\begin{align}
\det \lp - \Delta_{(s)} + M^2 \rp &=  \exp [ - \p_z\zeta_s(M,0) ] \,,
\end{align}
so as to get the familiar formula
\begin{align}
- \log \det \lp - \Delta_{(s)} + M^2 \rp &=  \int_0^\infty  \frac{dt}{t}  \, e^{-M^2 t} \int_{\bH^3} d^3x \sqrt g \,  K^{(s)}(t, x , x) \,.
\label{DetHKformal}
\end{align}

In the computation of the determinant, one encounters UV divergences, which are regularized in the computation of $\zeta$ by the analytical continuation in $z$, and IR divergences coming from integration over the infinite $\bH^3$ volume and which needs further regularization.

To compute the determinants on thermal AdS$_3$, the authors of \cite{Giombi:2008vd, David:2009xg} relied on the method of images, which expresses the heat kernel on the quotient space in terms of the heat kernel on $\bH^3$,
 \be\ba
 - \log \det \lp - \Delta_{(s)} + M^2 \rp &=  \int_0^\infty  \frac{dt}{t}  \, e^{-M^2 t}  K^{(s)}(t)\,, \cr
 K^{(s)}(t) &= \int_{\bH_3/\tau\bZ} d^3x \sqrt g \, \sum_{n \in \bZ}   K_{\bH_3}^{(s)}(t, x , \omega^n(x)) \,,
 \label{DetandIntegratedHK}
 \ea\ee
 where $\omega$ describes the action of the $\tau \bZ$ quotient \eqref{quotient} and $ K^{(s)}(t)$ is the heat kernel on $\bH^3_\tau$ at coincident points, integrated over $\bH^3_\tau$.
In \cite{David:2009xg}, the integrated heat kernel $K^{(s)}(t)$ was computed and expressed as an integral over the continuous parameter $\lambda \in \bR_{>0}$ labeling the Laplacian eigenvalues
\footnote{Here we correct a typo in formula (6.5) of \cite{David:2009xg}. Instead of having an extra factor of two, we have taken into account the contribution of the sum of the two characters $\chi_{s,\lambda}$ and $\chi_{-s,\lambda}$, as is explained in that paper. This produces the product of cosines in our formula. For $s=0$, there only one character $\chi_{0,\lambda}$ to sum over, resulting in the factor $\frac{2 -\delta_{s,0}}{2}$. }
\begin{align}
K^{(s)}(t) &=  K_0^{(s)}(t) +  \sum_{n \in \bZ \backslash \{0\}} \frac{1}{2^{\delta_{s,0}}}  \frac{\tau_2}{| \sin \pi n\tau |^2}  \int_0^\infty d\lambda \cos(2\pi  n s \tau_1) \cos(2\pi n\lambda\tau_2) \,  e^{-(\lambda^2 + s+1)t}  \, ,
\label{HK}
\end{align}
where the term $K_0^{(s)}(t)$ is the contribution from the $n=0$ sector, whose computation is carried out separately and leads to
\begin{align}
K_0^{(s)}(t) &= (1 + 2 s^2 t) \, \frac{e^{-(s+1)t}}{4 (\pi t)^{3/2}} \, \textrm{Vol}(\bH_3/\bZ) \, \frac{2 -\delta_{s,0}}{2} \,.
\end{align}
This term carries the infrared divergence of the determinant. It is proportional to the (infinite) volume of $\bH^3_\tau$ and needs to be regularized by adding appropriate counter-terms, however we will not perform this analysis here.

The determinants that we need to compute involve Laplacian operators $\wat\Delta_{(s)}$ which are covariantized with respect to the gauge connection \eqref{HKgauge}, $\scA= (\beta + \frac{\tau_1}{\tau_2}\alpha)d\chi$, flavor symmetry connection $v = \beta_F d\chi$ and R-symmetry connection $A =  -\frac{\bar\tau}{2\tau_2} d\tauc$. Therefore we need to modify the heat kernel computation to take into account these background connections. It is easy to see that the heat kernel $\wat K^{(s)}$ defined by the equation \eqref{HKdefeqn} with covariantized laplacian $\wat\Delta_{(s)}$ is related to the heat kernel without background connection $K^{(s)}$ by
\begin{align}
\wat K^{(s)}(t, x^\mu, y^\mu) = e^{i \mathfrak{s} (\chi_x - \chi_y)} K^{(s)}(t, x^\mu, y^\mu) \,,
\end{align}
where $\mathfrak{s} =  q_F\beta_F +  q_G.(\beta + \frac{\tau_1}{\tau_2}\alpha) - q_R \frac{\bar\tau}{2\tau_2}$, with $q_F$ the flavor charge, $q_G$ the gauge charge and $q_R$ the R-charge of the field for which we compute the determinant.
Following the method of images of \cite{David:2009xg} and using the modified heat kernel $\wat K^{(s)}(t, x, \omega^n x)= e^{-2\pi i   \mathfrak{s}  n \tau_2} K^{(s)}(t, x, \omega^n x)$ , we obtain the modified heat kernel 
\begin{align}
\wat K^{(s)}(t) &=  K_0^{(s)}(t) +  \sum_{n \in \bZ \backslash \{0\}}  \frac{\tau_2 \cos(2\pi  n s \tau_1) }{ 2^{\delta_{s,0}}  | \sin \pi n\tau |^2} \, e^{-2\pi i   \mathfrak{s}  n \tau_2} \int_0^\infty d\lambda \cos(2\pi n\lambda\tau_2) \,  e^{-(\lambda^2 + s+1)t}   \, .
\label{Keqn2}
\end{align} 

The computations in \cite{David:2009xg} proceed by the usual method of first evaluating the integral over $\lambda$ and then the integral over $t$ to obtain the determinant \eqref{DetandIntegratedHK}. This, of course, implicitly involves an inversion of the $t$ and $\lambda$ integral. 
This regularization method, applied to our problem, leads to a one-loop determinant where the mass parameter $m$ appears in the expressions as $|m + r-1|$ and $|m+r -1/2|$. We argue that this regularization must break supersymmetry. 
Indeed we notice that in the super-algebra \eqref{susyalgebrawithmass}  the parameters $\mm, \alpha, \beta, r, w$ appear only in the complex combination $\mu = \mm + i \beta_F + i w.\big(\beta + \frac{\tau_1}{\tau_2}\alpha\big) $, in the representations carried by the fields of the chiral multiplet. In the supersymmetric theory the one-loop determinant can be computed as the index of the operator $\scH = Q^2$, as explained in Section~\ref{ssec:IndexTheorem}, so that the final result must be a holomorphic function of $\mu$. This is not what we find by following the regularization method of \cite{David:2009xg}, therefore we must find a different regularization method.
 For this we write the zeta function, ignoring the divergent $n=0$ term,
\be
\zeta_s(M,\mathfrak{s},z) = \sum_{n \in \bZ \backslash \{0\}} \frac{\tau_2 \cos(2\pi  n s \tau_1) e^{-2\pi i   \mathfrak{s}  n \tau_2}}{ \Gamma(z) 2^{\delta_{s,0}}  | \sin \pi n\tau |^2}  \int_0^\infty dt \, t^{z-1} \int_0^\infty d\lambda \cos(2\pi n\lambda\tau_2) \,  e^{-(\lambda^2 + s+1 + M^2)t} \,,
\ee
and we perform first the integral over $t$, 
\be
\zeta_s(M,\mathfrak{s},z) = \sum_{n \ge 1} \frac{2 \tau_2 \cos(2\pi  n s \tau_1) \cos(2\pi   \mathfrak{s}  n \tau_2)}{ 2^{\delta_{s,0}}  | \sin \pi n\tau |^2}  \int_0^\infty d\lambda \cos(2\pi n\lambda\tau_2) \,  (\lambda^2 + M(s)^2)^{-z} \,,
\ee
with $M(s)^2 = s + 1 + M^2$. It will turn out that the expressions we will use for $M(s)^2$ are the square of linear combinations of the parameters $m$ and $r$, and we will define $M(s)$ as the corresponding linear combinations, with a positive coefficient for $m$.
We then propose a manipulation, which ensures an analytic result in $M(s)$: we first replace the integral over $(0,\infty)$ to an integral over $(-\infty,\infty)=\bR$ and divide by an extra factor of two -- this  does not change the expression  since the function of $\lambda$ is even -- and then we replace the factor $(\lambda^2 + M(s)^2)^{-z}$ by $(\lambda +i M(s))^{-z} + (\lambda -i M(s))^{-z}$, so that
\be\ba
& \zeta_s(M,\mathfrak{s},z) \cr
&= \sum_{n \ge 1} \frac{\tau_2 \cos(2\pi  n s \tau_1) \cos(2\pi   \mathfrak{s}  n \tau_2)}{ 2^{\delta_{s,0}}  | \sin \pi n\tau |^2}  \int_\bR d\lambda \cos(2\pi n\lambda\tau_2) \big[ (\lambda +i M(s))^{-z} + (\lambda -i M(s))^{-z} \big] \,.
\ea\ee
This last replacement might seem a strong modification at first sight, however one should remember that only the derivative at $z=0$ of $\zeta$ is relevant to the computation of the determinant, and we have $(\lambda^2 + M(s)^2)^{-z} = (\lambda +i M(s))^{-z}(\lambda -i M(s))^{-z} \simeq (\lambda +i M(s))^{-z} + (\lambda -i M(s))^{-z} - 1$ around $z=0$. The replacement that we propose does not change the value of $\p_z\zeta |_{z=0}$ formally and therefore may legitimately be considered.
The integral over~$\lambda$ can then be evaluated and analytically continued in $z$, using the formulas\footnote{The integral is well-defined for $Re(z)>0$, $a\in\bR$, $Re(c)\neq 0$ and $Im(c)<0$.}
\be\ba
  \int_{\bR} d\lambda \cos(a\lambda)(\lambda + i b)^{-z} &= \frac{\pi |a|^{z-1}}{\sin(\pi z)\Gamma(z)} \big[ (-ib)^z \sin\lp \frac{\pi z}{2} - i b |a| \rp + (ib)^z \sin\lp \frac{\pi z}{2} + i b |a| \rp \big] \,, \cr
  &= \frac{\pi}{|a|} e^{- b |a| } z + O(z^2) \,,
\ea\ee
leading to\footnote{We deform $M(s)$ by an infinitesimally small negative or positive imaginary part to evaluate the integrals.}
\be\ba
& - \log\det{}'(-\Delta_{(s)} + M^2) = \p_z \zeta_s(M,\mathfrak{s},0) \cr
& \qquad = \sum_{n \ge 1} \frac{\cos(2\pi  n s \tau_1) \cos(2\pi   \mathfrak{s}  n \tau_2)}{ 2^{\delta_{s,0}} n  | \sin \pi n\tau |^2}   \cosh(2\pi n \tau_2 M(s)) \,,
\ea\ee
where $\det{}'$ denotes the determinant without the $n=0$ contribution.
Defining $q = e^{2\pi i \tau}$, this can be re-written as
\be\ba
- \log\det{}'(-\Delta_{(s)} + M^2) = \sum_{n \ge 1} & \frac{(q\bar q)^{\frac n2}}{ 2^{1+\delta_{s,0}} n  (1-q^n)(1-\bar q^n)}  \big[ (q\bar q)^{\frac{i n\mathfrak{s}}{2}} + (q\bar q)^{-\frac{i n\mathfrak{s}}{2}} \big]   \cr
& \times  \Big[ \big( \frac{q}{\bar q} \big)^{\frac{ns}{2}}  +  \big(\frac{q}{\bar q} \big)^{-\frac{ns}{2}} \Big]   \big[ (q\bar q)^{\frac{n M(s)}{2}} + (q\bar q)^{-\frac{n M(s)}{2}} \big]  \,.
\label{DetSpinSfinal}
\ea\ee

After these preliminaries, we can extract the one-loop determinant of interest to us.
We want to compute the determinant over the complex scalar $\phi$ and the spinor $\psi$ associated to the Lagrangian \eqref{LagrangianChi},\footnote{We neglect the integration over the auxiliary field $F$ which evaluates to a number.}
\begin{align}
Z^{\rm one-loop}_{\text{chi}, \, w} &= \frac{\det \lp  i \gamma^\mu D_\mu + i m_{\psi} \rp  }{\det\lp - \Delta_{(0)} + m_\phi^2 \rp }
 \,,
\end{align}
with $m_\phi^2 = \lp \mm + r-1 \rp^2 - 1$ and $m_\psi = \mm + r - \frac 12$. We have set here $L=1$ (it can be recovered by rescaling $\mm \to \mm L$). The scalar determinant can be read off directly from \eqref{DetSpinSfinal} with $s=0$,
\be
\log\det{}' \big( - \Delta_{(0)} + m_\phi^2 \big) =  -\sum_{n \ge 1}  \frac{\big( (q\bar q)^{\frac{i n\sss_0}{2}} + (q\bar q)^{-\frac{i n\sss_0}{2}} \big)    \big( (q\bar q)^{\frac{n}{2}(m+r)} + (q\bar q)^{-\frac{n}{2}(m+r-2)} \big)}{2 n  (1-q^n)(1-\bar q^n)} \,,
\label{bosonDet}
\ee
with $\sss_0 = \beta_F + w.(\beta + \frac{\tau_1}{\tau_2}\alpha) - r \frac{\bar\tau}{2\tau_2}$.

The spinor determinant is less straightforward to extract, since we are looking for the determinant of the Dirac operator and not of the Laplacian.
First we notice that taking the fermion determinant $\det \lp  i \gamma^\mu D_\mu + i m_{\psi} \rp$ implicitly assumes the reality conditions $\ti\psi^\alpha = (\psi_\alpha)^\ast$, which is not the same as the reality conditions expressed in terms of the twisted variables $\ti C = C^\ast$, $\ti B = - B^\ast$. The difference can be interpreted as a deformation of the contour of integration in field space of the path integral and we will work under the assumption that this does not change the evaluation of the determinant.

We can then make use of the relation
\be\ba
& - \Delta_{(\frac 12)} - \frac 32 + m_{\psi}^2 =  - (\gamma^\mu D_\mu +  m_{\psi}) (\gamma^\mu D_\mu - m_{\psi}) \,, 
\ea\ee
which implies
\be\ba
&\log\det{}'(i\gamma^\mu D_\mu + i m_{\psi}) + \log\det{}'(i\gamma^\mu D_\mu - im_{\psi}) = \log\det{}'(- \Delta_{(\frac 12)} - \frac 32 + m_{\psi}^2) \cr
& =  -\sum_{n \ge 1}  \frac{(q\bar q)^{\frac n2} \big[ (q\bar q)^{\frac{i n}{2}\sss_{1/2}} + (q\bar q)^{-\frac{i n}{2}\sss_{1/2}}  \big]  }{ 2 n  (1-q^n)(1-\bar q^n)}     \big[ \big( \frac{q}{\bar q} \big)^{\frac{n}{4}}  +  \big(\frac{q}{\bar q} \big)^{-\frac{n}{4}} \big]   \big[ (q\bar q)^{\frac{n}{2}m_\psi} + (q\bar q)^{-\frac{n}{2}m_\psi} \big]    \,,
\label{DiracLaplacianRel}
\ea\ee
with $\sss_{1/2} = \beta_F + w.(\beta + \frac{\tau_1}{\tau_2}\alpha) - (r-1) \frac{\bar\tau}{2\tau_2} = \sss_{0} + \frac{\bar\tau}{2\tau_2}$, and the relation
\be\ba
& \det (i\gamma^\mu D_\mu +  i m_{\psi})|_{q} =   e^{\scF_{\rm CS}} \det (i \gamma^\mu D_\mu - i m_{\psi})|_{\bar q} \,,
\label{ParityRel}
\ea\ee
where $\det (\cdots )|_{q}$ denotes the determinant on the space $\bH^3_\tau$, with $q=e^{2\pi i \tau}$.  The identity follows from the action of parity $\scP: \varphi \to -\varphi$, $\psi \to i\gamma^3 \psi$, which reverts the sign of the mass term and the sign of $\tau_1$.  
If we extrapolate from flat space results, the parity transformation is anomalous and brings the factor  $e^{\scF_{\rm CS}}$, which denotes the contribution of Chern-Simons terms \cite{Aharony:1997bx}. With a fermion of charges $q_i$ under $U(1)_i$ symmetries, the parity transformation brings mixed $U(1)_i-U(1)_j$ Chern-Simons terms with level $k_{ij} = q_i q_j sign(m_\psi)$ in \eqref{ParityRel}. It is not clear whether the same phenomenon appears in hyperbolic space and therefore we will not provide an explicit expression for $e^{\scF_{\rm CS}}$.
Assuming that this possible Chern-Simons term is captured by the $n=0$ contribution, we can extract the determinant of the Dirac operator from \eqref{DiracLaplacianRel}, consistently with \eqref{ParityRel}, as\footnote{In checking the parity relation \eqref{ParityRel}, one should consider the gauge connection parameter $\sss_{1/2}$ as a fixed parameter, independent of $\tau$, since parity does not act on it (the gauge connections have no component along $\varphi$). } 
\be\ba
& \log\det{}' (i\gamma^\mu D_\mu +  i m_{\psi}) \cr
& = -\sum_{n \ge 1}  \frac{(q\bar q)^{\frac n2} \big[ (q\bar q)^{\frac{i n}{2}\sss_{1/2}} + (q\bar q)^{-\frac{i n}{2}\sss_{1/2}}  \big]  }{ 2 n  (1-q^n)(1-\bar q^n)}       \big[ \big(\frac{q}{\bar q} \big)^{\frac{n}{4}}  (q\bar q)^{-\frac{n}{2}m_\psi} +  \big(\frac{q}{\bar q} \big)^{-\frac{n}{4}} (q\bar q)^{\frac{n}{2}m_\psi} \big] \,.
\label{fermionDet}
\ea\ee
Note that there is an alternative identification of $\det{}' (i\gamma^\mu D_\mu +  i m_{\psi})$ consistent with the parity relation, which amounts to reversing the sign of $m_{\psi}$ in the right-hand side of \eqref{fermionDet}, however, this does not lead to cancellation with the bosonic determinant, and does not yield the holomorphicity in $\mu$, therefore it would be the wrong identification.

Combining \eqref{bosonDet} and \eqref{fermionDet} and using $(q\bar q)^{\frac{i n}{2}\mathfrak{s}_{1/2}} = (q\bar q)^{\frac{i n}{2} \mathfrak{s}_{0}} \bar q^{\frac n2}$, we observe spectacular cancellations,
\be\ba
\log Z^{\rm one-loop}_{\text{chi}, \, w} &= \scF + \log\det{}'(i\gamma^\mu D_{\mu} + im_\psi ) - \log\det{}'(-\Delta_{(0)}+m_\phi^2) \cr
&= \scF + \sum_{n\ge 1} \frac{(q \bar q)^{\frac n2 (m+r + i \sss_0)}  - q^n (q \bar q)^{-\frac n2 (m+r + i \sss_0)}  }{2 n (1- q^n)} \cr
&= \scF + \sum_{n\ge 1} \sum_{n'\ge 0} \frac{(q \bar q)^{\frac n2 (m+r + i \sss_0)}  - q^n (q \bar q)^{-\frac n2 (m+r + i \sss_0)}  }{2 n} \, q^{n n'} \cr
&= \scF - \frac 12 \sum_{n'\ge 0}  \log\big(1 - q^{n'}(q \bar q)^{\frac 12 (m+r + i \sss_0)} \big) - \log\big( 1 - q^{n' +1} (q \bar q)^{-\frac 12 (m+r + i \sss_0)} \big) \,,
\ea\ee
with $\scF$ denoting the $n=0$ contribution.
 We obtain the final evaluation
\be
Z^{\rm 1-loop}_{\rm chi} =  e^{\scF} \prod_{n \ge 0} \lp  \frac{1 - q^{ n+1- \frac{r}{2}} e^{2\pi\tau_2\mu} }{1 - q^{ n + \frac{r}{2}} e^{-2\pi\tau_2\mu}} \rp^{\frac 12} \,.
\ee
with the parameter $\mu= \mm+i \beta_F+iw.\big(\beta+\frac{\tau_1}{\tau_2}\alpha\big)$. 
Although the boson and fermion determinants are not separately holomorphic in $\mu$, their combination is holomorphic, as predicted from the super-algebra considerations. Moreover the final result is in perfect agreement with the index computation \eqref{ZchiFinal}. The prefactor $e^{\scF}$ is not easy to compute from the heat kernel method since it needs some extra regularization of infrared divergences. We assume that this can be done in supersymmetric fashion and that it would match the trivial prefactor in \eqref{ZchiFinal}.

\subsection{Vector multiplet }
\label{sec:VectorDet}

To compute the one-loop determinant we will make use of the twisted variables defined in 
Section~\ref{sssec:TVarVec}. These are bosonic fields $(X^-,X^0,X^+,\Sigma, D^0)$ of R charges (-2,0,2,0) respectively, and fermionic fields $(\Lambda^-,\Lambda^0,\Lambda^+, \Theta)$ of R charges (-2,0,2,0) respectively, all Lorentz scalars.  To these fields we must add the ghost fields $(c, \ti c, b)$ of vanishing R charge. 
Here we denote by the same name the fluctuation of the field around the localization locus.

As for the chiral multiplet, we decompose the vector multiplet one-loop determinant into the product over the contributions of the weights of the adjoint representation, which are labeled by the generators of the gauge algebra,
\be
Z^{\rm one-loop}_{\rm vec}  = Z_{\rm Cartan} \prod_{\gamma \in \mathfrak{g}} Z^{\rm one-loop}_{\rm vec, \, \gamma} \,,
\ee
where $Z_{\rm Cartan}$ denotes the contribution from the Cartan components and  $Z^{\rm one-loop}_{\rm vec, \, \gamma}$ denotes the contribution from the components ${\bf X}_\gamma$ of the fields, with $\gamma \in \mathfrak{g}$ running over the non-zero roots of the gauge algebra $\mathfrak{g}$. The Cartan contribution is independent of  the background flat connection, as well as the parameters of the theory, except the rank of the gauge group $N$, so it evaluates to constant (to the power $N$), which factorizes in the exact partition function and which we neglect by setting $Z_{\rm Cartan} =1$.

Focusing on the $\gamma$-component contribution, we observe that the fields decompose into
the set of Grassmann even scalar fields $(X^-,X^0,X^+)_\gamma$ of R-charge (-2,0,2) and the set Grassmann-odd scalar fields $(\Theta, c, \ti c)_\gamma$ of vanishing R-charge. The other fields $(\Lambda^-,\Lambda^0,\Lambda^+)_\gamma$ and $(D^0, \Sigma, b)_\gamma$ are their $\wat Q$ super-partners. Assuming that the one-loop determinant can be computed using the index theorem as for the chiral multiplet, we can directly extract the one-loop determinant by applying the formulas \eqref{DD'rel} and \eqref{ASindthm} to the above set of fields. The computation is further simplified by noticing that the contribution of these fields to the index computation matches the contribution of the twisted fields of a chiral multiplet of R charge $q_R=2$ and gauge charge $w=\gamma$ (and vanishing flavor charge), plus the contribution of a scalar and a Grassmann-odd scalar of vanishing R-charges, whose contributions cancel each other. Therefore we have
\be
Z_{\rm vec, \, \gamma}^{\rm one-loop} = Z_{\rm chi, \gamma}^{\rm one-loop}[q_R =2] \,.
\ee
Using the result of the chiral multiplet one-loop determinant we conclude, after simplification\footnote{There is a large cancellation between the $\gamma$ and $-\gamma$ contributions.},
\begin{align}
Z_{\rm vec}^{\rm one-loop} &=   \Delta^{-1} \prod_{\gamma >0} \pm 2\sin [\pi \gamma.(\tau_1 \alpha + \tau_2 \beta)] \,,
\label{Z1loopVec}
\end{align}
where $\pm$ denote the sign ambiguity coming from evaluating square roots. This sign ambiguity must be fixed by physical requirements (see Section~\ref{sec:summary}). In \eqref{Z1loopVec} we have inserted for consistency the factor $\Delta^{-1} = [\gamma.(\beta + \t_1\alpha/\t_2)]^{-2}$, which is the inverse of the Vandermonde determinant discussed in Section~\ref{sssec:VecLocus}. In the full partition function it cancels with the Vandermonde determinant, restoring invariance of the integrand of the matrix model under large gauge transformations in the Yang-Mills theory. 
We conjecture that this extra factor appears as a factor compensating for overcounting some fermionic unpaired eigenmodes of the fields which have vanishing R-charges. These unpaired eigenmodes have $n_1=n_2=0$ in \eqref{UnpairedModes}, corresponding to modes bounded at infinity but which do not go to zero and with eigenvalues $\gamma.(\beta + \t_1\alpha/\t_2)$. Indeed, in localization computations on compact space these modes are excluded based on normalizability condition (see~\cite{Hama:2011ea}). In our situation, where we have been led to count the contributions of diverging unpaired modes, it is not clear why we should exclude these fermionic modes. To confirm this result, it might be useful to carry out the heat kernel computation for the vector multiplet fields including ghosts.

\section{Exact partition functions and Wilson loops}
\label{sec:summary}

In this section we gather the results of the previous sections and write the complete exact partition functions and Wilson loop observables in theories with unitary gauge groups.

The partition function is expressed as a sum over flat connections $\scA_{\rm flat}$ on $\bH^3_\tau$,
\be
Z = \sum_{\scA_{\rm flat}} Z_{\rm cl}[\scA_{\rm flat}] \, Z_{\rm one-loop}[\scA_{\rm flat}] \,,
\ee
where $Z_{\rm cl}[\scA_{\rm flat}]$ contains the classical contributions of Chern-Simons and Fayet-Iliopoulos terms, while $Z_{\rm one-loop}[\scA_{\rm flat}]$ is the product of the one-loop determinant of the vector and matter multiplets around the background $\scA_{\rm flat}$.
The sum over flat connections is restricted by the asymptotics of the gauge field analysed in 
Section~\ref{ssec:BdryCond}.

At infinity the flat connections are given by
\begin{align}
& \scA_{\rm flat}^{\infty} = \alpha d\varphi + \beta d\chi  \,,
\end{align}
with $\alpha, \beta$ constant and valued in the Cartan subalgebra $\mathfrak{t} \subset \mathfrak{g}$. 
As explained in Section~\ref{sssec:VecLocus}, for $U(N)$ or $SU(N)$ gauge theories, we have
\be\ba
& \alpha = \text{diag}(a_1 ,a_2, \cdots, a_N) \,, \quad \beta = \text{diag}(b_1 ,b_2, \cdots, b_N) \,, \cr
& \text{with} \quad \{ a_i\}_{1\le i \le N} \in \bZ^N \,, \quad \sum_{i=1}^N a_i = 0 \,. 
\ea\ee
The results in the previous sections were given in terms of $\alpha$ and $\beta$. The final partition function is obtained by imposing the further restrictions associated to the choices of gauge field asymptotics in different theories. For the other fields, we have only considered decaying asymptotics, corresponding to square-normalizability.

First we consider the gauge field asymptotics \eqref{VecBdyCond1},
\be
\scA^{\infty}_{\bar z} \equiv \scA^{(0)}_{\bar z} =  0 \quad \Rightarrow \quad \alpha + i \beta = 0 \,.
\ee
The one-loop determinant depends on the combination $\tau_1 \alpha + \tau_2 \beta = \tau \alpha$ and the partition function becomes a holomorphic function of $q=e^{2\pi i \tau}$.
The partition function of the theory with $U(N)$ gauge group, Chern-Simons level $k$, and with $M$ chiral multiplets of R-charge $r_I$, in representations $\scR_I$ of $U(N)$, is given by
\be\ba
Z &=  \frac{1}{N!} \sum_{\{n_i\} \in \bZ^N \atop \sum_i n_i =0}    q^{ \frac k2  \sum_i n_i^2} 
\prod_{i < j} \big(q^{\frac{|n_i-n_j|}{2}} - q^{-\frac{|n_i-n_j|}{2}} \big) \, 
\prod_{I=1}^M \prod_{w_I \in \scR_I} \frac{\big( q^{1- \frac{r_I}{2} + w_I.n} y^{-Q_I} , q \big)_\infty^{1/2}}{\big( q^{\frac{r_I}{2} - w_I.n} y^{Q_I} , q \big)_\infty^{1/2} }  \,,
\label{ZFinalCS}
\ea\ee
where we have introduced possible deformations by $U(1)^K$ flavor background: $Q_I = (Q_{I,1}, \cdots, Q_{I,K})$ denote the flavor charges and $y^{Q_I} \equiv \prod_{k=1}^K y_k^{Q_{I,k}}$, with $y_k =  e^{-2\pi \tau_2 (m_k + i \beta_k)}$ the deformation parameters, corresponding to turning on real masses $m_k$ and background vector field $v_k = \beta_k d\chi$. Here we have re-introduced the flavor charges $Q_{I,k}$, compared to the result of 
Section~\ref{sec:OneLoopDet}.
The factor $\prod_{i < j} (q^{\frac{|n_i-n_j|}{2}} - q^{-\frac{|n_i-n_j|}{2}})$ comes from the one-loop determinant of the vector multiplet, where the sign ambiguty has been fixed by  $\prod_{i < j} \pm [2\sin(\pi \tau (n_i - n_j))] \to \prod_{i < j}  [2\sin(\pi \tau |n_i - n_j|)]$, ensuring that the factor is invariant under Weyl group gauge transformations $(n_i,n_j) \to (n_j,n_i)$ for each pair~$(i,j)$.
 We also dropped an overall factor $(-i)^{\frac{N(N-1)}{2}}$ in the full partition function. 
 
 Note that for the abelian Chern-Simons theory, the only flat connection compatible with these asymptotics is the trivial connection $\scA = 0$ and the partition function is given by a single term, carrying the contribution of the matter one-loop determinants. In particular it is independent of the Chern-Simons level. 
 
 As observed in \eqref{ZclassNonAb}, the contribution of the FI term vanishes in the supersymmetric Chern-Simons theory.
 
 The exact evaluation of supersymmetric Wilson loops, as defined in Section~\ref{ssec:WilsonLoops}, is obtained by including the Wilson loop factor \eqref{WilsonLoopFactor} in the summand in \eqref{ZFinalCS}. With the asymptotics $\alpha + i \beta=0$, the exact (un-normalized) vacuum expectation value of the BPS Wilson loop in the representation $\scR$ of $U(N)$ is
 \be\ba
\langle W_\scR \rangle =  \frac{1}{N!} \sum_{\{n_i\} \in \bZ^N \atop \sum_i n_i =0} & \  \tr_\scR \big[ q^{n} \big]  \  q^{ \frac k2  \sum_i n_i^2}  \cr
& \times \prod_{i < j} \big(q^{\frac{|n_i-n_j|}{2}} - q^{-\frac{|n_i-n_j|}{2}} \big) \, 
\prod_{I=1}^M \prod_{w_I \in \scR_I} \frac{\big( q^{1- \frac{r_I}{2} + w_I.n} y^{-Q_I} , q \big)_\infty^{1/2}}{\big( q^{\frac{r_I}{2} - w_I.n} y^{Q_I} , q \big)_\infty^{1/2} }  \,,
\label{WFinalCS}
\ea\ee
 where $\tr_\scR \big[ q^{n} \big] = \sum_{w\in \scR} q^{w.n} = \sum_{w\in \scR} q^{\sum_i w_i n_i}$, with $w$ running over the weights of~$\scR$. 
We observe that the partition function and the BPS Wilson loop are holomorphic functions in $q$.
This suggests that the answer may have a holographic interpretation as arising from a holomorphic current algebra.
This would be the supersymmetric analog of the results of~\cite{Giombi:2008vd, David:2009xg, Gaberdiel:2010ar}. 
Indeed the non-supersymmetric AdS$_{3}$ partition functions computed in those papers exhibits the 
phenomenon of holomorphic factorization in the part of the Hilbert space of the theory which had an interpretation 
as a boundary current algebra (\emph{e.g.}~graviton, gauge fields, or higher spin fields). In our case we have a purely 
holomorphic result, which suggests that the only contribution to the BPS observable that we compute comes from
holomorphic currents in the boundary theory.

The result for an $SU(N)$ gauge group is identical, since the diagonal $U(1)$ does not support flat connections (with this choice of asymptotics).  
We can also consider a pure Yang-Mills theory with the same choice of asymptotics and the result is as above, simply with vanishing Chern-Simons level $k=0$.

A second choice of asymptotics is \eqref{VecBdyCond2},
\be
\scA^{\infty}_{z} \equiv \scA^{(0)}_{z} =  \frac C2 \quad \Rightarrow \quad \alpha - i \beta = C \,,
\ee
with $C=$diag$(c_1,c_2, \cdots, c_N)$ a constant Cartan-valued matrix. Let us set $C=0$ for simplicity. In this case the partition function is analogous to \eqref{ZFinalCS}, but with $q$ replaced by $\bar q^{\, -1}$ in several places, 
\be\ba
Z &=  \frac{1}{N!} \sum_{\{n_i\} \in \bZ^N \atop \sum_i n_i =0}  \bar q^{ \, - \frac k2  \sum_i n_i^2} 
\prod_{i < j} \big(\bar q^{\, -\frac{|n_i-n_j|}{2}} - \bar q^{\frac{|n_i-n_j|}{2}} \big) \, 
\prod_{I=1}^M \prod_{w_I \in \scR_I} \frac{\big( q^{1- \frac{r_I}{2}} \bar q^{\, - w_I.n} y^{-Q_I} , q \big)_\infty^{1/2}}{\big( q^{\frac{r_I}{2}} \bar q^{w_I.n} y^{Q_I} , q \big)_\infty^{1/2} }  \,.
\label{ZFinalCS2}
\ea\ee
The exact vacuum expectation value of the supersymmetric Wilson loop  is obtained in this case by adding the factor $\tr_\scR \big[ \bar q^{\, - n} \big] = \sum_{w\in \scR} \bar q^{\, - w.n}$ in the summand.
The partition function and the Wilson loops in this case are not  holomorphic, nor anti-holomorphic, in $q$. 
Note that, because $|q|<1$, the expression \eqref{ZFinalCS} seems ill-defined for $k <0$, since the sum diverges. Conversely, with the second choice of asymptotics, the sum in \eqref{ZFinalCS2} is then divergent for $k > 0$.

The reason for the asymmetry between $q$ and $\bar q$ in the results \eqref{ZFinalCS} and \eqref{ZFinalCS2} is to be attributed to the initial choice of supersymmetric background geometry, which selects a preferred complex coordinate $z$ on the torus slices, transverse to the radial coordinate. The supersymmetry preserved by the background we studied has the anti-holomorphic translation generator $\bar L_0 = \p_{\bar z}$ appearing in the super-algebra \eqref{susyalgebra}, but not the holomorphic conterpart $\p_z$.

\medskip

Finally, in the pure Yang-Mills theory\footnote{As discussed in Section~\ref{sssec:YMandCSBdyCond}, we refer to the pure Yang-Mills theory as the theory with zero bare \emph{and} effective Chern-Simons coupling.}, we can consider the choice of asymptotics~(1) of \eqref{AandsigmaExpansions}, for which the asymptotic values of the gauge field $\scA_{z}^{(0)}$ and $\scA_{\bar z}^{(0)}$ are fluctuating. In this case $\alpha$ and $\beta$ are independent. The final result is obtained by integrating over a certain middle dimensional contour in the space of complex flat connections.  Let us consider the abelian theory for simplicity. In this case flat connections are given by $\alpha=0$ and $x \equiv e^{-2\pi i \tau_2 \beta} \in \bC$. The partition function of the $U(1)$ gauge theory, with $M$ chiral multiplets of R-charge $r_I$ and gauge charge $w_I$, is given by
\be
Z =  \int_{\scC} \frac{dx}{2\pi i x}  \,  \prod_{I=1}^M \frac{\big( q^{1- \frac{r_I}{2}} x^{- w_I} y^{-Q_I} , q \big)_\infty^{1/2}}{\big( q^{\frac{r_I}{2}} x^{w_I} y^{Q_I} , q \big)_\infty^{1/2} }  \,,
\label{ZFinalYMabelian}
\ee
where $y^{Q_I}$ is defined as above and $\scC$ is a  one-dimensional  integration contour in $\bC$. Taking $\scA_\mu$ hermitian corresponds to $\scC$ being the unit circle. Taking $\scA_{z}$ and $\scA_{\bar z}$ hermitian corresponds to $\scC$ being the imaginary axis.

The contour of integration in general is not specified by the localization computation and must be chosen a priori. The contour corresponds to the choice of integration over field space in the definition of the path integral and different choices may lead to different path integrals. Often a particular choice is the most relevant in the sense that it leads to interesting observables (see, for instance, \cite{Beem:2012mb,Benini:2013nda,Gadde:2013ftv}). In the case at hand, this choice must be compatible with the asymptotics of the fields. We leave this analysis for future work.

The vacuum expectation value of the supersymmetric Wilson loop with charge $q_W$ is computed by the above integral, with the addition of the factor $e^{2\pi i q_W \tau_2\beta} = x^{-q_W}$ in the integrand,
\be
\langle W(q_W) \rangle =  \int_{\scC} \frac{dx}{2\pi i x} \, x^{-q_W} \,  \prod_{I=1}^M \frac{\big( q^{1- \frac{r_I}{2}} x^{- w_I} y^{-Q_I} , q \big)_\infty^{1/2}}{\big( q^{\frac{r_I}{2}} x^{w_I} y^{Q_I} , q \big)_\infty^{1/2} }  \,.
\label{WFinalYMabelian}
\ee

The corresponding expression for the partition function of the non-abelian $U(N)$ pure Yang-Mills theory is given by
\be
Z =   \frac{1}{N!} \int_{\{\scC_i \}} \prod_{i=1}^N \frac{dx_i}{2\pi i x_i} \,  \prod_{i<j} \bigg( \sqrt{\frac{x_i}{x_j}} - \sqrt{\frac{x_j}{x_i}}  \, \bigg) \,   \prod_{I=1}^M \prod_{w_I \in \scR_I}\frac{\big( q^{1- \frac{r_I}{2}} x^{- w_I} y^{-Q_I} , q \big)_\infty^{1/2}}{\big( q^{\frac{r_I}{2}} x^{w_I} y^{Q_I} , q \big)_\infty^{1/2} }  \,,
\label{ZFinalYMNA}
\ee
with the same notations as above, \emph{e.g.}~$x^{w_I} = \prod_i x_i^{w_{I,i}}$, and $\{\scC_i \}$ are the contours of integration in $\bC^N$  of the complex variables $x_i$, which remain to be determined. The Wilson loop factor is $\tr_\scR x^{-1} = \sum_{w\in \scR} x^{-w}$.
In the presence of an FI term, the asymptotics \eqref{BdryCondFI} require the product $\prod_i x_i$ to be constant, reducing effectively the partition function to that of an $SU(N)$ gauge theory.

The presence of square roots in the integrands \eqref{ZFinalYMabelian}, \eqref{WFinalYMabelian}, \eqref{ZFinalYMNA}, introduce branch cuts on the $x_i$ planes, which require some extra care when defining the contour of integration. We hope these issues will be addressed in the future.

\section{Discussion }
\label{sec:Discussion}

In this paper we have computed the exact partition function 
and the expectation value of certain BPS Wilson loops 
of ${\cal N}=2$ supersymmetric gauge theories, defined on a non-compact 
quotient  of the hyberbolic space $\IH^3$. Our results rely on the method of supersymmetric localization, applied to this unexplored domain.  We hope that the findings of this paper will pave 
the way for extending the localization technique to a broader class of theories and geometries, some of which we will allude to in the remainder of this section. 

Our set-up differs from previous localization calculations in the literature because the background geometry that we consider is hyperbolic and  non-compact. The intuition arising from holography 
suggested to deal with this situation by working on a ``chopped'' space, including a boundary at a large radial distance from the center, and  sending this to infinity at the end of the calculations. In this 
way, we could study systematically various supersymmetric actions, necessary for implementing the localization technique, which generically comprise both bulk and boundary terms. At the same time, again following the ideas 
of holography, we have discussed boundary  conditions for the fields, namely their asymptotic expansions at infinity.

We found that a careful treatment  of this problem is much more complicated than  in the context of analogous computations in compact spaces. In this paper we have attempted a comprehensive analysis. There remain, however, some puzzling issues related to the boundary conditions.  For example, it appears that the modes that contribute to the one-loop determinants around the BPS locus have unphysical asymptotic behaviour -- we will comment more on this momentarily. Another issue that we have not settled
is the choice of integration contour for the (complexified) gauge field in the case of Yang--Mills theories; on the other hand, the presence of Chern-Simons terms selects a natural prescription for this contour. 

As in more standard situations, the localization arguments imply that the path integral is computed exactly by the one-loop determinants about the BPS locus. However, for the problem we considered, 
it was \emph{a priori} not obvious which approach 
would be the most appropriate to evaluate these determinants. In the context of hyperbolic space, one-loop determinants of fields with different spins were computed previously  using the method of the heat kernel 
\cite{Camporesi:1992tm,Camporesi:1994ga,Camporesi:1995fb}.  However,  in the existing literature supersymmetry was not taken into account, in particular even computations in the context of supergravities were performed in backgrounds breaking supersymmetry  \cite{Giombi:2008vd,David:2009xg}. On the other hand, in most of the localization computations, two methods have been utilised to compute  one-loop determinants efficiently: the pairing of (bosonic end fermionic) eigenvalues 
(see \emph{e.g.} \cite{Hama:2011ea,Alday:2013lba,Closset:2013sxa}) or some version of the index theorem (see \emph{e.g.} \cite{Pestun:2007rz,Gomis:2011pf,Drukker:2012sr,Benini:2012ui,Nishioka:2014zpa}). We have shown that in our set-up all three methods yield the same results, provided a number of caveats are appropriately taken into account.

Perhaps the most elegant and succinct method is the one of the index theorem. This method begins by using off-shell
supersymmetry to pair up all the fields of the theory in doublets of the supercharge~$Q$, and then looks for another
pairing~$D_{10}$ of the doublets themselves. The super-determinant computation is captured by an index of this 
operator~$D_{10}$ which, quite remarkably reduces to a simple quantum-mechanical computation at the set of 
fixed points of the~$U(1)$ action generated by~$Q^{2}$. 
The method of pairing of eigenvalues is based on the idea that supersymmetry pairs up most of the bosonic and fermionic eigenmodes, leaving a net contribution arising from ``unpaired'' 
modes that obey some ``shortening condition''. We have implemented this method by using a set of twisted variables analogous to those introduced in~\cite{Closset:2013sxa}. However, we have found some key novelties: on one hand, 
 by explicitly solving for the unpaired eigenmodes, we have observed that after requiring that they are regular in the bulk, we cannot require that they are appropriately normalizable at infinity. This phenomenon may be analogous to the one discussed in~\cite{Keeler:2016wko}. 
 On the other hand, the eigenmodes contributing to the final result did not arise in pairs of complex conjugate modes, but rather as isolated ``holomorphic" or ``anti-holomorphic" modes: this is a crucial difference with respect to what happens for example on $S^3$~\cite{Hama:2011ea,Alday:2013lba} or $S^1\times S^3$~\cite{Closset:2013sxa}, and ultimately is responsible for  the appearance of the square root in the formula of the partition function, as we discussed in Section~\ref{ssec:UnpairedEigenmodes}.
Finally, to carry out the technique of the heat kernel we had first to incorporate appropriately the effect of the 
various background gauge fields (R-symmetry, flavor symmetry, and the localized dynamical gauge field), and most importantly we proposed a recipe to regularize the formal  integrals as to respect holomorphy of the final result. We interpret this as strong evidence  that our regularization method does not break supersymmetry. 

We briefly discussed, in Section~\ref{sec:summary}, 
an interpretation of our results for the one-loop determinant as indicating the presence of 
holomorphic currents in a putative holographic boundary theory. Of course the observables that we compute here 
is not meant to be the holographic computation of any boundary SCFT$_{2}$ directly -- such a computation would require
the inclusion of the supergravity fields in AdS$_{3}$. Nevertheless, it is tempting to think of our results 
as a piece of the full answer in such a holographic computation.

We expect that it will be possible to refine our results, tying up some loose ends, and  that there will be a number of extensions that could be explored in the future. In the concluding part of this paper, we will make some 
comments on a few problems that we have not discussed so far. 

One the issues that we think should be addressed more carefully is the derivation of the localization locus. The hyperbolic nature 
of our space here leads to non-positive-definite localizing actions, as we discussed at some length in Section~\ref{sec:Asymptotics}. 
As a consequence, we could not prove that the localization locus coincides exactly with the space of solutions of the off-shell 
BPS equations. A closely related issue is that of the choice of reality conditions on the various fields. 
We suspect that the first-principles construction of Euclidean off-shell supergravities~\cite{deWitReys} may help elucidate these issues.

We would like to make some remarks about a very close relative to the case that we studied in this paper, namely the hyperbolic space $\IH^3$, without any quotient. In this case, the background is supersymmetric, without the need to include any background R-symmetry gauge field and the space has the topology of the three-ball, with a (round) $S^2$ at the conformal boundary\footnote{In this case it is more convenient to use a different coordinate system, see \emph{e.g.} \cite{Camporesi:1994ga}.}. In principle, all the ideas and methods that we used to study the case of $\IH^3_\tau$ can be adapted to this case. However, it is not difficult to convince oneself that in this case the one-loop determinants will  be the exponential of simple polynomial functions of the parameters (more precisely of the masses of the various kinetic operators), up to a divergent factor proportional to the volume of the space. For  scalar and vector fields
the results can be found for example in \cite{Giombi:2008vd}. We have checked that incorporating supersymmetry does not alter this generic feature, and  for this reason, we have not pursued all the details here. In principle, the one-loop determinant of the chiral multiplet can be extracted from the limit of large $\tau_2$ of our expressions, leading to a trivial factor. The contributions associated to the vector multiplet should be revisited after studying the new supersymmetric asymptotics.

It will not escape the attention of the reader that in this paper we have not discussed the use of our results to test  non-perturbative dualities between different field theories.  For example, in 
\cite{Kapustin:2010xq, Jafferis:2011ns} it has been checked (either analytically or numerically) that, upon an appropriate mapping of parameters, the localized partition function on the three-sphere match between pairs of dual theories. We have looked at number of simple cases which are known to work for the case of the partition function on $S^3$, and checked that our partition functions (\ref{ZFinalCS}) do not match on the two sides. There could be different (speculative) reasons for this:  one option is that on spaces such as the one we considered, in order to 
test dualities, one needs to consider more general boundary conditions, including degrees of freedom living on the asymptotic boundary (see \emph{e.g.}~\cite{Gaiotto:2008ak}) .
Another possibility is that the dualities will hold only after choosing appropriate integration contours, in the spirit of \cite{Beem:2012mb}. 
It would be very interesting to shed light onto this conundrum. 

Finally, let us mention some promising extensions of our results. It was shown in \cite{Dumitrescu:2012ha} that $\IH^3\times S^1$ is a  supersymmetric background of four-dimensional rigid new minimal supergravity (in fact, preserving four supercharges). Based on the results of our paper, and on those in \cite{Assel:2014paa}, we expect that it should be straightforward to compute the localized partition function of four dimensional supersymmetric 
gauge theories on $\IH^3_\tau\times S^1$. Moreover, it is known that  supersymmetric field theories may be defined on AdS$_4$ (see \emph{e.g.}  \cite{Festuccia:2011ws}), suggesting that 
another likely-looking case to study is the partition function of ${\cal N}=2$ supersymmetric field theories on AdS$_4$, perhaps following in the footsteps of \cite{Pestun:2007rz}.

We also hope that our work will be useful towards the more ambitious goal of computing the exact  partition function of supergravity theories defined on spaces containing an
 $\IH^3_\tau$ factor.

\subsection*{Note added} While we were about to submit this paper to the arXiv,  the paper \cite{David:2016onq} appeared. 
It discusses localization of ${\cal N}=2$ supersymmetric Chern-Simons  theory on the non-compact space AdS$_2\times S^1$. 
While there may be interesting relations, this space is different from the one discussed in our paper, and there is no evident overlap between the two papers.

\subsection*{Acknowledgments}

We wish to thank Ofer Aharony, Cyril Closset, Stefano Cremonesi, Tudor Dimofte, Rajesh Gopakumar, Rajesh Gupta, 
Cynthia  Keeler, Mauricio Romo, and Cristian Vergu for discussions.
B.~A., D.~M.,  and D.~Y.~are supported by the ERC Starting Grant N.~304806, ``The Gauge/Gravity Duality and Geometry in String Theory''. S.~M.~is supported by the EPSRC First Grant UK EP/M018903/1, and by 
the ERC Consolidator Grant N.~681908, ``Quantum black holes: A macroscopic window into the microstructure of 
gravity''.

\appendix

\section{Conventions and useful identities}
\label{conventions:appendix}

We adopt the conventions of \cite{Closset:2012ru}, except for the spin connection which we take with the opposite sign compared to them. 
Spinors indices are raised and lowered acting on the left with $\epsilon_{\alpha\beta}$ and $\epsilon_{\alpha\beta}$ with $\epsilon^{12} = \epsilon_{21} = 1$. 
Spinor bilinears are defined as 
\begin{align}
\psi \chi &= \psi^\alpha \chi_\alpha \, .
\end{align}
The gamma matrices are
\begin{align}
\gamma^1 & = \sigma^3 \, , \quad \gamma^2  = -\sigma^1 \, , \quad \gamma^3  = -\sigma^2 \, , 
\end{align}
with $\sigma^a$ the Pauli matrices.
The spin connection is defined by 
\begin{align}
de^a + \omega^a{}_b \wedge e^b = 0 \, ,
\end{align}
and the covariant derivative on spinors is 
\begin{align}
\nabla_{\mu} \zeta = \p_\mu \zeta + \frac{i}{4} \omega_\mu {}^{ab} \epsilon_{abc} \gamma^c \zeta \, . 
\end{align}
We define the hermitian conjugation on spinors as
\begin{align}
\lp \psi^\dagger \rp^\alpha &= \lp \psi_\alpha \rp^\ast \, .
\end{align}
Fierz identities for commuting  (Grassmann-even) spinors
\begin{align}
& (\chi_1 \sigma^a \chi_2)(\chi_3 \sigma_a \chi_4) =  2 (\chi_1\chi_4)(\chi_3  \chi_2) - (\chi_1\chi_2)(\chi_3  \chi_4) \label{Fierz1}\\
& 2 (\chi_1 \sigma^a \chi_2)(\chi_3 \sigma^b \chi_4) - (\chi_3 \sigma^a \chi_2)(\chi_1 \sigma^b \chi_4) - (\chi_1 \sigma^a \chi_4)(\chi_3 \sigma^b \chi_2) \label{Fierz2} \\
& =  -2 \delta^{ab} (\chi_1\chi_2)(\chi_3  \chi_4) + 2 \delta^{ab} (\chi_1\chi_4)(\chi_3  \chi_2) + i \epsilon^{abc} (\chi_1 \sigma_c \chi_4)(\chi_3  \chi_2) - i \epsilon^{abc} (\chi_1 \chi_4)(\chi_3 \sigma_c \chi_2) \, , \no
\end{align}
with $ a, b \in \{ 1,2,3 \}$ in the second identity.

The Killing vectors defined in Section~\ref{sec:Background} as bilinear of the spinors $\zeta, \ti\zeta$ obey the following identities:
 \begin{align}
 & K^\mu K_\mu = - \frac 12 \, P^\mu \ti P_\mu = 1 \, , \quad K^\mu P_\mu = K^\mu \ti P_\mu = P^\mu P_\mu = \ti P^\mu \ti P_\mu = 0 \,, \no\\
 &  i \epsilon^{\mu\nu\rho} P_\nu K_\rho =  P^\mu \,, \quad  i \epsilon^{\mu\nu\rho} \ti P_\nu K_\rho =  - \ti P^\mu \,, \quad  i \epsilon^{\mu\nu\rho} P_\nu \ti P_\rho =  2 \, K^\mu  \,, \no\\
 & 2 \, K_{[\mu}  P_{\nu]} = i \epsilon_{\mu\nu\rho} P^\rho  \,, \quad  2 \, K_{[\mu}  \ti P_{\nu]} = - i \epsilon_{\mu\nu\rho} \ti P^\rho  \,, \quad  2 \, P_{[\mu} \ti P_{\nu]} = - 2i \epsilon_{\mu\nu\rho} K^\rho \no\\
 & D_\mu X_\nu = - D_\nu X_\mu = \frac{i}{L} \epsilon_{\mu\nu\rho}  X^\rho \quad  \textrm{for} \quad X_\mu = K_\mu, P_\mu, \ti P_\mu \,.  
 \end{align}

\section{Supersymmetry transformations of twisted fields}
\label{app:SusyTwistedFields}

In this appendix we provide the supersymmetry transformations in the language of the twisted variables.
We introduce the complex parameters $u, \ti u$ to parametrize a generic supersymmetry transformation $\delta \equiv u \delta_\zeta + \ti u \delta_{\ti\zeta}$. 

The supersymmetry transformations of the vector multiplet twisted fields $X^{\pm}, X^0$, $\Sigma$, $\Lambda^{\pm}$, $\Lambda^0$, $\Theta$, $D^0$ are 
\be
\begin{aligned}
& \delta X^+ = 2 \ti u \Lambda^{+} \,, \quad  \delta X^- = 2 u \Lambda^{-} \,, \\
& \delta \Lambda^+ = -i u \lp \wat\scL_K X^+ + \wat\scL_P \Sigma - [\Sigma,X^+] - \frac 2L X^+ \rp \,, \\
& \delta \Lambda^- = -i \ti u \lp \wat\scL_K X^- + \wat\scL_{\ti P} \Sigma - [\Sigma,X^-] + \frac 2L X^- \rp \,, \\
& \delta X^0 = u (\Lambda^0 - i \Theta ) + \ti u (\Lambda^0 + i \Theta) \,, \quad \delta \Sigma = 0 \,, \\
& \delta (\Lambda^0 - i \Theta ) = -2 i \ti u \lp D^0 - \frac 12 \wat\scL_K(\Sigma - X^0) - \frac 12 [\Sigma,X^0] \rp\,, \\ 
& \delta (\Lambda^0 + i \Theta ) = 2 i u \lp D^0 + \frac 12 \wat\scL_K(\Sigma - X^0) + \frac 12 [\Sigma,X^0] \rp \,, \\ 
& \delta \lp D^0 - \frac 12 \wat\scL_K(\Sigma - X^0) - \frac 12 [\Sigma,X^0] \rp = u \lp \wat\scL_K(\Lambda^0 - i\Theta) -[\Sigma,\Lambda^0 - i\Theta] \rp  \,, \\
& \delta \lp D^0 + \frac 12 \wat\scL_K(\Sigma - X^0) + \frac 12 [\Sigma,X^0] \rp = - \ti u \lp \wat\scL_K(\Lambda^0 + i\Theta) -[\Sigma,\Lambda^0 +i\Theta] \rp \,,
\end{aligned}
\ee
where $\wat\scL_Y \equiv Y^\mu(\nabla_\mu - i q_R A_\mu)$ is \emph{not} covariant with respect to the gauge connection $\scA_\mu$.

The supersymmetry transformations of the chiral multiplet twisted fields are
\be\ba\label{TwistSusyTransfoChi1}
\delta \phi  &=  u \sqrt 2 \, C \,, \cr 
\delta B &=   u\sqrt 2 \, F + i \ti u \sqrt 2 \scL_{\ti P} \phi  \,,\cr
\delta  C  &=    i \ti u \sqrt 2 \Big[ \big( \mm + \frac{r}{L} \big) \phi  -   \scL_{K} \phi \Big] \,, \cr
\delta  F  &=      i \ti u \sqrt 2 \Big[ \big( \mm + \frac{r-2}{L} \big)  B -   \scL_{K} B   -  \scL_{\ti P} C \Big] \,,
\ea\ee
and 
\be\ba\label{TwistSusyTransfoChi2} 
\delta \ti\phi  &=  \ti u \sqrt 2 \, \ti C  \,,  \cr
  \delta\ti B  &=  i u \sqrt 2   \scL_{P} \ti\phi  +  \ti u\sqrt 2 \, \ti F  \,, \cr
\delta \ti C  &=  -  i u\sqrt 2  \Big[ \big( \mm + \frac{r}{L} \big)  \ti\phi  + \scL_{K} \ti\phi   \Big] \,, \cr
\delta\ti F  &=   -  i u \sqrt 2 \Big[ \big( \mm + \frac{r-2}{L}\big)  \ti B + \scL_{K} \ti B   + \scL_{P} \ti C \Big] \,.
\ea\ee
The supercharge used for localizing the theory is $Q = \frac 12 (\delta_\zeta + \delta_{\ti\zeta})$, corresponding to $u=\ti u=\frac 12$.

\section{Supersymmetry computations}
\label{app:susycomputations}

We provide here a few intermediate computations leading to the relations \eqref{deltaV1vec}:
\begin{align}
& \dd_{\zeta} (\tr \ \ti\lambda \lambda) = \tr \ - \frac i2 \epsilon^{\mu\nu\rho}( \ti\lambda \gamma_\rho \zeta) \scF_{\mu\nu} + i (\ti\lambda \zeta) (D+ \sigma H) - i (\ti\lambda \gamma^\mu \zeta) D_\mu \sigma , \no\\
& \dd_{\zeta} ( \tr \ 2i D \sigma) = \tr \ 2i (\zeta\gamma^\mu D_\mu \ti\lambda) \sigma -  i (\zeta \ti\lambda) \lp 2 D -  \sigma H \rp  ~,
\no\\
& \dd \scF_{\mu\nu} = 2 i \lp \zeta \gamma_{[\mu} D_{\nu]}\ti\lambda + \ti\zeta \gamma_{[\mu} D_{\nu]} \lambda  \rp + \epsilon_{\mu\nu\rho} H \lp \zeta \gamma^\rho \ti\lambda + \ti\zeta \gamma^\rho \lambda \rp  \, .
\end{align}
We provide also intermediate computations leading to the relations \eqref{deltaV1}:
\begin{align}
 \delta_\zeta \delta_{\ti \zeta} \left( -\frac{1}{2} \ti BB \right)
 &= -\CL_P \ti\phi \CL_{\ti P}\phi -\ti FF  \nonumber\\
 &\quad +i \CL_K\ti B B +i \CL_{P}\ti CB -i\ti B \CL_{\ti P} C +i\left(m+\frac{r-2}{L}\right) \ti BB  ~, \\
 \delta_\zeta \delta_{\ti \zeta} \left( -\frac{1}{2} \ti CC \right)
 &= -\CL_K\ti\phi\CL_K\phi +\left(m+\frac{r}{L}\right) \CL_K\ti\phi\phi
 -\left(m+\frac{r}{L}\right)\ti\phi \CL_K\phi +\left(m+\frac{r}{L}\right)^2 \ti\phi\phi  \nonumber\\
 &\quad  +i \ti C\CL_K C -i\left(m+\frac{r}{L}\right)\ti CC ~   \\ 
 \delta_\zeta \delta_{\ti \zeta} \left( i\ti \phi \CL_{K} \phi \right)
 &= 2\CL_K\ti\phi\CL_K\phi +2\left(m+\frac{r}{L}\right)\ti\phi \CL_K\phi -2i\ti C \CL_K C  ~, \\
 \delta_\zeta \delta_{\ti \zeta} \left( -\frac{i}{L} \ti \phi \phi \right)
 &= -\frac{2}{L} \left(m+\frac{r}{L}\right) \ti\phi\phi -\frac{2}{L} \CL_K\ti\phi\phi +\frac{2i}{L}\ti CC ~ .
\end{align}

\bibliography{Newbib}
\bibliographystyle{JHEP}

\end{document}